\documentclass[a4paper,fleqn]{cas-dc}

\usepackage[english]{babel}

\usepackage[T1]{fontenc}
\usepackage{newunicodechar}

\usepackage{balance}

\usepackage[square,numbers,sort]{natbib}
\newcommand*{\doi}[1]{\href{http://dx.doi.org/#1}{#1}}

\usepackage{tikz}
\newcommand{\circled}[1]{\tikz[baseline=(myanchor.base)] \node[circle,fill=.,inner sep=1pt] (myanchor) {\color{-.}\bfseries\footnotesize #1};}

\newcommand{\squared}[1]{%
\tikz[baseline=(myanchor.base)] 
\node[rectangle,fill={rgb,255:red,255;green,175;blue,175},inner sep=2pt,minimum size=10pt,align=center] (myanchor) {\color{black}\bfseries\footnotesize #1};}

\usepackage{pifont}% http://ctan.org/pkg/pifont

\usepackage{float}
\usepackage{xparse}
\makeatletter
\DeclareDocumentEnvironment{appendixfig}{}{
   \par\medskip\noindent
   \begin{minipage}{\textwidth}
   \def\@captype{figure}
   \centering
}
{
\end{minipage}
\par\bigskip
}
\makeatother

\usepackage{longtable}
\usepackage{footnotehyper}
\makesavenoteenv{longtable}
\usepackage{geometry}

\usepackage[section]{placeins} % Prevent figures from being moved across sections

\usepackage[hypcap=false]{caption} 

\usepackage{subcaption}
\usepackage{nameref}

\usepackage{afterpage}

\usepackage{url}
\RequirePackage{hyperref}
\PassOptionsToPackage{hyphens}{url}\usepackage{hyperref}
\hypersetup{
 colorlinks=true,
 linkcolor=blue,
 filecolor=magenta, 
 urlcolor=cyan,
 pdftitle={Overleaf Example},
 pdfpagemode=FullScreen,
 }
\usepackage{xurl}
\hypersetup{pdfauthor=whatever}

\usepackage{amsmath}
\usepackage[scr=esstix,cal=boondox]{mathalfa} 
\usepackage{pifont}
\usepackage{amssymb}

\usepackage[inline]{enumitem}
\usepackage[normalem]{ulem} 

\usepackage{tabularx}
\usepackage{bm}
\usepackage{mathtools} 

\usepackage{booktabs, makecell}
\usepackage{adjustbox}

\usepackage{setspace}

\usepackage{multirow}
\usepackage{array,colortbl}
\usepackage{diagbox}
\usepackage{siunitx}
\usepackage{graphicx}
\usepackage{stfloats}

\usepackage{savesym}
\savesymbol{fax}
\savesymbol{Hermaphrodite}
\usepackage{marvosym}
\restoresymbol{MARV}{fax}
\restoresymbol{MARV}{Hermaphrodite}

\usepackage{threeparttable, tablefootnote}
\usepackage{arydshln}
\usepackage[capitalize,noabbrev]{cleveref} 

\usepackage{soul}

\begin{document}
\let\WriteBookmarks\relax
\def\floatpagepagefraction{1}
\def\textpagefraction{.001}

\shorttitle{Designing the virtual CAT: A digital tool for algorithmic thinking assessment in compulsory education}

\shortauthors{G Adorni et~al.}

\title [mode = title]{Designing the virtual CAT: A digital tool for algorithmic thinking assessment in compulsory education}                      

% \tnotetext[1]{This document is the results of the research
%   project funded by the National Science Foundation.}
%
%\tnotetext[2]{The second title footnote which is a longer text matter
%   to fill through the whole text width and overflow into
%   another line in the footnotes area of the first page.}


\author[1]{Giorgia Adorni}[type=editor,
%                          role=Researcher,
                           orcid=0000-0002-2613-4467]
\cormark[1]
\ead{giorgia.adorni@usi.ch}
%\ead[url]{www.cvr.cc, cvr@sayahna.org}
\credit{Conceptualization, Methodology, Software, Validation, Formal analysis, Investigation, Resources, Data curation, Writing -- original draft \& review \& editing, Visualization, Supervision}
\affiliation[1]{organization={Università della Svizzera italiana (USI), Dalle Molle Institute for Artificial Intelligence (IDSIA)},
    addressline={Polo universitario Lugano - Campus Est, Via la Santa 1}, 
    postcode={6962},
    city={Lugano-Viganello},
    country={Switzerland}}

\author[2]{Alberto Piatti}[style=chinese]
\affiliation[2]{organization={University of Applied Sciences and Arts of Southern Switzerland (SUPSI), Department of Education and Learning (DFA)},
    addressline={Piazza S. Francesco 19}, 
    postcode={6600}, 
    city={Locarno},
    country={Switzerland}}
\ead{alberto.piatti@supsi.ch}
\credit{Conceptualization, Validation, Writing -- review \& editing, Supervision, Project administration, Funding acquisition}

\cortext[cor1]{Corresponding author}
%\cortext[cor2]{Principal corresponding author}

%\fntext[fn1]{This is the first author footnote.}
%\fntext[fn2]{Another author footnote, this is a very long footnote and
%  it should be a really long footnote. But this footnote is not yet
%  sufficiently long enough to make two lines of footnote text.}

%\nonumnote{This note has no numbers. In this work we demonstrate $a_b$
%  the formation Y\_1 of a new type of polariton on the interface
%  between a cuprous oxide slab and a polystyrene micro-sphere placed
%  on the slab.
%  }

\begin{abstract}
% \textcolor{red}{REWRITE}
% In today's digital era, holding algorithmic thinking (AT) skills is crucial, not only in computer science-related fields. 
% These abilities enable individuals to break down complex problems into more manageable steps and create a sequence of actions to solve them.
Algorithmic thinking (AT) is a critical skill in today's digital society, and it is indispensable not only in computer science-related fields but also in everyday problem-solving. 
As a foundational component of digital education and literacy, fostering AT skills is increasingly relevant for all students and should become a standard part of compulsory education. However, successfully integrating AT into formal education requires effective teaching strategies and robust and scalable assessment procedures.

In this paper, we present the design and development process of the virtual Cross Array Task (CAT), a digital adaptation of an unplugged assessment activity aimed at evaluating algorithmic skills in Swiss compulsory education.
The development process followed iterative design cycles, incorporating expert evaluations to refine the tool's usability, accessibility and functionality.

A participatory design study played a dual role in shaping the platform. 
First, it gathered valuable insights from end users, including students and teachers, to ensure the tool's relevance and practicality in classroom settings. 
Second, it facilitated the collection and preliminary analysis of data related to students' AT skills, providing an initial evaluation of the tool's assessment capabilities across various developmental stages.
This was achieved through a pilot study involving a diverse group of students aged 4 to 12, spanning preschool to lower secondary school levels.

% A participatory design study further informed the tool's evolution by gathering insights from end users, including students and teachers, to ensure the platform meets practical classroom needs.
% To assess the platform's effectiveness, the pilot study conducted within the participatory sessions aimed at gathering also informations regarding  reliably assess AT skills across various developmental stages.
% It involved a diverse group of students aged 4 to 12, spanning preschool to lower secondary school levels. 
% [chiarificare che il participatory design ha una doppia funzione: la prima gathering insights from end users, la seconda analizzare i dati relative all'assessment di AT degli studenti e fare un analisi preliminre...]

The resulting instrument features multilingual support and includes both gesture-based and visual block-based programming interfaces, making it accessible to a broad range of learners. 
Findings from the pilot study demonstrate the platform's usability and accessibility, as well as its suitability for assessing AT skills, with preliminary results showing its ability to cater to diverse age groups and educational contexts. Additionally, the CAT has proven capable of handling large-scale, automated assessments, offering a scalable solution for integrating AT evaluation into education systems.

\end{abstract}

% Keywords
% Each keyword is separated by \sep

\maketitle
% \linenumbers

% \tableofcontents
% \clearpage

\section{Introduction}\label{sec:intro}
% Provide an overview of the topic: assess the development of computational thinking skills 
% \hl{Rewrite and shorten by focusing more on the problem and not talk about computational thinking as if it were something new, but take it for granted. Particularly shorten the first paragraph.}

% \hl{1. Definition of CT}
% Start with a brief definition and overview of CT as a concept, explaining its significance in education.
Computational thinking (CT) is a cognitive process that involves the coordinated application of interrelated skills, including breaking down complex problems, identifying recurring patterns, abstracting relevant information, and formulating algorithms to devise effective solutions \citep{wing2006computational,Wing2008,wing2011research,adorni2023,DENNING2021,denning2019computational,Lodi2021,Tedre2016,papert1980,Papert2000}. 
In the academic context, this structured approach fosters systematic and logical reasoning, positioning CT as a widely acknowledged 21st-century skill that equips students to tackle interdisciplinary challenges and navigate the demands of a rapidly evolving digital landscape.

% \hl{2. Definition of AT}
% Define AT, emphasising that it's a subset of CT focused on creating and following step-by-step procedures
Algorithmic thinking (AT), as a fundamental aspect of the broader concept of CT \citep{Syso2015,futschek2006algorithmic,Lodi2021}, focuses on designing step-by-step procedures or algorithms that computers or humans can execute to solve problems and achieve specific outcomes systematically \citep{seehorn2011csta,barr,poulakis2021computational,futschek2006algorithmic,adorni2023}. 
In educational settings, the integration of AT into curricula aims to enhance students' CT skills by fostering procedural problem-solving capabilities.

% \hl{3. Differences between CT and AT}
% Clarify the distinction between CT and AT, highlighting that while CT covers a broad range of cognitive processes, AT specifically deals with procedural problem-solving
% \hl{% [REVISE THIS...]
% While CT encompasses a broad range of cognitive processes applicable across various domains, AT specifically emphasizes the logical structuring and execution of tasks by developing algorithms} \citep{Syso2015,futschek2006algorithmic,Lodi2021}. 
% \hl{% This distinction is crucial for educational practitioners, as a nuanced understanding of both CT and AT can inform curriculum design, enhancing students' problem-solving capabilities while fostering a more profound comprehension of computational methodologies. Integrating both concepts into educational frameworks not only equips students with essential skills for the digital age but also cultivates an environment conducive to interdisciplinary learning.
% }

% \hl{4. Importance of AT}
% Discuss the significance of AT in education, explaining why it's crucial for students to develop these skills, especially in a digital age
AT is becoming an indispensable skill in today's digital era, transcending its origins in computer science (CS) \citep{yadav2014computational, yildiz2020ideal,amini2016bedside,jocz2023motivating,wing2006computational}. 
In the context of education, AT has gained significant importance as it empowers individuals to excel in various personal and professional domains by enhancing problem-solving abilities, logical reasoning, and creativity \citep{kules2016computational,olkhova2022development}. 
The relevance of AT is evident in its integration into educational systems worldwide, where it is increasingly recognised as a foundational skill for understanding essential concepts like algorithms and data structures \citep{korkmaz2019adapting,oluk2016comparing,dagiene2016it,kong2019introduction}. 
This shift signifies a broader understanding of education that extends beyond traditional subjects, emphasising the need for students to acquire skills relevant to an increasingly technology-driven society.

The growing recognition of AT in education also underscores the necessity for reliable assessment instruments to measure students' development in this area. 
Such tools can facilitate the identification of learning gaps and allow educators to tailor instructional methods to meet the diverse needs of learners \citep{ezeamuzie2021computational,scherer2019the,pilotti2022is}. 
Moreover, the inclusion of AT in curricula is increasingly aligned with global educational standards, reflecting an urgent need for curricula that prepare students for future challenges, including those posed by automation, artificial intelligence, and rapidly evolving technology.

% \hl{5. Focus on AT in the context of this article}
% Conclude the introduction by narrowing the focus to AT, setting the stage for the specific research and discussion in the paper, the need of a comprehensive assessment tool
In response to the evolving educational landscape, which increasingly prioritises AT as a critical competency, a comprehensive assessment tool to evaluate students' proficiency in this domain is compellingly necessary \citep{martinez2022assessing}. 
Prioritising the assessment of these skills is essential to monitor their ongoing establishment as a foundational component of compulsory education.

This article aims to present a tool for assessing AT that can be universally applied across different educational settings and age groups and provides automated, large-scale assessments.
% \textcolor{red}{TO BE REVISED IF THE STRUCTURE OF THE ARTICLE IS CHANGED}
The paper starts with a review of literature on global, European, and Swiss trends in integrating CT and AT in education, as well as assessment approaches for AT. It then introduces the Cross Array Task (CAT), discussing its transition from an unplugged to a virtual format. The methodology covers the process from defining objectives and developing prototypes to expert evaluations and participatory design. A technical overview of the implementation is provided, followed by details on prototype development. 
Finally, the paper presents preliminary data analysis and concludes with a discussion of study limitations and future work.

\section{Related works}\label{sec:related_work}

% \textcolor{red}{THIS IS A COMPLETELY NEW SECTION}

% \hl{1. Historical evolution of AT}
% Start with the historical development of AT, explaining how it emerged and evolved within educational contexts
% Within educational contexts, AT has evolved significantly over the past few decades. 
As the world continues to shift towards digitalisation and technology becomes increasingly integrated into everyday life, AT has emerged as a crucial skill in modern education\citep{wing2006computational,yadav2014computational}.
Once a niche concept within computer science (CS), AT is now recognised globally as essential for equipping students with the problem-solving and logical reasoning skills required in the 21st century \citep{wing2011research,Wing2008,wing2014computational}. 
This shift and the global recognition of AT's importance have spurred efforts to formally integrate it into educational systems, leading to its inclusion in numerous national curricula \citep{bocconi2016developing,bocconi2022reviewing,weintrop2021assessing}. 
By equipping students with skills beyond traditional subjects, AT prepares future generations for technology-driven careers and fosters adaptability in a rapidly evolving world \citep{Voogt2015}.

This related works section explores AT's role in education and provides a foundation for understanding its assessment across various educational settings.

% \hl{2. Implementation and integration of AT in worldwide educational institutions}
% Discuss how AT has been incorporated into education systems around the world, highlighting different approaches and the global importance of AT.

\subsection{Worldwide integration of CT and AT in education}

Across the globe, nations have embraced digital technologies and AT within their curricula, underscoring the growing demand for CT skills in education.

\subsubsection{Global trends}
In the \textit{United States}, the emphasis on CT began in the early 2000s, particularly through initiatives like the Next Generation Science Standards (NGSS)\citep{schwarz2017helping,Bybee2014,wilkerson2017using,ngss} and the CS for ALL Students initiative \citep{csforall,obama}, which aimed to ensure CS education is accessible to all students from early education onward.
Other countries are also advancing in this area.
\textit{New Zealand}, \textit{Australia}, \textit{South Korea} and \textit{Japan} from 2015 have started integrating digital technologies, CT and AT across STEM subjects at all educational levels, focusing on themes like algorithms and problem-solving, making programming a compulsory subject \citep{newzeland,japan,Australia,bocconi2016developing,park2016preparing}.
Similarly, \textit{Singapore}, under the 2014 Smart Nation initiative led by the Prime Minister to promote early programming exposure \citep{Hoe2016,smartnation}, launched a CT framework in 2016, introduced a computing subject focused on programming and algorithms in secondary schools by 2017, and mandatory CT and coding program for upper primary students by 2020 \citep{singapore,bocconi2016developing,bocconi2022reviewing}.
In \textit{Canada} (British Columbia) CT has been incorporated into middle school subjects, with plans for broader application at the secondary level \citep{BC, bocconi2016developing}.

\subsubsection{Progress in Europe}
% \subsection{Implementation and integration of AT in Europe}
Several European countries have significantly advanced in integrating CT and AT into their compulsory education systems. 
While some have incorporated these skills across all compulsory educational levels, others have focused primarily on secondary education.
The degree of integration and the scope of the curricula reforms vary widely across the continent, with some countries adopting a holistic, cross-curricular approach, while others emphasise CS or technology education as separate subjects \citep{bocconi2016developing,bocconi2022reviewing}.
% In Europe, the integration of AT has progressed at varying paces. 
% A study by Bocconi et al. (2022) \citep{bocconi2022reviewing} revealed that 25 out of 29 European countries have included CT and AT in their compulsory education curricula, primarily between 2016 and 2021.

The pioneers in integrating CT and AT across both primary and secondary levels have significantly influenced the approaches of subsequent nations. 
Among them, \textit{England} was one of the earliest to make CT mandatory, incorporating it into its national curriculum in 2014 as a separate subject \citep{uk2013national}. %This early adoption set a precedent for other countries
\textit{France} followed closely, integrating CT within existing subjects such as mathematics and technology in 2015 \citep{france}.
\textit{Finland} incorporated CT and AT in 2016 as a cross-curricular theme, later extending its integration within subjects like mathematics, crafts, and environmental studies by 2022 \citep{finnish}.

In the years following these initial pioneering efforts, several other countries have embraced CT and AT, albeit at different rates and in various formats. 
Countries such as \textit{Malta}, \textit{Slovakia}, \textit{Poland}, \textit{Portugal}, \textit{Croatia}, \textit{Greece}, \textit{Austria}, and \textit{Hungary} have integrated CT/AT as a separate subject, primarily through informatics or CS courses, emphasizing the importance of computational skills as a distinct area of study with dedicated instructional time \citep{poland,Croatia,grece,hungary,Malta,Slovakia2024,slovakia,AUSTRIA,austria2024}.
In contrast, \textit{Sweden}, \textit{Norway} and \textit{Lithuania} have opted to embed CT within existing subjects, such as mathematics, science, and the arts, promoting an interdisciplinary model that fosters CT across various academic domains \cite{norway,Lithuania,sweden}.
In \textit{Cyprus}, \textit{Luxembourg}, and \textit{Serbia}, CT is integrated into primary education primarily within other subjects, while in secondary education, it is structured as a separate subject, reflecting a flexible and context-specific approach to embedding CT across different educational levels \citep{cyprus,Luxembourg,serbia}.  

Despite notable advancements in various countries, several have achieved only partial integration of CT and AT. Specifically, \textit{Ireland}, \textit{Romania}, and \textit{Scotland} have incorporated these skills into secondary education, while formal integration at the primary level continues to be lacking \citep{ireland2,Ireland,Romania,scotland}.

Additionally, several countries have made little to no progress in integrating CT and AT into their educational systems. In \textit{Denmark}, \textit{Slovenia}, \textit{Italy}, the \textit{Czech Republic}, the \textit{Netherlands}, and \textit{Spain}, the situation varies, with most of these countries at the drafting stage of curricula or strategic plans for future actions \citep{denmark,Italy,Czech,slovenia,Netherlands,bocconi2016developing,bocconi2022reviewing}.
For instance, Denmark has yet to integrate CT but has initiated a pilot program \citep{werner2012children}, while Italy recognizes CT as a key topic but lacks formal integration in its national curriculum.

The situation in Belgium further illustrates this complexity, as integration depends on specific regions. In Flanders, CT has been integrated as part of a separate subject, while Wallonia plans to address it as a compulsory subject for primary and lower secondary schools \citep{belgium}.

\subsubsection{The Swiss approach}
% \hl{3. AT as part of the Swiss curriculum}
% Focus on the specific case of Switzerland, explaining how AT is integrated into the national curriculum and its importance in Swiss education.

% Switzerland integration of AT into the national curriculum reflects a strategic effort to prepare students for the demands of a digital society. 

The Swiss educational system has progressively integrated AT and CT into its curriculum, adapting to the specific needs of its diverse linguistic regions. These skills are embedded within various subjects, such as mathematics and CS, through activities like coding, algorithmic exercises, and robotics projects, ensuring that students acquire essential skills from an early age \citep{bocconi2016developing,bocconi2022reviewing}.

%   SWITZERLAND FROM 
In the German-speaking region, the integration of CT began around 2014, with competencies such as coding and programming incorporated into the curricula of primary and lower secondary schools. At the upper secondary level, these skills are formalised within the national curriculum framework for non-vocational schools, ensuring a comprehensive acquisition of computational skills throughout the educational journey \citep{des2016lehrplan}.
In the French-speaking region, CT is taught through the Plan d'études romand (PER) under the subject MITIC (Média, Image, Technologie de l'Information et de la Communication), implemented since 2015 \citep{per}. Within the framework, students engage in activities that require them to analyse problems, devise solutions, and implement basic programs, reinforcing CT skills from early education onward. Additionally, the subject ``Media and Informatics'' introduces CT as a core component, fostering logical thinking and problem-solving abilities.
Similarly, the Italian-speaking region incorporates AT and CT into subjects like mathematics and CS, with a particular emphasis on coding, problem-solving, and robotics from an early age. This approach promotes interdisciplinary learning, enabling students to apply these skills across various subjects and ensuring their preparedness for the technological demands of future academic and professional environments \citep{cantone2015piano,cantone2022piano}.
% Several initiatives have been established to attract girls into CS, such as "Programming Workshops for Girls" and "Swiss Tech Ladies". These efforts reflect Switzerland’s commitment to developing future generations equipped to engage with the complexities of an increasingly digital world, ensuring that students not only acquire technical skills but also develop the critical thinking and problem-solving abilities essential for success in the 21st century.
% \hl{Overall, the Swiss educational approach provides consistent exposure to CT and AT across all educational levels, aligning with international standards while catering to the unique linguistic and cultural diversity within the country.}

% \hl{5. Variations in the approaches to teach and assess AT}
% Discuss different methods used globally to teach and assess AT, including unplugged activities, coding exercises, and more.

\subsection{Assessment approaches for AT}
% ------
The growing integration of CT and AT into compulsory education has led to increased efforts to develop effective assessment tools, but there remains a scarcity of research specifically targeting the assessment of AT, posing challenges for educators and policymakers \citep{tikva_mapping_2021,tang2020assessing,grover2014assessing}.
The ambiguity and variety of definitions for CT and AT make it challenging to develop effective assessment instruments, as standardised definitions are essential for accurately measuring these skills across diverse educational settings \citep{piatti_2022,pisa,Fraillon2020,ezeamuzie2021computational,grover2017assessing,pilotti2022is,scherer2019the}.

% spieghiamo un po quali sono le defnizioni di ct in letteratura, quale noi adottiamo
The literature on CT offers various frameworks to define and measure its components.
Wing popularised CT as a problem-solving approach involving algorithms, abstraction, and recursion, applicable beyond computing \citep{wing2006computational}.
Grover and Pea emphasise CT's cognitive processes like problem-solving, abstraction, and decomposition decomposition, central to computer science but transferable across domains \citep{grover2013computational}.
Despite multiple frameworks for defining CT, most overlook developmental factors like age and competence, limiting understanding of how CT evolves over time \citep{shute2017demystifying,tikva_mapping_2021}.
Moreover, since AT is a specific subcomponent of CT, it often receives insufficient focus within these broader frameworks, further complicating the assessment and integration of these skills in educational contexts.

% e in particolrare quale definizione di AT adottiamo
% ricordiamo che la nostra definizione é nel CT-cube
To address this, Piatti et al. (2022), with the CT-cube framework, propose a developmental definition of CT that highlights its iterative nature -- problem setting, algorithm creation, and solution assessment -- while considering age, competence, social context, and available tools (``artefactual environment'') \cite{piatti_2022}. 
They define AT as the process of designing step-by-step an algorithm for implementation by human, artificial, or virtual agents \citep{piatti_2022,adorni2023}.

% With this operationalisation of AT we aim to create an instrument that enhances assessing students' AT skills across various educational contexts, allowing educators to effectively measure the development of AT competencies and ensure that assessments are aligned with developmental factors.

% \subsubsection{Categories of assessment tools}

% \textcolor{red}{THIS SECTION IS STILL WORK IN PROGRESS}

% \hl{In practical terms, AT is assessed through tasks that require students to plan, develop, and test algorithms, whether in programming environments or in unplugged activities that simulate the process.
% }

Various pedagogical methods have been adopted globally to teach and assess AT and can be broadly categorised into unplugged and virtual tools \citep{adorni2023}.
% While robotics-based activities also provide a valuable avenue for engaging students in hands-on, real-world problem-solving, they have been excluded from this discussion due to their specific resource requirements, such as access to physical robots and specialised training for educators. 
In this discussion, we focus on scalable, widely accessible approaches that can be integrated into diverse classroom environments without needing specialised equipment.

% \subsubsection{Unplugged tools}
Unplugged tools involve activities that do not require the use of computers, making them practical for introducing algorithmic concepts to young learners \citep{adorni2023,brackmann2017development,del2020computational}. 
These include traditional methods, such as closed-ended questions and multiple-choice tests, commonly used in education and training to assess students' ability to understand and recall key concepts related to AT. 
% Closed-ended questions elicit brief, direct responses, while multiple-choice tests require selecting answers from predefined options. 
While effective for evaluating basic knowledge, these methods often prioritise rote memorisation and basic knowledge recall, which has led to criticism for their inability to adequately capture the depth of students' problem-solving abilities or their capacity for critical thinking \citep{garcia2022application,oyelere2022developing,simmering2019what,csernoch2015testing,campbell2012exploring}.
% 
% \textcolor{red}{MAYBE REMOVE OPEN-ENDED QUESTIONS}
% Open-ended questions, on the other hand, provide more flexibility by allowing students to demonstrate their reasoning and approach to solving algorithmic problems. These methods encourage deeper reflection, as students are required to articulate their thought processes and solutions. However, grading these responses can be time-intensive and may introduce inconsistencies due to subjective interpretation \citep{csernoch2015testing}.
% 
Hands-on tasks within unplugged activities focus on helping students understand and apply algorithmic principles without computers. For instance, students might follow or design step-by-step procedures to simulate algorithms using physical objects or paper-based tasks. These activities support the internalisation of algorithmic logic while fostering creativity and collaboration as students work through tangible representations of CT \citep{WEIGEND2019,Chen2023}.

% \subsubsection{Virtual tools}
Virtual tools leverage digital environments to engage students in more dynamic assessments of AT.
They typically include programming assignments, coding challenges, and gamified exercises that allow students to apply AT concepts in real-world scenarios \citep{Rijo2022,noone2018visual,shin2014visual}.
By working directly with coding environments, students can demonstrate their understanding of algorithms through problem-solving and project-based learning.
Virtual tools are particularly effective for engaging students with immediate, hands-on experiences \citep{Wang2022,tsarava2017training}
Coding platforms and virtual labs allow learners to explore core computational concepts in a controlled and interactive setting. 
While these activities allow for deeper engagement with AT, grading them manually can be resource-intensive and may require significant effort from educators \citep{sun2021comparing}.

The assessment of AT in education is challenging due to the limited technical expertise of general educators, who may struggle with evaluating complex aspects of AT \citep{Ukkonen2024,mason2019preparing,yadav2014computational}. While some schools use collaborative models with IT professionals, these experts are often unavailable, particularly in under-resourced areas. 
As a result, educators rely on standardised tools that may oversimplify the evaluation process \citep{DENNING2021}. 
To address this, automatic assessment tools are essential, as they can be easily administered by teachers without specialised training, providing scalable and reliable evaluations of AT skills across diverse contexts.
Automatic assessment systems are increasingly used for unplugged and virtual tools, offering rapid, real-time feedback while maintaining consistency and scalability in assessments, addressing challenges related to standardisation and subjectivity in evaluation \citep{qian2018using,romero2017computational}. 
In unplugged settings, they simplify grading through tests and quizzes, while in virtual environments, they assess programming tasks by analysing correctness and code efficiency.
% 
% These systems reduce the time required for manual grading and ensure objective assessments of each student's performance. Automated tools also track students' progress over time, offering insights into their learning trajectory. 
% 
Although these systems are particularly effective in large-scale educational programs, research is ongoing to improve these systems' ability to assess complex AT more comprehensively, particularly in detecting nuanced problem-solving strategies and higher-order thinking skills \citep{stanja2022formative}.

\subsubsection{Comparison of AT assessment instruments}
Despite the variety of existing assessment methods, many overlook important facets such as developmental aspects, social and environmental contexts, and the availability of appropriate educational resources \citep{brennan2012new,roman2017complementary,yadav2014computational,korkmaz2019adapting,korkmaz2017validity,martinez2022assessing,roman2019combining,tsai2020the,roman2017cognitive,polat2021a}.
These gaps highlight the need for a comprehensive, reliable, and objective assessment tool that can be broadly applied and scaled to accommodate diverse age groups and educational settings \citep{grover2013computational,barr,voogt2012,giannina2020pisa}.

% \textcolor{red}{A comparative table between existing instruments and the CAT tool could provide valuable context and clarity. HOW}
\cref{tab:tools} provides an overview of existing instruments for assessing AT in unplugged and virtual domains, focusing on key characteristics relevant to their functionality and suitability for educational assessment purposes.
The variables included in the table were carefully selected based on established frameworks for categorising CT problems (CTPs).
Specifically, as aforementioned, we adopt the developmental definition of CT proposed by Piatti et al. \citet{piatti_2022}, emphasising the importance of available artefacts, competence levels, and age ranges.

A critical characteristic for our analysis is the artefactual environment. 
We categorised instruments as unimodal, offering a single interaction method, or multimodal, providing multiple options for engagement. 
Unimodal tools can be limiting, as a single method may not suit all students, potentially affecting their performance. 
In contrast, multimodal tools offer greater flexibility, allowing students to choose the interaction method they find most intuitive. 
By focusing on multimodal tools, we aim to enhance engagement and ensure more accurate assessments by offering diverse interaction methods that cater to a wider range of student needs and abilities.

Another key characteristic is the level of skills targeted by each tool. To differentiate the existing instruments, we considered the hierarchical progression of CT skills outlined by Gouws et al. \citet{CTF}, ranging from foundational levels, such as recognising and understanding algorithmic concepts, to more advanced stages, like applying and assimilating them \citep{bloom1956taxonomy}.
This framework helps evaluate how effectively each tool fosters algorithmic skills across these four levels. 
By focusing on tools that engage multiple levels of learning, we aim to identify instruments that assess various skill levels and support students' progression in AT, providing meaningful feedback to guide their development.

The target age range is another crucial characteristic in evaluating AT tools, as it determines the versatility and inclusivity of the instrument. 
Instruments designed for a broad age range can accommodate students with varying developmental levels and educational needs, making them more adaptable to diverse classroom settings. 
By focusing on tools with a wide age range, we aim to identify those capable of supporting learners at different stages, facilitating longitudinal assessments, and promoting scalability across educational contexts. 
This ensures that the tools are effective for specific groups and adaptable for use in mixed-age classrooms or across multiple educational levels.

The final characteristic we focus on is whether the instruments support automated assessment. 
Automated assessment tools offer significant advantages in scalability, efficiency, and consistency. 
They can provide real-time feedback, enabling students to receive immediate insights into their performance, which can be crucial for reinforcing learning. 
Moreover, automated systems reduce the burden on educators, allowing them to focus on personalised teaching. 
By focusing on tools with automated assessment capabilities, we aim to identify instruments that can facilitate large-scale, data-driven evaluations while maintaining reliability and objectivity in measuring students' AT skills.

\begin{table}[!ht]
\footnotesize
\renewcommand{\arraystretch}{1.4}
\setlength{\tabcolsep}{1pt}
\centering
\caption{\textbf{Overview of AT assessment tools.} 
A comparison of instruments used to assess AT skills, categorised into unplugged and virtual tools. 
For each of them, the artefactual environment (unimodal, multimodal), the skill level targeted (RC = recognise, UN = understand, AP = apply, AS = assimilate), the target age range of users, and the type of assessment (M = manual, A = automated, MX = mixe) are outlined.}\label{tab:tools}
\begin{subtable}[t]{0.49\textwidth}
\centering
\caption{Unplugged Tools}\label{tab:unplugged_tools} 
\begin{tabular}{>{\raggedright\arraybackslash}m{2.3cm}>{\raggedright\arraybackslash}m{1.7cm}>{\raggedright\arraybackslash}m{2.2cm}>{\centering\arraybackslash}m{.8cm}>{\centering\arraybackslash}m{1cm}}
\toprule
\textbf{Instrument}  & \textbf{Artef. env.} & \textbf{Skill level} & \textbf{Age} & \textbf{Assmt.} \\ \midrule 
Bebras Challenge \citep{Dagien}             & unimodal   & RC, UN, AP     & 11-16 & A \\ 
BPAt \citep{Mhling2015}                     & unimodal   & RC, UN         & 12-15 & M \\  
CTt \citep{roman2015computational}          & unimodal   & RC, UN         & 12-14 & M \\ 
BCTt \citep{ZapataCaceres2020,zapata2021}   & unimodal   & RC, UN         &  5-12 & M \\ 
cCTt \citep{ElHamamsy2022cctt}              & unimodal   & RC, UN         &  7-9  & M \\ 
CAT \citep{piatti_2022}                     & multimodal & RC, UN, AP, AS &  3-16 & M \\
\bottomrule
\end{tabular}
\end{subtable}%
\\\bigskip
\begin{subtable}[t]{0.49\textwidth}
\centering
\caption{Virtual Tools}\label{tab:virtual_tools}
\begin{tabular}{>{\raggedright\arraybackslash}m{2.3cm}>{\raggedright\arraybackslash}m{1.7cm}>{\raggedright\arraybackslash}m{2.2cm}>{\centering\arraybackslash}m{.8cm}>{\centering\arraybackslash}m{1cm}}
\toprule
\textbf{Instrument}  & \textbf{Artef. env.} & \textbf{Skill level} & \textbf{Age} & \textbf{Assmt.} \\ \midrule 
{Alice} \citep{cooper2000developing,werner2012children}  & unimodal & RC, UN, AP     & 10-14 & M  \\
Code.org \citep{Blockly,PZ,AB}                           & unimodal & RC, UN, AP, AS &  5-17 & M  \\ 
Scratch \citep{maloney,moreno2015dr,GroverPeaCooper2015} & unimodal & RC, UN, AP, AS &  8-16 & MX\\ 
\bottomrule
\end{tabular}
\end{subtable}%
% \\\vspace{0.5em}
% \begin{minipage}{.9\linewidth}
% $^a$ Artefactual environment: \\
% \setlength{\parindent}{1.5em}\indent Um = unimodal, Mm = multimodal\\
% $^b$ Skill level: \\
% \setlength{\parindent}{1.5em}\indent RC = recognise, UN = understand, \\
% \setlength{\parindent}{1.5em}\indent AP = apply, AS = assimilate\\
% $^c$ Assessment:\\
% \setlength{\parindent}{1.5em}\indent  M = manual, A = automated, MX = mixed
% \end{minipage}
\end{table}

% \textcolor{red}{Comment on the table}

From the analysis of the table, it emerges that, regarding the artefactual environment, the only multimodal tool is the CAT, which provides a broad range of interactive and cognitive artefacts supporting various learning styles.

In terms of skill levels, virtual tools like Code.org \citep{Blockly,PZ,AB} and Scratch \citep{maloney,moreno2015dr,GroverPeaCooper2015}, along with the unplugged CAT \citep{piatti_2022}, cover a broader range from recognition to assimilation, making them more suitable for tracking progress in AT. 
In contrast, tools like the Basic programming abilities test (BPAt) \citep{Mhling2015}, the Computational Thinking test (CTt) \citep{roman2015computational}, the Beginners Computational Thinking test (BCTt) \citep{ZapataCaceres2020,zapata2021} and the competent Computational Thinking test (cCTt) \citep{ElHamamsy2022cctt} focus on foundational skills such as recognition and understanding, limiting their effectiveness for more advanced assessments.

Tools like the CAT, Code.org, and Scratch that cover a wide age range are ideal, as they accommodate diverse learners. However, despite frameworks like Code.org and Scratch being suitable for various age groups, the selected activities are often tailored for specific age groups, whereas the CAT is a single activity suitable for all learners. 
Other instruments targeting narrower age groups, such as BPAt and cCTt, limit their applicability in projects with a broader student demographic.

Automated or mixed tools like Bebras Challenge \citep{Dagien} and Scratch are preferred for scalability and efficiency, as they allow for quick data collection and timely feedback. Manual assessments, like Alice \citep{cooper2000developing,werner2012children} or unplugged instruments, while providing more detailed insights, are less scalable and time-consuming, making them less ideal for large-scale projects. However, many of these tools could be automated in the future.
 
In conclusion, while the tools presented offer a variety of methods to assess AT across different domains and age ranges, the integration of automatic assessment remains a key area for future development.
The CAT stands out as an ideal compromise, combining an inclusive artefactual environment, multi-level skill development, and adaptability to a wide range of age groups. This makes it effective in supporting learners at different cognitive stages. Its design also enables digital adaptation, addressing both scalability and the need for automatic assessment in large-scale educational settings.

\section{The Cross Array Task (CAT)}\label{sec:cat}

\begin{figure*}
    \centering
    \begin{subfigure}[t]{0.48\linewidth}
        \centering
        \includegraphics[height=7cm]{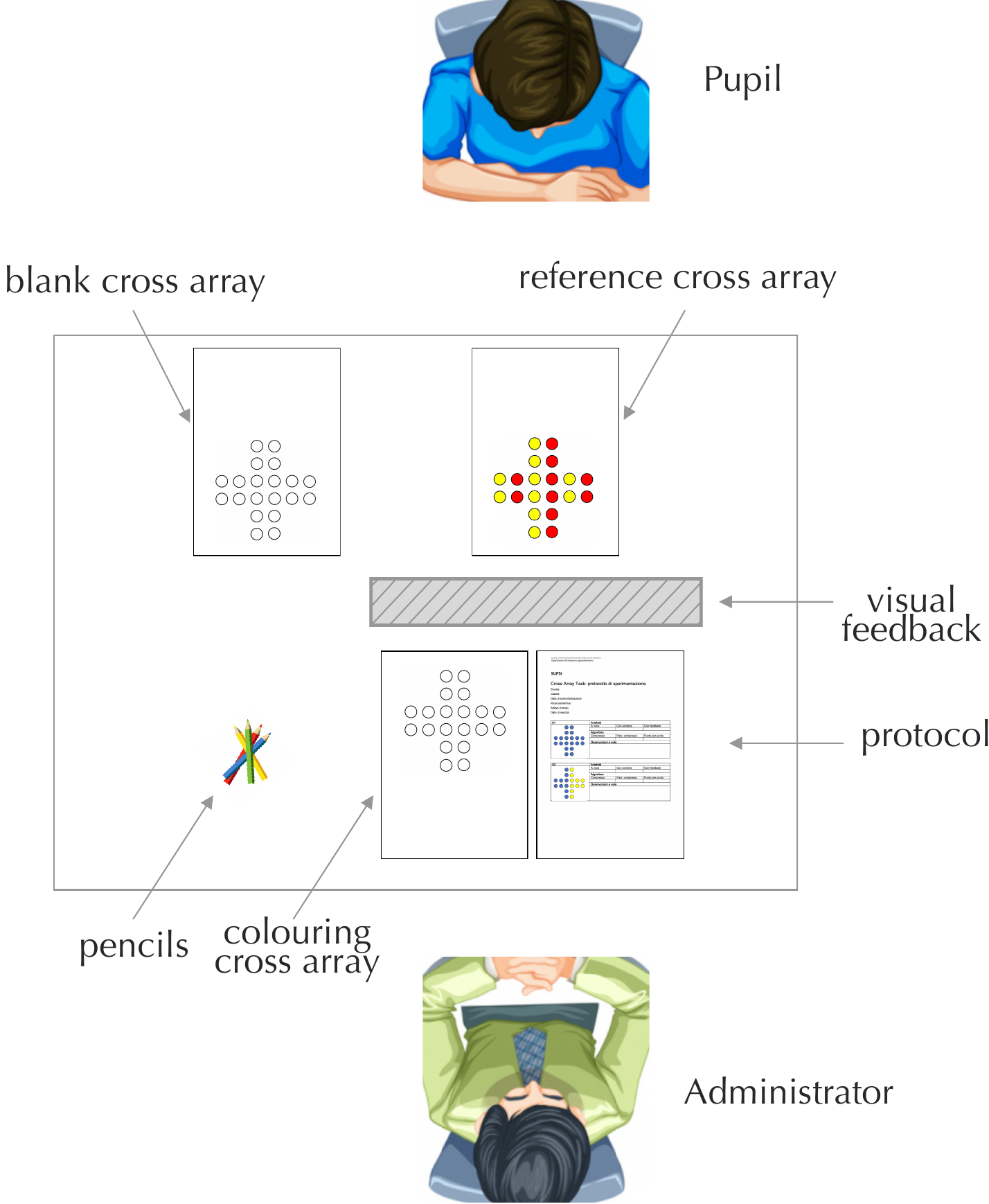}
        \caption{\textbf{The unplugged {CAT}.}
                 While the student uses voice or gestures on a blank schema to instruct the administrator to reproduce a reference pattern on a colouring schema, he interprets and records the algorithm in a protocol.
                 Visual feedback can be enabled by removing the physical barriers separating them.
                 }
        \label{fig:unpluggedCAT}
    \end{subfigure}
    \hfill
    \begin{subfigure}[t]{0.48\linewidth}
        \centering
        \includegraphics[height=7cm]{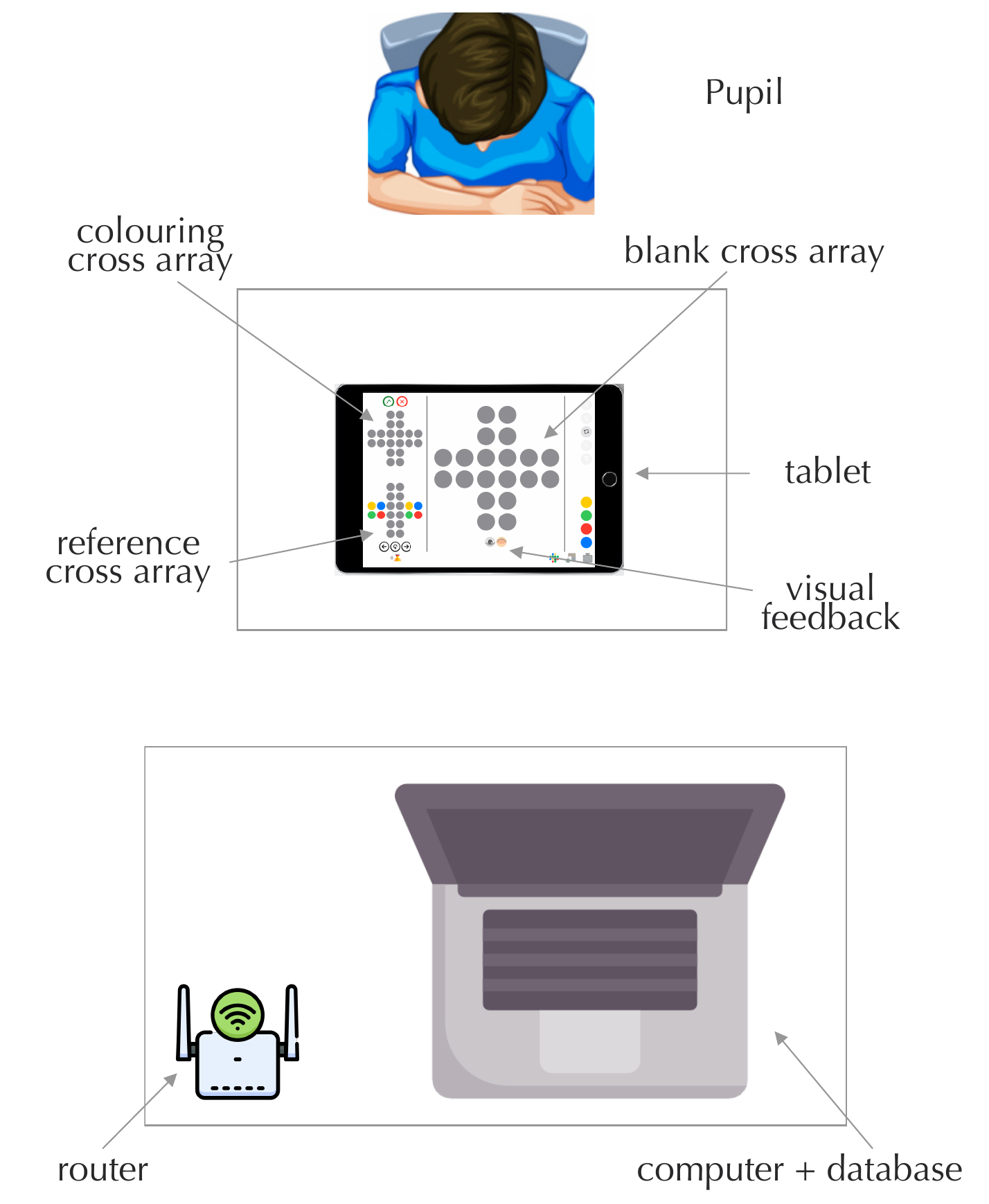}
        \caption{\textbf{The virtual {CAT}.}
                 While the student reproduces a reference pattern using either a gesture-based interface with a blank schema or a visual block-based programming interface, the system automatically interprets all actions and records the algorithm.
                 Visual feedback can be easily activated or deactivated using a button.
                 }
        \label{fig:virtualCAT}
    \end{subfigure}
    \caption{\textbf{Comparison of the setting between the unplugged and virtual CAT.}}
    \label{fig:CATsettings}
\end{figure*}

The Cross Array Task (CAT) is an educational activity focused on algorithm development. % or breaking down complex processes into more straightforward instructions that a human or an artificial agent can execute to solve the problem. 
It was designed to assess students' AT skills in a way that aligns with the distinctive educational landscape of Switzerland \citep{piatti_2022}.
% The CAT's focus on these characteristics complements the Swiss educational philosophy, which strongly emphasises diversity and integration. Swiss schools are known for accommodating students with diverse backgrounds, capabilities, and linguistic proficiencies, and the CAT's flexibility enables it to support these goals effectively.

In this task, students are asked to replicate a cross array reference pattern. It consists of a cross-shaped array of $20$ dots arranged in the shape of a 2-thick cross of coloured dots. %, with colours chosen from a set of $k$ colours, usually yellow, green, blue, or red.
Each student receives 12 reference schemas -- coloured cross arrays with increasing complexities and exhibiting different regularities. The student's task is to devise a set of instructions, known as algorithm, to replicate these patterns on a blank cross array.
These instructions must then be conveyed to an agent who will interpret and execute them, replicating the colouring pattern on the blank cross array, also known as a colouring schema.

To communicate the algorithm, students can use a variety of artefacts, such as gestures, depending on the version of the CAT administered.
Typically, a visual barrier is placed between the student and the colouring schema to increase the task's challenge, however, it can be removed if needed, allowing to rely on visual feedback of the agents actions during the colouring process. 

For a detailed discussion of the specific tasks within the CAT, including their structure, objectives, and the competencies they are designed to assess, please refer to our previous work \citep{piatti_2022}, which provides a comprehensive explanation of the theoretical foundation, task design, and rationale behind the selection of these tasks, which are crucial for understanding how the CAT assesses AT.

AT is assessed by the complexity of operations in the algorithm, starting from basic tasks like colouring dots individually (0D), to more advanced tasks such as colouring multiple dots in patterns like rows, diagonals, or squares (1D), and the most complex tasks involving intricate patterns with alternating colours, repetition, or mirroring (2D).
The final classification of a student's AT is based on the highest level of complexity demonstrated in their solution.

The assessment goes beyond AT by considering aspects of situated cognition often overlooked in the literature. These include the context in which the task is carried out, particularly the student's level of autonomy or degree of independence, and the artefactual environment, which refers to the cognitive tools and resources the student uses to solve the task. 
Therefore, these factors influence the student's performance and are integral to the assessment.

% so... the CAT assessment process is designed to evaluate pupils based on these dimensions: algorithm complexity, artefact choice, and autonomy level.
% Specifically, 
% In particular, the algorithm complexity is determined by the type of operations that comprise it.
% Zero-dimensional operations correspond to the fundamental Colour-One-Dot (COD) operation, which colours individual dots.
% One-dimensional operations correspond to the Colour-Several-Dots (CSDs) operation, where multiple dots are coloured with the same colour based on patterns such as rows, diagonals, and squares. 
% Two-dimensional operations correspond to CSDs operations with alternating colours and repetition or mirroring of COD or CSDs operations. 
% The final classification of an algorithm is determined by the highest-dimensional class of operation employed.

The task is considered successful if the student creates a complete and correct algorithm, regardless of the algorithm's complexity, the artefactual environment, or the level of autonomy. 

A comprehensive metric called the CAT score is used to quantify this multi-facet performance, consisting of the algorithm and interaction dimensions -- a combination of artefact used and autonomy level.
Each component is assigned a numerical score, with a higher score indicative of a student who has navigated the complexities of challenging artefacts, assumed an autonomous role, and/or conceived a higher-dimensional algorithm.

% The CAT was originally conceived as an unplugged activity, meaning it does not rely on digital tools, which we present here. However, the goal of this paper is to explore a digital version of the CAT, addressing the limitations of the unplugged version by enabling more flexibility and scalability.

\subsection{Overview of the original unplugged version}

The CAT was conceived as an unplugged activity, characterised by a face-to-face interaction between the student and a human administrator. 
This setup is illustrated in \Cref{fig:unpluggedCAT}, visually representing the task environment and interaction dynamics.

In this variant, students can use two types of representational artefacts to convey their algorithms: they can either verbally communicate instructions using natural language or enhance their voice instructions with physical gestures, such as pointing to specific dots on an empty cross array to illustrate their commands visually.
As aforementioned, a physical barrier prevents the student from observing the administrator colouring the empty cross array.
The administrator listens to and interprets the student's instructions, records them in a protocol, and then uses them to colour the cross array.

While the CAT proved effective for evaluating AT abilities in K-12 Swiss students, it had several limitations.
Designed as a one-on-one activity, it is time-consuming and unsuitable for simultaneous administration.
Moreover, the reliance on a human administrator introduced potential inconsistencies in interpreting and executing instructions, further hindering its scalability and efficiency for large assessments.

\subsection{Transition to the virtual version}\label{sec:transition}

This section focuses on the conceptual choices involved in transitioning from the unplugged activity to the virtual platform. Details on the development and design process and final implementation are discussed from \cref{sec:method,sec:prototypes,sec:platform_development}.

\begin{table*}[!ht]
\footnotesize
\caption{\textbf{Comparison of unplugged and virtual CAT.} The table compares the unplugged and virtual CAT, highlighting key differences in interaction types, available artefacts, autonomy settings, algorithm classification and data collection approaches.}\label{tab:unplugged_virtual}
\centering
\begin{tabular}{l>{\raggedright\arraybackslash}m{6.5cm}>{\raggedright\arraybackslash}m{6.5cm}}
\toprule
 & \textbf{Unplugged CAT} & \textbf{Virtual CAT} \\
\midrule
\textbf{Interaction type} & Face-to-face (problem solver \& human agent) & Face-to-device (problem solver \& virtual agent)  \\
\\[-5pt]
% \hline
\textbf{Embodied artefact} & S: hand gestures on a schema & G: gesture interface \\
\\[-5pt]
% \hline
\textbf{Symbolic artefact} & V: voice & P: visual programming interface \\
\\[-5pt]
% \hline
\textbf{Autonomy} & F (or not): a removable physical barrier to enable visual feedback & F (or not): a button to enable visual feedback \\
\\[-5pt]
% \hline
\textbf{Algorithm classification} & A human agent interprets instructions and manually codifies the algorithm & A virtual agent interprets actions and codifies the algorithm into a formal programming language \\
\\[-5pt]
% \hline
\textbf{Data collection} & Manual & Automatic \\
\bottomrule
\end{tabular}
\end{table*}

The transition from the unplugged to the digital version of the CAT aimed to address the limitations of the original unplugged setup while enhancing scalability, efficiency and accessibility. 
The virtual CAT retains the core elements of the original activity: 
(i) the \textit{task} for students is to devise a set of instructions or an algorithm to replicate coloured patterns on a cross-shaped grid;
(ii) the \textit{artefactual environment} includes a variety of cognitive artefacts based on both embodiment and perception as well as symbolic representation;
(iii) the students' \textit{autonomy} reflects their level of independence during the activity, ranging from those who do not engage with the task to those who rely on visual feedback up to those who provide clear instructions independently.

% but leverages technology to streamline the process and expand its capabilities. 

The virtual CAT supports multiple languages, including Italian, French, and German, to accommodate Switzerland's diverse linguistic landscape while offering an English version to broaden its potential application.
This multilingual support ensures the tool is accessible across Switzerland and in countries where these languages are spoken, expanding its usability to a broader range of educational institutions (see Figure~\ref{fig:language}).

The transition to a digital version required adjustments, particularly in interaction methods.
Where the unplugged CAT involved direct communication with a human administrator, the virtual CAT introduced a virtual agent that interprets and executes the algorithms devised by students, removing face-to-face interaction, eliminating potential human errors and inconsistencies in instruction interpretation and ensuring a more standardised assessment experience. This setup is illustrated in \Cref{fig:virtualCAT}. 

% \begin{figure*}[!h]
%     \centering
%     \includegraphics[height=9.5cm]{images/s2_text_partial.pdf}
%     \caption{\textbf{Example of usage of the CAT-VPI with textual commands.} }
%     \label{fig:s2_text_partial}
% % \end{figure*}
% % \begin{figure*}[h]
% \vspace{10pt}
%     \centering
%     \includegraphics[height=9.5cm]{images/s2_gesture_full_feedback.pdf}
%         \caption{\textbf{Example of usage of the CAT-GI.}}
%         \label{fig:s2_gesture_full_feedback}
% \end{figure*}

Additionally, this tool enables simultaneous assessment of multiple pupils, overcoming the logistical impracticalities of one-on-one interactions in the unplugged version, as each student interacts with the task individually on their device, making it feasible to administer the activity to entire classrooms or larger groups without additional human resources. 

As for the artefacts available in the virtual environment, two interfaces are provided to accommodate diverse learning styles and preferences.
We did not provide a modality of interaction based on verbal instruction, but we provided an alternative symbolic language, a visual block-based programming interface (CAT-VPI).
This decision was driven by several technical challenges of implementing speech recognition, particularly in a multilingual classroom context with young students \citep{AiminZhou2024,Wilpon1996}.
Most speech recognition systems are trained on adult voices, making them less accurate for children, whose speech differs in pitch and tone \citep{GurunathShivakumar2020,Potamianos2003,Yeung2018,GurunathShivakumar2022}. 
Additionally, there isn't enough data available to improve the accuracy of children’s voices \citep{Claus2013,Fainberg2016}. 
While there are techniques to adjust children's voices to be more like adult voices, they don't work well enough \citep{Ahmad2017,Shahnawazuddin2017,Kathania2018,Kathania2018b}. 
For these reasons, we chose a visual programming interface instead, which is more reliable in this context, and its implementation requires less effort.

The CAT-VPI is designed to make coding accessible to K-12 students, including beginners with no prior programming experience. 
It allows students to construct colouring algorithms using drag-and-drop programming blocks, which mirror the instructions observed in the unplugged version of the activity.
Nevertheless, these blocks are customisable, enabling users to adjust parameters like colour and pattern choices.
This intuitive and flexible approach reduces the likelihood of syntax errors, potentially improving the overall learning experience.
For a visual representation of this interface, refer to \Cref{fig:programming_final}, with further details on its development provided in \cref{sec:third_prototype}.
% An example of usage of this interface is illustrated in \Cref{fig:s2_text_partial}, where a student starts constructing an algorithm for the second reference schema by placing a block command to select the starting point for colouring and using another block command to colour a blue square.

The CAT-GI is designed to emulate the hand gestures observed in the unplugged CAT activity, providing a tactile experience similar to interacting with the physical cross array that ensures continuity in the interaction type while leveraging digital capabilities.
Users can build the colouring algorithm by selecting colours, tapping on individual dots, dragging across multiple dots to create patterns, or using icons to perform more advanced actions, such as repeating instructions or mirroring patterns.
For a visual representation of this interface, refer to \Cref{fig:gestures_final}, with further details on its development provided in \cref{sec:third_prototype}.
% \Cref{fig:s2_gesture_full_feedback} illustrated a usage example of this interface, where a student is completing the task started with the CAT-VPI by colouring the remaining dots of the cross with a yellow pattern.

% Details on the iterative development process leading to this version are provided in \cref{sec:third_prototype}.

A key feature of the unplugged CAT was the limited visual feedback provided to students during the task. The problem solver could not observe the agent executing their instructions, encouraging clearer, more precise communication. 
This challenge was preserved in the digital version by restricting students from seeing the virtual agent’s progress unless they explicitly choose to enable it. This maintains the original difficulty and autonomy requirements of the unplugged task while allowing some flexibility regarding feedback, which can be toggled if necessary.

The agent maintains the role of interpreting the student's instructions. 
A programming language interpreter translates gesture interaction and visual blocks into a formal programming language that mimics the operations observed in the unplugged activity, which we assumed the student would reuse in the virtual version. 
The algorithms are thus automatically recorded by the virtual agent. %providing a basis for evaluation and analysis.
All session data are simultaneously acquired and managed on a central computer with a configured database, ensuring constant accessibility and data integrity.

Pupils also have the flexibility to choose their preferred interaction mode, navigate between tasks, restart them, confirm completion, or skip tasks as needed.
Upon completing all tasks, pupils are directed to a results dashboard that comprehensively summarises their performance, including visual representations of attempted tasks, scores, completion status, and time taken.

These options ensure the task remains engaging and accessible to pupils with varying levels of familiarity with technology.

\Cref{tab:unplugged_virtual} summarise the principal differences between the two versions of the CAT activity.

\section{Method}\label{sec:method}
% \hl{Detailed tables and figures illustrating the task design, theoretical framework would greatly enhance its appropriateness and usefulness to readers.}

% \textcolor{red}{THIS IS A REVISITING OF THE OLD DESIGN SECTION. I REMOVED THE YELLOW HIGHLIGHTS FOR READABILITY.}

In this section, we present the methodology employed to develop our tool, guided by the User Experience (UX) design life cycle. This structured approach, involving the systematic collection of data on user behaviours, preferences, and requirements, ensures the development of a user-centred product aligned with their actual needs \citep{hartson2018ux}.

\begin{figure}[h]
    \centering
    \includegraphics[width=.9\linewidth]{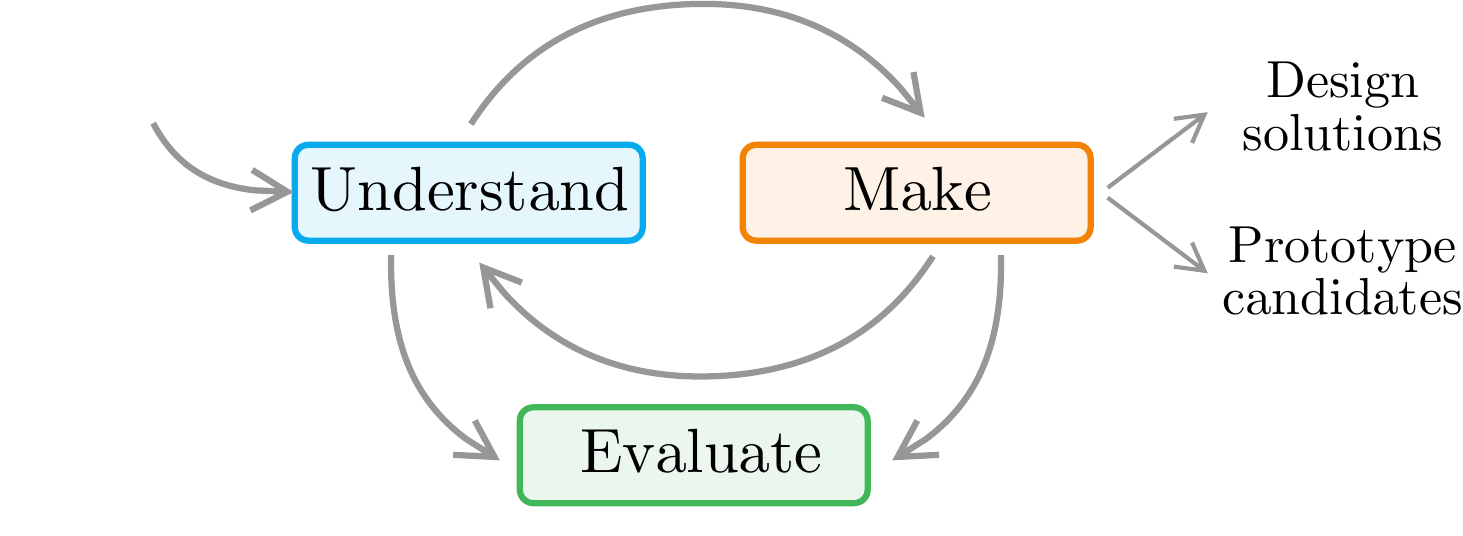}
    \caption{\textbf{UX design life cycle.}
    The iterative UX design life cycle encompasses understanding user needs (U), making (M) -- or designing and prototyping solutions --, and evaluating them through user and expert feedback (E).
    These phases repeat cyclically, with evaluation insights leading to refinements in earlier stages.
    }
    \label{fig:design_process}
\end{figure}

The UX design life cycle is an iterative process that encompasses three main phases \citep{hartson2018ux}: (1) \textit{understand} (U) -- gathering insights into user needs and problem domains; (2) \textit{make} (M) -- designing and prototyping solutions based on the understanding phase; (3) \textit{evaluate} (E) -- testing prototypes and solutions through user feedback and expert analysis.
The cyclic flow between understanding, making, and evaluating emphasises the iterative nature of this process. Insights gained from evaluations often lead to revisiting earlier phases to refine and enhance the design. 
While the specific phases and iterations of the UX design life cycle can vary in the literature, this particular iterative process has been adopted for our work to ensure a structured and user-centred approach (see Figure~\ref{fig:design_process}).

In the context of product development, two types of evaluations are commonly employed to guide and assess design: formative and summative evaluation.
\textit{Formative evaluation} takes place throughout the iterative design process, helping to refine and improve the product before it reaches its final form. It ensures that continuous user feedback informs ongoing refinements of the design, intending to improve usability, functionality, and overall user experience. 
\textit{Summative evaluation}, on the other hand, is conducted once the product has been fully developed, aiming to assess its overall effectiveness and impact. It occurs after the design has been finalised. It focuses primarily on evaluating the effectiveness and impact of the final product, typically through large-scale studies, and does not involve redesign or further iterations \citep{scriven1967methodology,dick2005systematic,williges1984evaluating,CARROLL1992}.

In this article, we focus on formative evaluation, which is integral to the iterative design process and encompasses all phases of the UX design cycle, while the summative evaluation is detailed in a separate publication \citep{adorni_chbr}.

% \textcolor{red}{INSERIRE GRAFICO DEL PROCESSO DI DESIGN E ASSESSMENT}

% \subsection{Design and evaluation process}\label{sec:design}
% The design process was shaped by the input of users, experts, and educators to ensure the tool's relevance and usability. Through iterative cycles of feedback and refinement, we aimed to align the design with both user needs and expert insights. 

\begin{figure*}[h]
    \centering
    \includegraphics[width=1\linewidth]{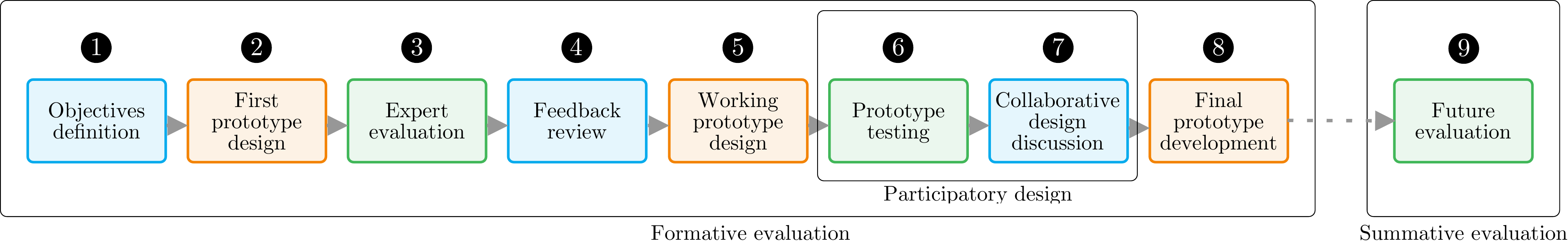}
    \caption{\textbf{Design and evaluation process overview.}
    This diagram illustrates the various stages of the design process, from defining objectives to developing and evaluating prototypes. Different colours represent the phases of the UX design lifecycle: blue for the understand (U) phase, orange for the make (M) phase, and green for the evaluate (E) phase.}
    \label{fig:method}
\end{figure*}

Our design and evaluation process, illustrated in \cref{fig:method}, followed a structured and iterative procedure aligned with the phases of the UX design cycle of \cref{fig:design_process}.
The process spanned 17 months, from February 2022 to June 2023.
The first prototype was developed from February to July 2022 (5 months), the second from July 2022 to March 2023 (8 months), and the final version from March to June 2023 (4 months).

\subsection{Objectives definition}\label{sec:objectives_definition}
This approach began with the objectives definition stage \circled{1}, where we set the instrument's goals and decided how to adapt the unplugged activity to the virtual format.
As discussed in \cref{sec:transition}, we focused on preserving the pedagogical value of the original task by identifying key components to maintain and determining how to translate them into the digital environment. This step was crucial for ensuring the educational objectives were upheld, laying the foundation for the entire process.

\subsection{Initial prototype development}\label{sec:initial_prototype}
Following is an iterative design process to develop the first digital prototype \circled{2}, prioritising user experience accessibility and usability \citep{hartson2018ux}. 
In educational technology, accessibility centres on crafting solutions to meet users' needs from various backgrounds, regardless of their physical or cognitive abilities \citep{kali2011teaching}.
In contrast, usability focuses on the user experience, aiming at delivering an intuitive and effective learning environment \citep{nikahmad2021usability,gould1985usability}. 
% While accessibility and usability assessments often overlap, combining both aspects results in a more comprehensive evaluation, unveiling hidden nuances that may hinder some users while enriching the overall experience for all.
To achieve these goals, we made a series of strategic decisions, including considerations such as the choice of compatible devices, language support, and the design and layout of various interfaces, following guidelines and best practices \cite{hartson2018ux}.

The development of the initial prototype was grounded in the architectural and technical decisions outlined in \cref{sec:platform_development}, ensuring the prototype' functionality and alignment with the educational objectives defined in \cref{sec:objectives_definition}.

In this phase, we sketched interface layouts, selected appropriate technologies, and built interactive prototypes to simulate user interactions. Throughout this process, we continuously evaluated the user experience, ensuring the prototype was intuitive for students and teachers. Additionally, expert consultations were integrated to validate the design choices, ensuring the prototype met usability and pedagogical standards.

Due to time constraints and limited access to schools and children, we skipped certain prototyping stages, such as producing paper prototypes, and directly developed a functional prototype. This streamlined approach was also necessary because of the age of the children involved in the participatory design, who may struggle with abstract reasoning and therefore require a more accessible prototype \citep{hanna1997,Guran2020,markopoulos2002compare,druin1999}. 
% These practical challenges reiterate the pivotal role of usability in shaping our design approach as we strive to ensure that our application is accessible and effective for our target users.

The resulting first prototype is detailed in \cref{sec:first_prototype}.
% This prototype served as a preliminary version of the tool, which would be tested and refined later. 

\subsection{Expert evaluation and prototype redesign}\label{sec:expert_evaluation}

Following the development of our first prototype, we conducted an expert evaluation \circled{3} to assess its design, usability and accessibility \citep{hartson2018ux}. This step involved the participation of experts in both UX design and educational technology, who examined the prototype and provided detailed feedback on various aspects of the platform, particularly focusing on interface clarity, functionality, and alignment with educational objectives.

% \textcolor{purple}{Dare info sugli esperti e il metodo usato. Stiamo parlando di uno o più esperti? Presentare subito in grandi linee il profile, in general, con tutti gli esperti.}

The first expert consulted, recruited through our institutional network, is an interaction design teacher-researcher with a background in educational technology and user-centred design.
The expert was provided with a brief description of the platform's intended use before independently exploring the application. 
His feedback was collected during a collaborative session, in which he shared detailed observations after testing the platform. 

In addition, three pedagogical experts with experience in computer science education were invited to review the prototype. These professionals were selected based on recommendations and their known contributions to technology-enhanced learning.
After being introduced to the platform, each expert independently tested the application and subsequently shared their observations during a feedback session. 

The feedback from these experts was instrumental in guiding the next steps. Entering a reflective phase \circled{4}, we carefully analysed and prioritised the proposed changes, ensuring that the adjustments aligned with both usability principles and pedagogical goals. 

The prototype redesign \circled{5} incorporated these changes and included the development of key technical features, such as a virtual interpreter and the infrastructure necessary for real-time interaction and data processing. 
% The updated prototype result was a completely working prototype that was significantly more user-friendly and technically robust and ready for further evaluation and testing.

The feedback received, the modifications decided upon, and the resulting updated prototype are documented in \cref{sec:second_prototype}.

\subsection{Participatory design}\label{sec:pilot_method}
Following the expert evaluations and the creation of the second prototype, the next phase of the design process focused on participatory design and the development of the final application.
This phase began with engaging the target users, students and teachers, in testing the prototype in real-world settings and providing feedback on its usability and effectiveness \circled{6}. 
The goal was to integrate their insights and preferences to refine the platform, ensuring it met their educational needs and user requirements.

% \subsubsection{Pilot study}\label{sec:pilot_method} % SOTTOSEZIONE IN CUI DIAMO INFORMAZIONI GENERICHE SUL COME AVVIENE IL PILOT STUDY

Our pilot study was designed as a participatory process involving three key roles: a researcher from our team, students and teachers \citep{schuler1993participatory}.

\paragraph{Selection and participation}

The study was conducted in March 2023 and took place during regular class sessions, providing a real-world educational setting for students to engage with the tool.
A total of 31 students, 21 girls and 10 boys, aged 4 to 12 (see Table~\ref{tab:students_analysis}), as well as 5 teachers, participated in the study.

Participating classes were selected through our network of contacts to ensure a representative sample. 
To achieve this, we randomly selected two schools in the Ticino canton, including a preschool class (ages 4-6) and two low secondary classes (ages 11-12), thus covering opposite ends of the compulsory education system in Switzerland.
% This aimed to demonstrate the platform's effectiveness for diverse school types, which can be extended to cover compulsory education.
The teachers participating in the study were those present in the classrooms during the activity.

\begin{table}[h]
\footnotesize
\caption{\textbf{Demographic analysis of students by school type, age category and gender.} The table shows the gender distribution and mean age for each school and age category. %ranges. The Female and Male columns represent the number of female and male students, respectively, while the Total column displays the combined count. The mean age ($\mu$) and standard deviation ($\pm$) are presented for each age range.
}\label{tab:students_analysis}
\centering
\setlength{\tabcolsep}{2mm}
\begin{tabular}{m{1.7cm}m{2.3cm}ccc}
\toprule
 \textbf{School}   & \textbf{Age}   &   \textbf{Female} &   \textbf{Male} &   \textbf{Total} \\
\midrule
 Preschool  & 4-6 years old\newline($\mu$ = 5.0 $\pm$ 1.0)    & 3 &   4 & 7 \\[.3cm]
 Low secondary school      & 11-12 years old\newline($\mu$ = 11.3 $\pm$ 0.6) & 18 &   6 &   24 \\
 \cmidrule(lr){1-2}\cmidrule(lr){3-3}\cmidrule(lr){4-4}\cmidrule(lr){5-5}\textbf{{{Total}}}     &      & 21 &  10 &   31 \\
\bottomrule
\end{tabular}\end{table}

Given the participation of young students, we strictly adhered to ethical guidelines and maintained transparent communication with pupils, parents, teachers and schools \citep{aebi2021code,petousi2020}. 
Initially, we obtained authorisation from school directors and teachers to conduct the research within their schools and classes.
Next, we provided parents with an exhaustive document explaining the research project, data collection and storage procedures, and the personnel involved. We also requested their consent for their children's participation and publishing the collected data.
Teachers obtained informed consent from parents without recording pupils' full names to safeguard privacy, ensuring data anonymity from the outset.
The activity was conducted only with pupils willing to participate and whose guardians had explicitly consented.

The pilot study was divided into two main sequential phases: training and validation, which occur on the same day.

\paragraph{Training module}
To familiarise students with the assessment tool, we designed a training module within the app, which introduces the activity and the platform, including interface and functionalities, before engaging in the evaluation tasks (see Figure~\ref{fig:mode}).

During this phase, the researcher leads a session guiding students through the app's features using 15 sample cross-array schemas for practice. This enables students to effectively navigate the tool in a collaborative setting, with teachers supporting the researcher and students as needed.

Training sessions typically last 30-45 minutes, and the presentations are adapted to the students' developmental level. 
For example, kindergarteners are guided to use only the CAT-GI, with simplified instructions and pacing to ensure comfort and understanding \citep{hanna1997}. 
% No data collection is conducted during this phase, allowing students to focus solely on becoming familiar with the tool and its functionalities.

% No data collection occurred during this phase, allowing students to concentrate on becoming comfortable with the tool and its functions.

\paragraph{Validation module}
After completing the training, students move on to the validation session, during which data collection takes place. This module mirrors the original unplugged activity, where students solved 12 cross-array schemas.

To begin the validation process, session and student details must be manually input into the app (see Figures~\ref{fig:session} and \ref{fig:student}). 
Session information includes the date, canton, school, and class's grade level, while student details are limited to gender and date of birth. Each student is assigned a unique identifier to ensure anonymity and privacy protection.

The app logs timestamped actions performed by the students during the tasks, capturing information relevant to their performance assessment, such as the algorithm and its complexity, the interaction modes given by the type of artefact used and the level of autonomy.
% This included recording the timestamp and type of operation carried out, such as adding, confirming, removing, or reordering commands, updating command properties like colours or directions, resetting the algorithm, changing the mode of interaction or visibility, confirming task completion, or surrendering. 
% This comprehensive data collection was instrumental in assessing pupil performance, providing valuable insights for the study's analysis and findings.
These data are compiled into a dataset, which is pseudonymised to remove any potentially identifiable information (e.g., school and class) in alignment with open science practices in Switzerland \citep{snfOpenScience}.

\paragraph{User feedback elicitation}
During the validation phase, we collected feedback from both students and teachers to evaluate their experience with the tool. It served as a foundation for identifying usability issues and refining the tool to meet user needs better. 

% PUPILS
Pupils were at the heart of the study, and their interactions with the platform were crucial for assessing the tool's usability and identifying new user requirements \citep{druin2002role,kujala2003,valguarnera2023,valguarnera2023challenge,schuler1993participatory}. 
We actively engaged children as informants and evaluators, enabling us to design with their needs and preferences in mind \citep{read2004,read2006,greig2007doing,greig2012doing}.
Their evolving thoughts and reflections, shared during testing activities, provided real-time insights into how they perceived and interacted with the tool \citep{yarosh2011,iversen2013,iversen2012,hourcade2007interaction}.

This participatory approach empowered children to take ownership of the tool's development while fostering critical thinking about its features \citep{iivari2018,kinnula2021,iversen2017,kinnula2017cooperation}. 
It also ensured the process remained enjoyable and rewarding for them, aligning with principles of co-design and participatory research \citep{MullerKuhn1993,MullerWildmanWhite1993,MullerWildmanWhite1994,bogdan1997qualitative}.

% TEACHERS
Teachers played an essential role by facilitating classroom activities, guiding students, and testing the platform themselves \citep{druin1999,scaife1997designing,kafai1997children,borgers2000,greenbaum2020design,nielsen1994,nielsen1995,baecker2014,holtzblatt1997contextual}.
Their observations highlighted how students engaged with the tool, identified areas of difficulty, and noted moments of success. 
Teachers' feedback was invaluable in refining the platform to balance educational goals with practical usability and address both pedagogical and logistical challenges in the classroom.

% RESEARCHERS
During the study, the researcher documented students' interactions with the tool, focusing on their behaviours, verbal feedback, and non-verbal cues. Key observations included moments of confusion, problem-solving strategies, and how students navigated specific features \citep{Guran2020}.
Real-time note-taking captured recurring patterns and usability issues, providing valuable insights into user experience. This structured approach ensured a detailed understanding of the tool’s strengths and areas for improvement, directly informing subsequent design iterations \citep{hanington2019universal,greig2007doing,hartson2018ux}.
These techniques, grounded in Human-Computer Interaction (HCI) and UX design principles, enabled us to triangulate data sources and derive actionable insights for iterative improvements \citep{hanington2019universal,hartson2018ux,druin2000design,lehnert2022Child,hourcade2007interaction,markopoulos2003interaction}.

% TODO
% By integrating feedback from diverse stakeholders and applying rigorous data collection techniques, the iterative design process ensured that the tool evolved to meet the practical and educational needs of its users effectively.

% SOTTOSEZIONE IN CUI DIAMO INFORMAZIONI SUL NUMERO DI PARTECIPANTI...
% non dare le informazioni dettagliate perche quelle le mettiamo poi nelle sezioni successive, ma piuttosto parliamo del meotod, cioe come abbiamo stability quanti participant....

% \textcolor{red}{information about the number of participants involved, the invitation method employed, as well as the location where these interactions took place.}

\paragraph{Collaborative design}

At the end of this process, we conducted a collaborative session with the users \circled{7}, where design proposals were presented based on the notes and feedback collected by the researcher.
Users were also asked for additional input. Based on their insights, modifications were proposed and discussed, allowing users to actively contribute to refining the prototype and ensuring the design is better aligned with their needs.
In \cref{sec:third_prototype}, we highlight the feedback received based on this collaborative session.

\subsection{Final application development}
In the final phase, the prototype is redesigned \circled{8} in response to the feedback and suggestions from the collaborative session, leading to the final working version, which is documented in \cref{sec:third_prototype}.
The redesign adhered to standard mobile application design principles to enhance usability and accessibility \citep{nayebi2012state,coursaris2011meta,tidwell2010designing,hartson2018ux}.

To create an interface familiar to user, we incorporated common elements, like a top bar and a left-side menu list.
Legibility and readability were prioritised using large font sizes and ensuring a high contrast between text and background. 
Accessibility considerations were central to the redesign. The interface included a colour-blind mode, high-contrast visuals, and a text-to-speech feature to accommodate users with visual impairments.

Consistency was maintained by using uniform names and labels for similar objects and functions, avoiding synonyms to ensure clarity and reduce cognitive load. % and employing precise wording in menus, icons, and data fields to enhance clarity. We avoided synonyms to ensure an intuitive user experience.
Frequently used features were placed in easily accessible locations, aligning with common mobile application conventions.
% Furthermore, the platform was designed to be responsive, adapting seamlessly to different devices for optimal usability.
By adhering to these principles, the final application aimed to provide a user-friendly and inclusive experience for a diverse range of users.

% \subsection{Future evaluations}
While this paper focuses on the formative evaluation, the summative evaluation \circled{9}, which follows the formative one, has already been conducted and is discussed in a separate paper \citep{adorni_chbr}. 
This evaluation assesses the effectiveness and impact of the final working prototype on a larger scale.

\section{Platform development}\label{sec:platform_development}

In this section, we present the technical components and architectural choices behind the development of the virtual CAT platform, covering its framework, data management, programming language formalisation, and interpreter implementation.
The latest version of the application, including its full source code and comprehensive documentation, is openly accessible online \citep{adorni_virtualCATapp_2023}.
For a detailed explanation of the system's data infrastructure, including collection and transfer protocols, application features, development process, and intended use cases, refer to the dedicated software paper \citep{adorni_softwarex}.

\subsection{Technical overview}\label{sec:technical_overview}
We focused on iPads as the target platform, aiming to create a user-friendly interface that enhances engagement, particularly for students in K-12 educational contexts. 
The touchscreen interaction offered by this device aligns closely with the intuitive and interactive learning experiences we sought to foster, ensuring a pedagogically effective and enriching experience \citep{scaife1997designing}. 
Additionally, iPads' portability and wide adoption in educational settings made them an ideal choice for our application, ensuring accessibility for a broad range of students.

The application was developed using the Flutter framework, chosen for its extensive capabilities and benefits \citep{flutter}.
Its cross-platform support, which includes Android, iOS, Linux, macOS, Windows, and web, enabled the creation of a single codebase that operates seamlessly across multiple platforms. This approach significantly reduced the time and effort required for platform-specific development.

The framework's hot reload feature played a key role in enhancing the development workflow by allowing real-time previews of code changes, streamlining the iterative design process and improving productivity. Additionally, Flutter provides a rich library of pre-built widgets and tools, simplifying the creation of visually engaging and interactive user interfaces.

Although the application was primarily designed for use on iPads, its responsive design ensures a consistent and visually appealing user experience across various devices and screen sizes. 
This adaptability highlights its flexibility without compromising functionality or aesthetics.

% \subsection{Server and data handling}\label{sec:server_and_data_handling}
% Considering the often limited availability of secure networks in educational settings, we implemented a technical framework to prioritise participant privacy and responsible data management \citep{adorni_virtualCATdatainfrastructure_2023}.
% % Ensuring compliance with data protection regulations and ethical standards for secure data transmission in Switzerland involved several steps.  
% To use this system, a local network infrastructure should be established using a router to connect all devices involved in the activity to a designated computer, which serves as the data collection point. 
% Within this network, a database should be configured to securely receive and store the acquired data from the iPads. Afterwards, all collected data can be transferred from the local database to a dedicated repository through a private network connection.

\subsection{Formal definition of the CAT programming language}\label{sec:commands}

% \textcolor{red}{This section provides a comprehensive textual explanation; however, I acknowledge the need for visual examples to enhance the understanding of concepts such as Cross Representation, Moves, Basic Colouring, Repetition-Based Colouring, and Symmetry-Based Colouring. The inclusion of illustrative images of these elements can offer a more visual and concrete understanding, contributing to clarity and the absorption of concepts by readers. I would recommend incorporating figures that exemplify each of these components, providing a more comprehensive and elucidate visual representation. }

To establish a standardised set of instructions that users could employ within the application interfaces to design the algorithm, we defined the CAT programming language, which codifies and formalises all the commands and actions observed during the original experimental study with the unplugged CAT.

\begin{figure}[h]
    \centering
    \includegraphics[height=3.8cm]{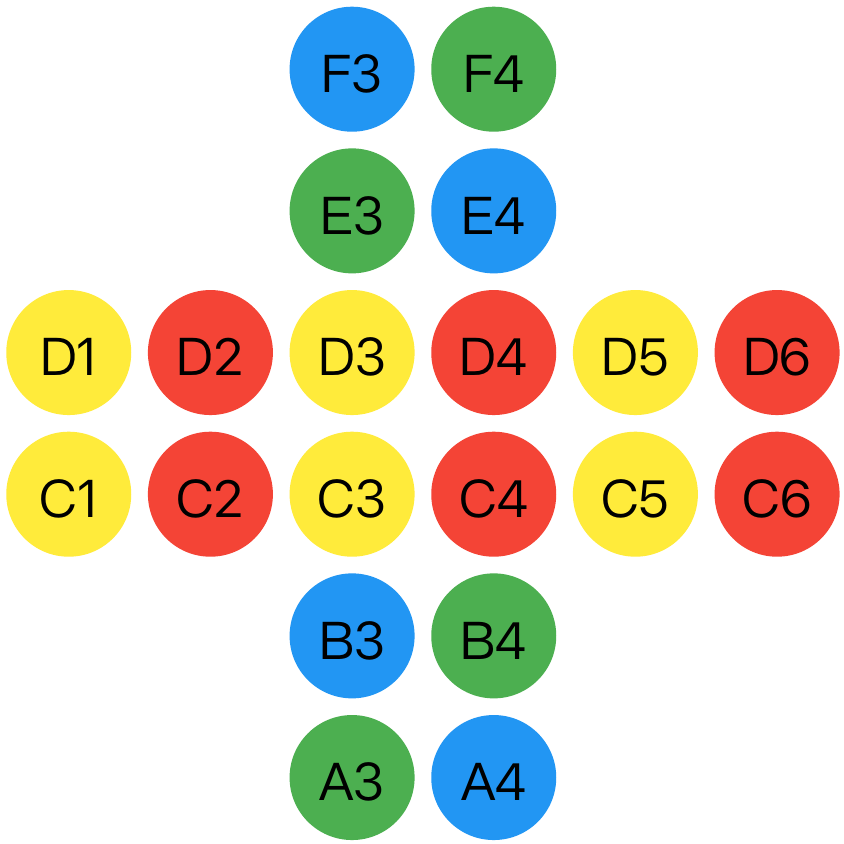}
    \caption{\textbf{Example of a cross-board with coordinate labels.}}
    \label{fig:crossboard}
\end{figure}

% \textit{Cross representation}
The cross-board dots are manipulated and referenced using a coordinate system (see \cref{fig:crossboard}), where rows are labelled from bottom to top using letters (A-F), and columns are numbered from left to right (1-6). 
% Combining a letter and a number allows a specific dot on the board to be easily identified.

% The cross can be constructed by initialising the grid as a $2 \times 2$ matrix of zeros, representing the cross-shaped array. Alternatively, the \texttt{Cross.fromList} constructor allows creating a board from an existing matrix. 
% The \texttt{Cross} class has several properties and methods.
% The \texttt{colors} in the \texttt{Styles} class property defines the possible colours of the dote, \texttt{(WHITE, GREEN, RED, BLUE, YELLOW)}, defined . 
% The \texttt{getGrid} method can be used to retrieve the colours of the cross array dots. On the other hand, the \texttt{grid} method allows direct access to the grid property of the array, allowing modification and manipulation of the grid, such as updating dot colours or checking the status of specific dots.

\begin{figure}[h]
    \centering
    \includegraphics[height=3.8cm]{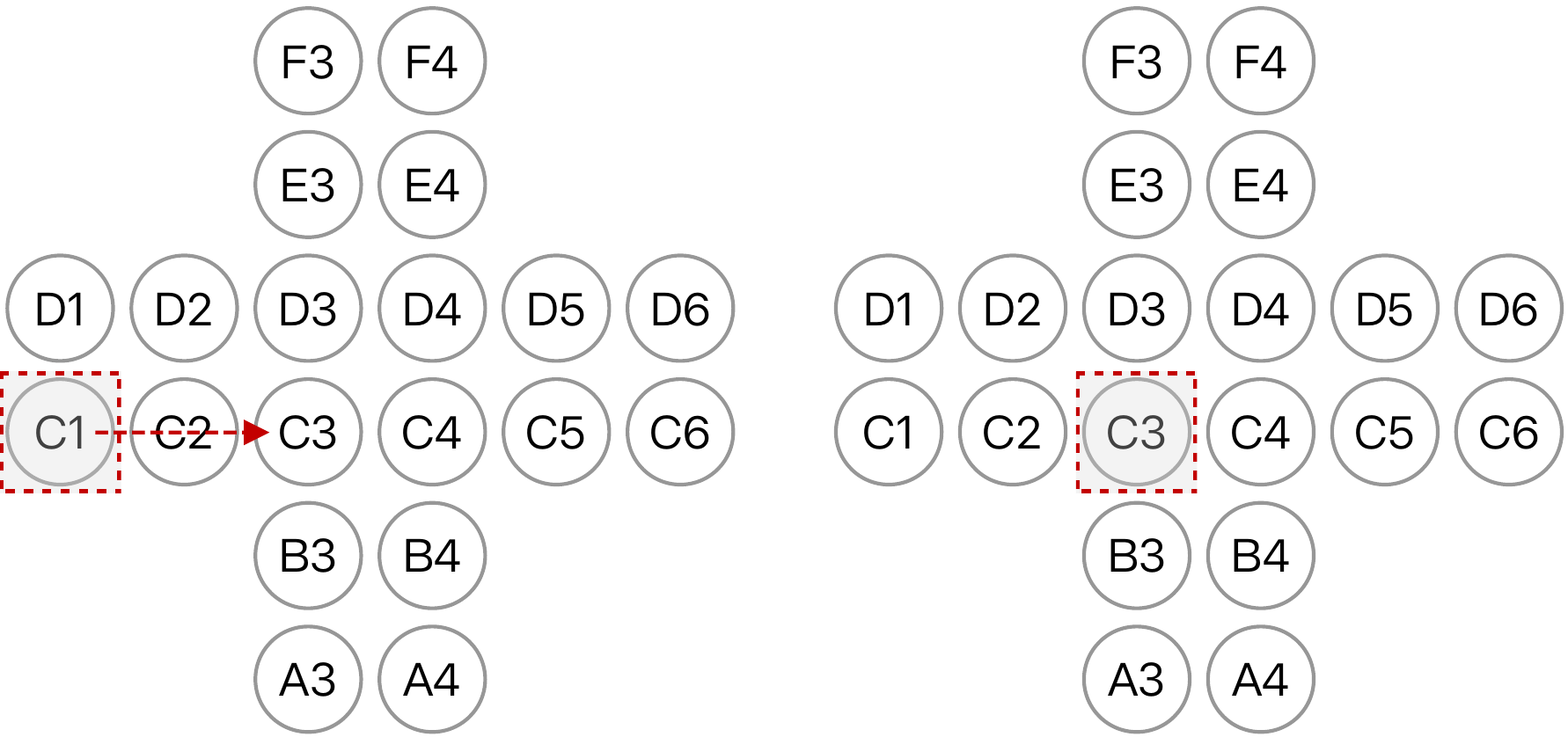}
    \caption{\textbf{Example of movement on the cross-board.} 
    Starting from C1, the destination cell C3 can be reached either by using the \texttt{goCell(C3)} command or by traversing two steps to the right using the \texttt{go(right,2)} command.}
    \label{fig:go}
\end{figure}

% \paragraph{{Moves}}
Moving around the cross-board can be done in two ways (see \cref{fig:go}). 
The \texttt{goCell(cell)} method allows jumping directly to a specific coordinate.
Alternatively, the \texttt{go(move, repetitions)} method allows traversing a certain number of dots in one of the eight available directions (either cardinal or diagonal) to reach the desired destination.
% The available directions are cardinal directions (\texttt{down}, \texttt{left}, \texttt{right}, \texttt{up}) or diagonal movements (\texttt{diagonalDownLeft}, \texttt{diagonalDownRight}, \texttt{diagonalUpLeft}, \texttt{diagonalUpRight}). 

\begin{figure}[h]
    \centering
    \includegraphics[height=3.8cm]{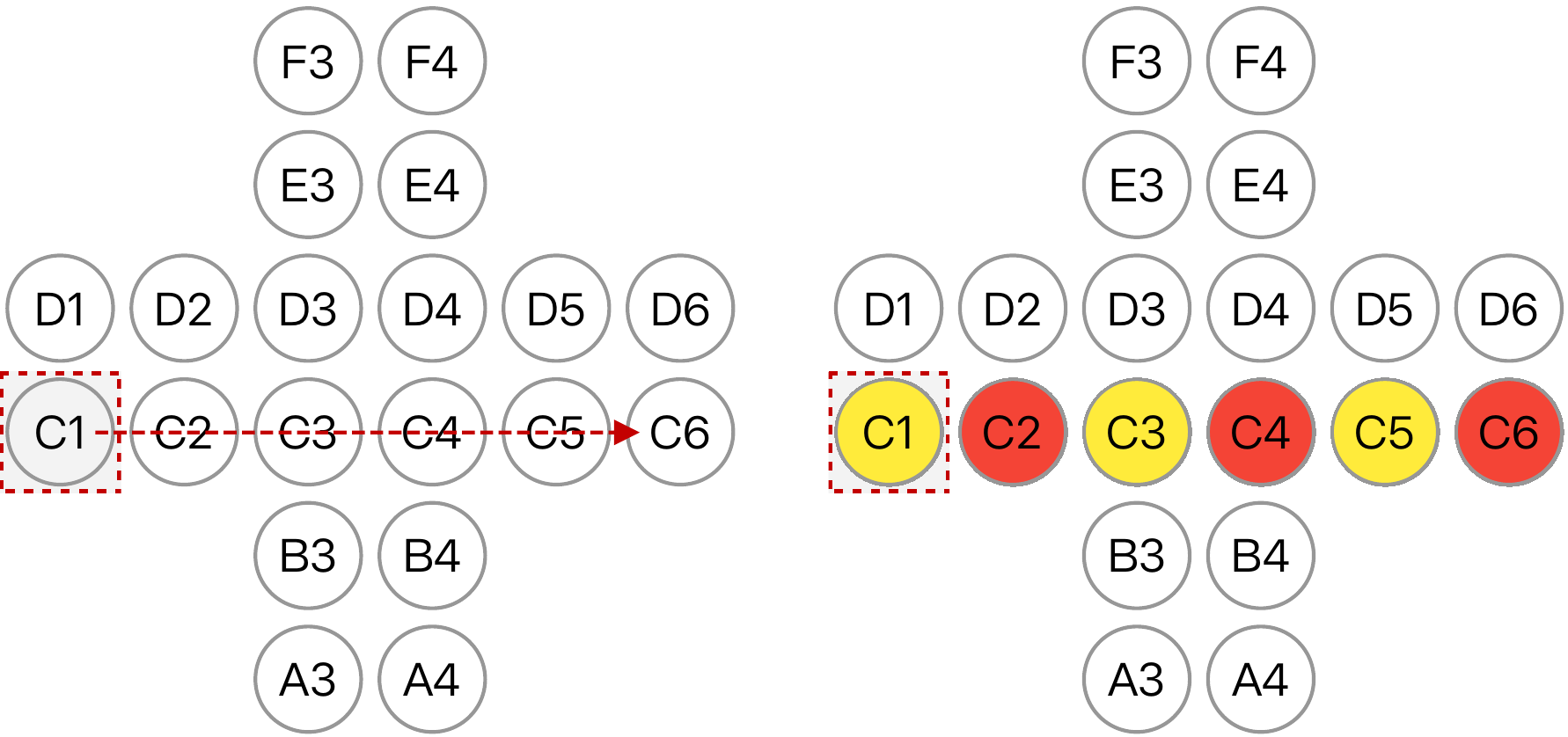}
    \caption{\textbf{Example of colouring a row of six dots.}
    Starting from C1, the row is coloured alternating yellow and red using either the \texttt{paintPattern(\{yellow,red\},6,right)} or the \texttt{paintMultipleCells(\{yellow,red\},\{C1,C2,C3,C4,C5,C6\})} command.}
    \label{fig:paint}
\end{figure}

% \paragraph{{Basic colouring}}
Colouring the board is a fundamental aspect of the CAT application, and we offer various methods to achieve this (see \cref{fig:paint}). 
% The process of colouring the board can be achieved using a range of methods that facilitate the application of the four possible colours, \texttt{(GREEN, RED, BLUE, YELLOW)}, to different patterns on the board. 
The \texttt{paintSingleCell(color)} method allows colouring the dot they are currently positioned on with a single colour. 
The \texttt{paintPattern(colors, repetitions, pattern)} method allows colouring multiple dots according to predefined patterns. 
A sequence of colours can be specified, which will alternate following the selected pattern. 
% The \texttt{repetitions} parameter determines the number of dots to be coloured, following the specified pattern.
% Finally, the \texttt{pattern} parameter specifies the pattern to be used.
Additionally, users can choose from five pattern types (cardinal, diagonal, square, L, zigzag), each with various directions. 
The \texttt{paintMultipleCells(colors, cellsPositions)} method enables colouring multiple dots with custom patterns, defined by specifying the coordinates of the cells to be coloured.
The \texttt{fillEmpty(color)} method colours all the uncoloured dots on the board with the same colour. 

\begin{figure}[h]
    \centering
    \includegraphics[height=3.8cm]{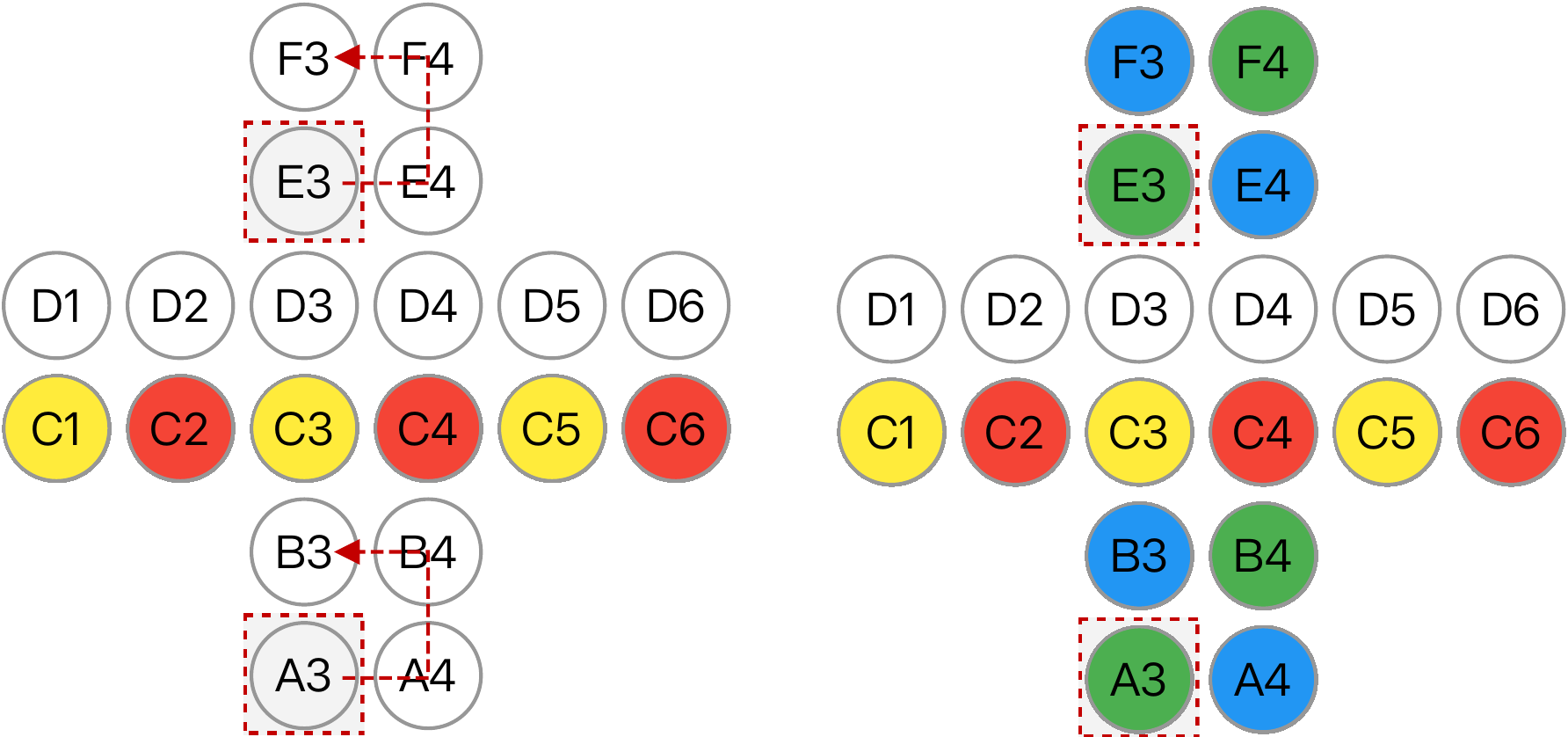}
    \caption{\textbf{Example of repetition of a square pattern.}
    On cells A3 and E3 is coloured a square pattern with green and blue using the \texttt{repeatCommands(\{paintPattern(\{green,blue\},4, square\_right\_up\_left)\},\{A3,E3\})} command.}
    \label{fig:repeat}
\end{figure}

% \paragraph{{Repetition-based colouring}}
Moving beyond the basics, other methods allow for more complex operations, like repetitions (see \cref{fig:repeat}). 
The \texttt{repeatCommands(commands, positions)} method allows specifying a sequence of commands (e.g., a series of \texttt{go} and \texttt{paint} operations) and applying them to specific coordinates.
The \texttt{copyCells(origin, destination)} method copies the colours from origin coordinates to destination coordinates.

\begin{figure}[h]
    \centering
    \includegraphics[height=3.8cm]{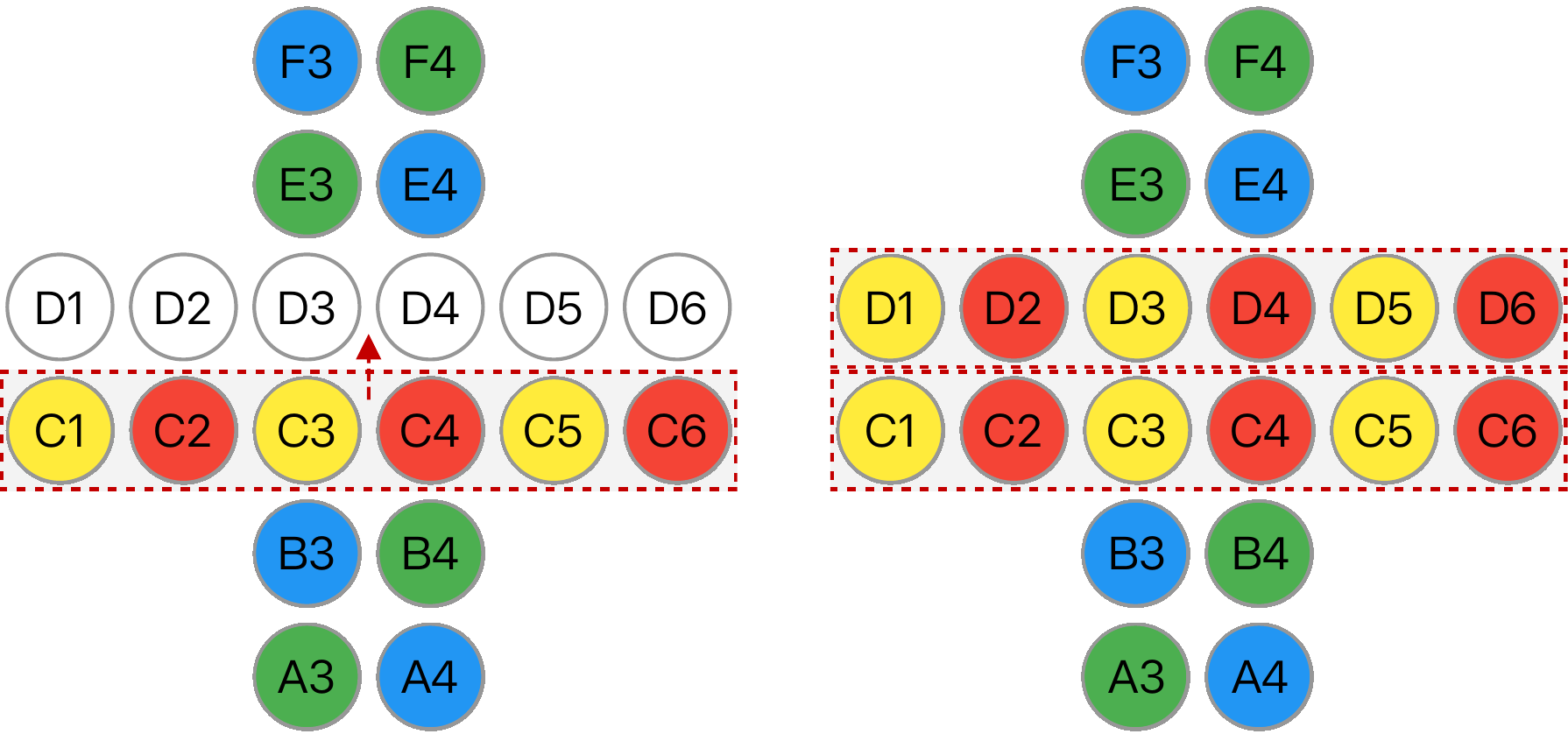}
    \caption{\textbf{Example of cells mirroring.}
    Starting from C1, the row is mirrored upwards along the horizontal axis using the \texttt{mirrorCells(\{C1,C2,C3,C4,C5,C6\},horizontal)} command.}
    \label{fig:mirror}
\end{figure}

% \paragraph{{Symmetry-based colouring}}
Finally, symmetrical colouring approaches are available (see \cref{fig:mirror}). 
The \texttt{mirrorBoard(direction)} method, which reflects the coloured dots on the board onto the non-coloured ones, following the principle of symmetry. This mirroring can be done horizontally on the x-axis or vertically on the y-axis.
The \texttt{mirrorCells(cells, direction)} method performs similar mirroring operations but on a specified set of dots.
The \texttt{mirrorCommands(commands, direction)} method applies the mirroring to a list of commands.

\subsection{Implementation of the virtual CAT interpreter}\label{sec:interpreter}
The virtual CAT programming language interpreter \citep{adorni_virtualCATinterpreter_2023} is a dedicated Dart package that can be integrated into any Flutter project, in our case, the virtual CAT app \citep{adorni_virtualCATapp_2023}.
It translates student actions, including gesture interactions and arranged visual programming blocks, into executable machine-readable instructions.
It analyses the user's input, converting actions into a formal algorithm specified using the CAT programming language.

Each command that composes the algorithm, such as colour selections and other operations, undergoes a validation process to identify and address semantic errors. 
Notably, the interface's design, featuring predefined programming blocks and buttons, obviates the need for syntax checking, as it inherently eliminates the possibility of such errors, significantly streamlining the process.
However, semantic errors can still occur during command execution, for instance, when users attempt to move outside the board boundaries using invalid directions or apply an inappropriate pattern for a colouring command.

Upon validation, the code is executed, and real-time feedback is provided to the user, including the display of current progress on the colouring cross and the CAT score. 
If the interpreter detects errors, it handles them and provides users with error notifications and potential suggestions for correction.

\section{Prototypes}\label{sec:prototypes}

In this section, we present the three prototypes developed throughout the study: the initial prototype, the version refined after expert evaluation, and the final version of the application following the participatory study.

 \subsection{First prototype}\label{sec:first_prototype}

\begin{figure*}[!ht]
    \centering
    \includegraphics[height=9.5cm]{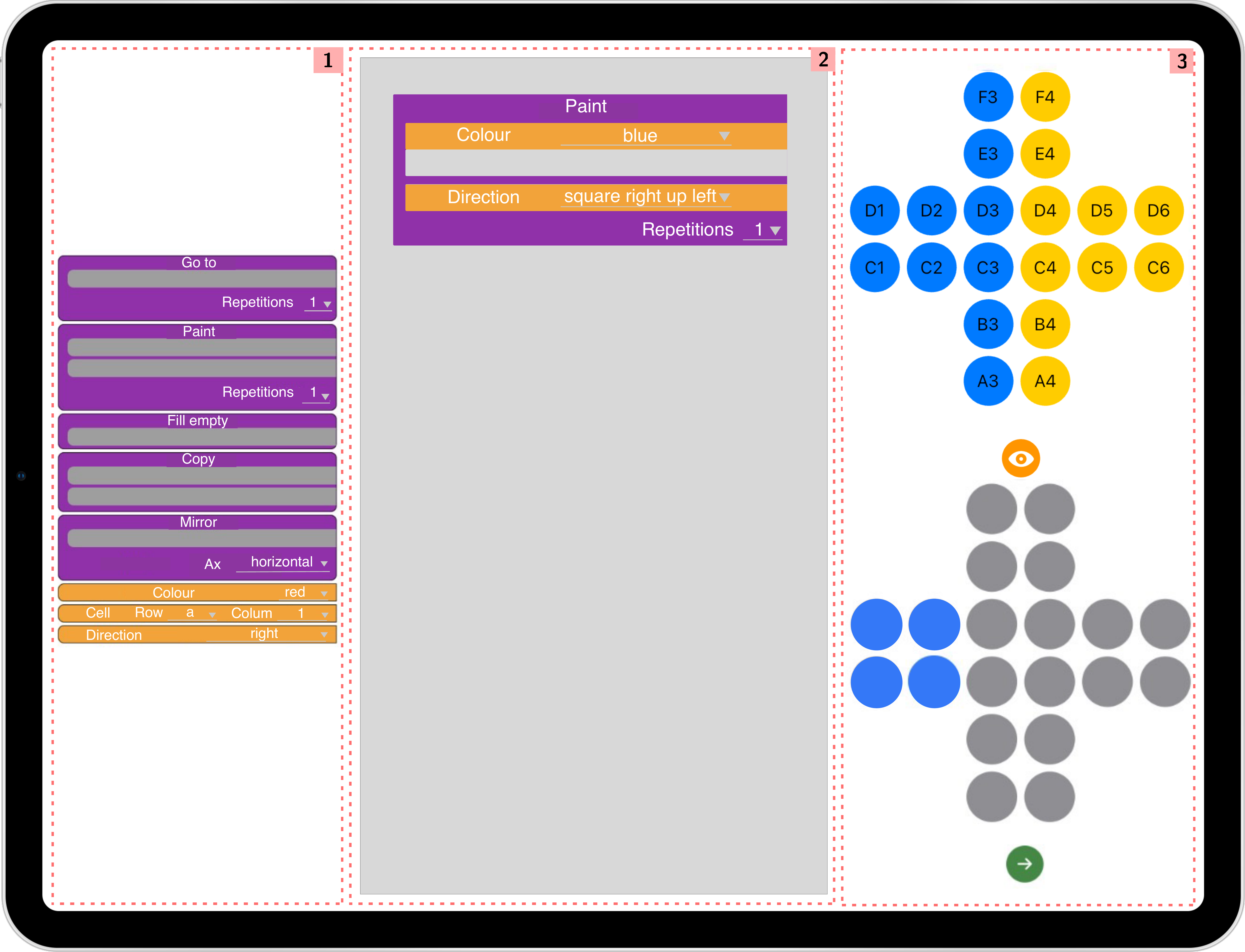}
    \caption{\textbf{First prototype of the CAT-VPI.}}
    \label{fig:1st_prototype_CAT-VPI}
% \end{figure*}
% \begin{figure*}[b]
\vspace{10pt}
    \centering
    \includegraphics[height=9.5cm]{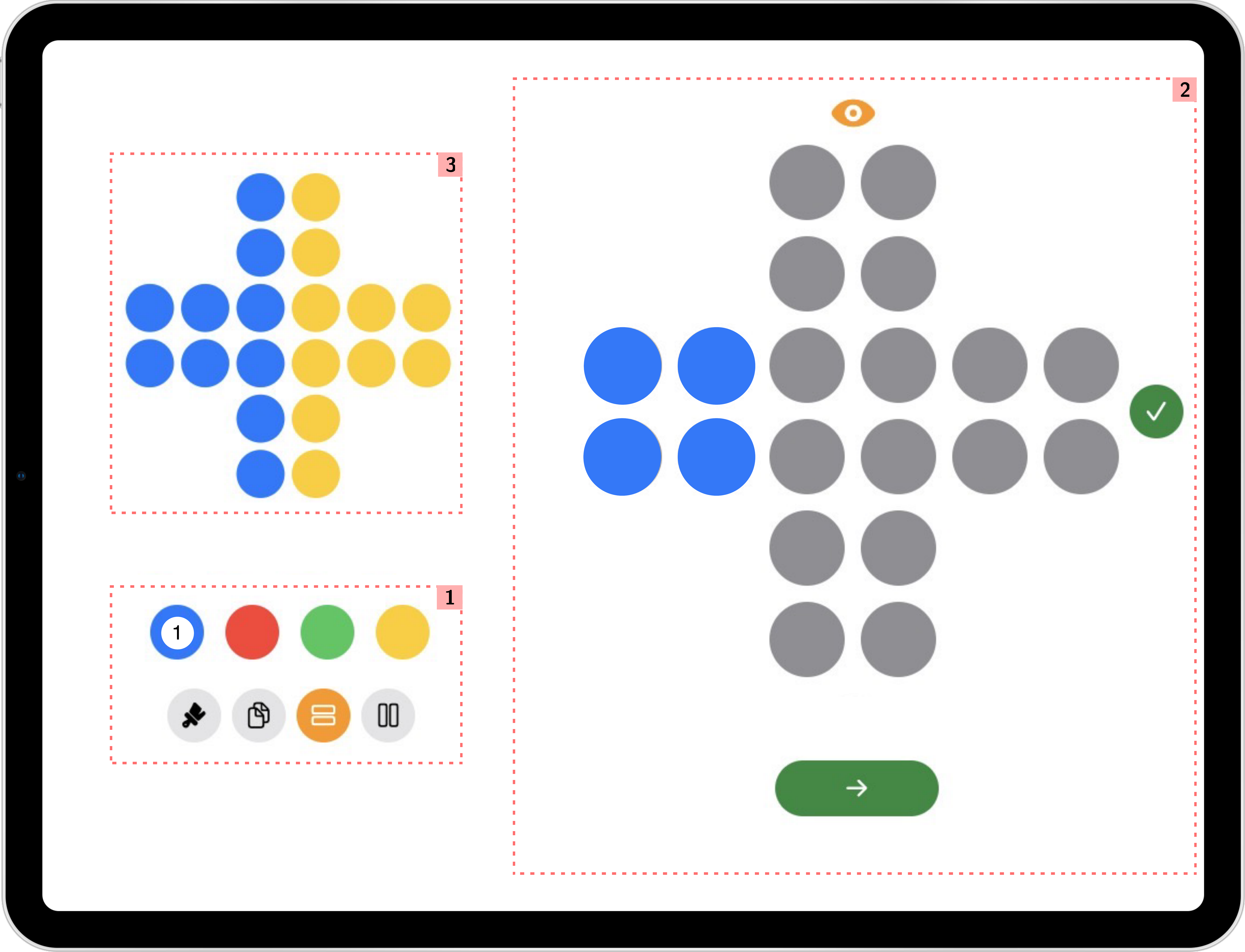}
    \caption{\textbf{First prototype of the CAT-GI.}}
    \label{fig:1st_prototype_CAT-GI}
\end{figure*}

The first prototype of the application was developed to explore and test the system's core functionalities, with the goal of creating a foundational version that experts could evaluate to gather feedback for improving its design and usability.

The CAT-VPI, illustrated in \cref{fig:1st_prototype_CAT-VPI}, features a three-column layout. 
The left column \squared{1} includes predefined code blocks, divided into two types: containers, in purple, are the commands defined in \cref{sec:commands} (i.e., go to, paint, fill empty, copy, and mirror); and components, in orange, that are the inputs for container blocks, such as the colour to be used, the cell to move to or to colour, or the direction for movement or colouring. %Components cannot be used independently. [merge this sentence to the previous and simplify to avoid repetitions]
The central column \squared{2} is the main workspace where users interact with and assemble code blocks.
The right column \squared{3} displays the reference schema to be replicated on top and the colouring schema on the bottom. This section also includes an eye icon to activate visual feedback and a green arrow to proceed to the next schema.

The CAT-GI, illustrated in \cref{fig:1st_prototype_CAT-GI}, presents a different layout.
The bottom left section \squared{1} contains buttons for interaction, including four selectable colours and the key commands defined in \cref{sec:commands} (i.e., fill empty, copy, and two types of mirror).
The right section \squared{2} is the main workspace, featuring a large cross array that students interact with after selecting colours and/or commands. To the right of the cross is a green tick to confirm the completion of a colouring action, an eye icon above it to activate visual feedback, and a green arrow at the bottom to proceed to the next task.
Finally, the top left section \squared{3} displays the reference schema to be replicated.

% User Interfaces
% The platform provides users with two distinct methods to engage with it - a gesture interface (CAT-GI) and a visual programming interface (CAT-VPI). 

\subsection{Second prototype}\label{sec:second_prototype}

% \textcolor{purple}{RISCRIVI NELL'OTTICA: L'INTERAZIONE CON LESPERTO HA PORTATO A DETERMINATE MODIFICHE
% Quando descrivi il feedback del primo esperto ti riferisci a lui un po' al singolare e un po' al plurale. Indicare quali features sono frutto di che tipo di indicazione di chi.}

Following expert evaluation and detailed feedback from UX and pedagogical experts, a series of modifications were made to improve the initial prototype.

% The iterative design process involved close collaboration with interaction design and pedagogical experts. Their insights directly informed several key modifications to the platform, enhancing its usability, accessibility, and educational value. [menzionare che piu info sugli esperti sono data nel metodo...]

\subsubsection{UX expert feedback}
% DESIGNERS
The feedback from the UX expert provided valuable insights into both the strengths and areas for improvement in the prototype.
On the positive side, the expert highlighted the overall clarity of the platform's purpose and its potential to engage users with minimal prior experience. He also appreciated the visual layout, particularly the effectiveness of the workspace design in fostering user engagement.

However, the expert identified two key areas for improvement. 
% 1
First, the interface lacked consistency across different screens, which could confuse users. 
He recommended restructuring the interface to create a more uniform and cohesive design across interaction modalities. 
% 2
Second, he suggested adopting more intuitive icons to enhance the visual clarity and usability of the interface, particularly for users with limited prior experience in using such tools.

\subsubsection{Pedagogical experts feedback}
% PEDAGOGIES
The feedback from the pedagogical experts provided valuable insights into how the platform aligns with educational goals, highlighting areas for improvement to enhance its pedagogical effectiveness. 
The experts praised the platform's visual engagement and the thoughtful integration of AT concepts, acknowledging the design’s clarity and its potential for supporting learning. 
However, they also identified several areas for refinement to further align the platform with best practices in computer science education.

The experts emphasised the educational benefits of allowing students to experiment, make mistakes, and learn from failures. They believe this iterative ``trial and error'' process fosters deeper learning and understanding. To support this, they recommended incorporating a mechanism that encourages ``trial and error'' preventing students from becoming discouraged by early failures.
% 2
While ``trial and error'' can be a valuable strategy, the experts raised concerns about students potentially getting stuck in a loop without making meaningful progress. 
Thus, they also recommended complementing this mechanism with a system to detect when students repeatedly fail or remain inactive for long periods. This would provide targeted hints or guidance to help students reflect on their approach and adjust their strategies, ensuring they continue to move forward without becoming discouraged.
% 3
Additionally, one of the recommendations was to ensure the platform is suitable for all ages. In particular, they suggested revisiting the CAT-VPI interface, as it could be too complex for younger students due to the programming blocks and the amount of text to read.

% \subsection{Prototype refinement and technical implementation} \label{sec:interface}
\subsubsection{Prototype revision}

\begin{figure*}[h]
    \centering
    % Prima figura (a sinistra)
    \begin{minipage}[t]{0.6\textwidth}
        \centering
        \includegraphics[height=8.5cm]{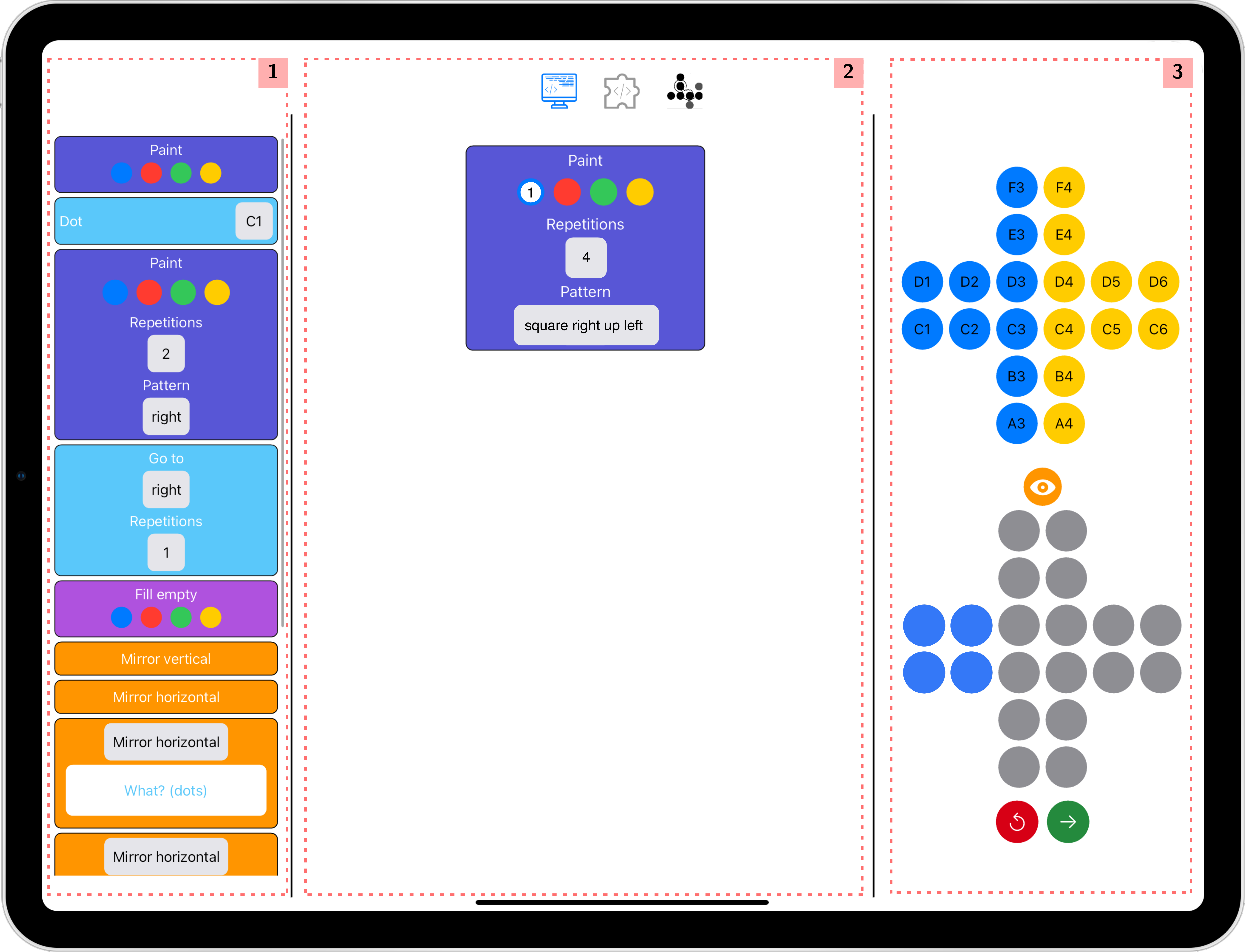}
        % \captionsetup{width=0.6\textwidth}
    \end{minipage}%
    \hfill
    % Seconda figura (a destra con le subfigure)
    \begin{minipage}[t]{0.17\textwidth}
        \centering
        \includegraphics[height=8.5cm]{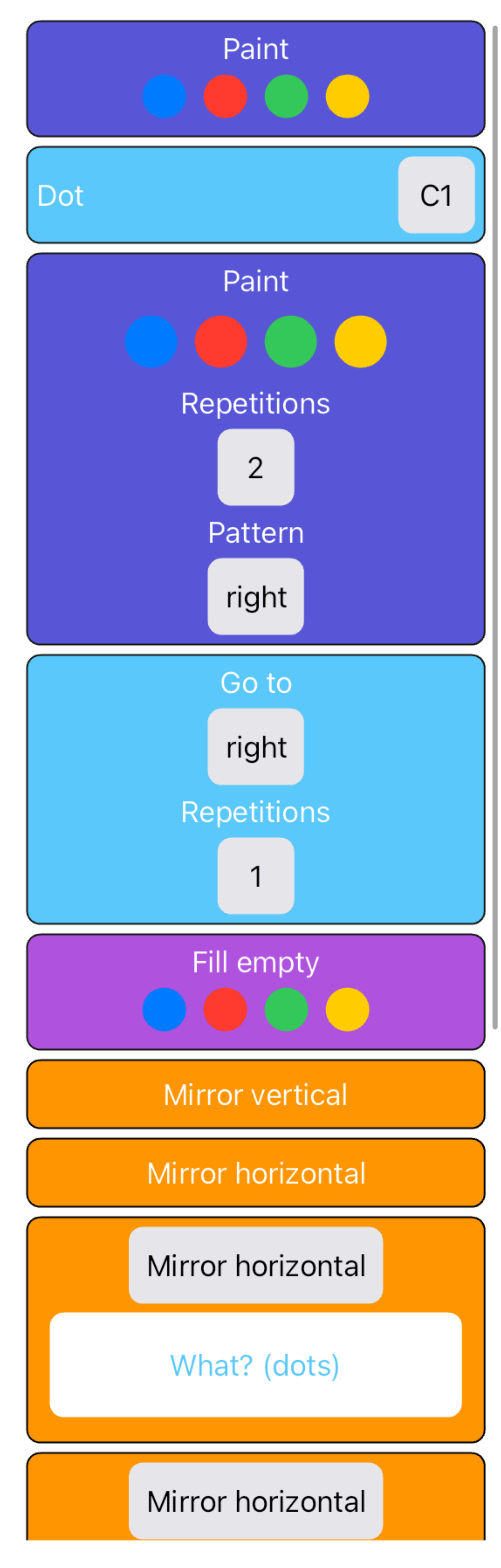}
        % \captionsetup{width=0.15\textwidth}
        % \caption{\textbf{Textual.}}
        % \label{fig:text}
     \end{minipage}
    \begin{minipage}[t]{0.17\textwidth}
        \centering
        \includegraphics[height=8.5cm]{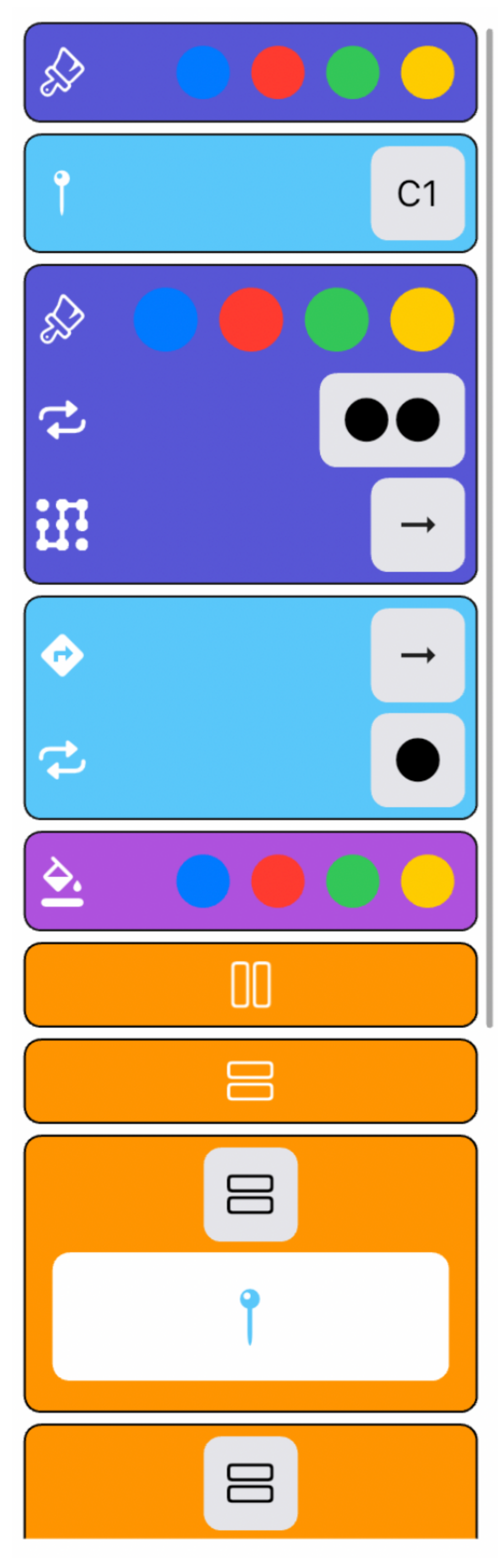}
        % \captionsetup{width=0.15\textwidth}
        % \caption{\textbf{Symbolic.}}
        % \label{fig:symbols}
        % \caption{\textbf{CAT-VPI blocks.}}
        % \label{fig:blocks}    
    \end{minipage}
    \caption{\textbf{Second prototype of the CAT-VPI.}}
    \label{fig:2nd_prototype_CAT-VPI}
        
    \vspace{0.5cm}
    
    % Terza figura sotto le prime due
    \centering
    \includegraphics[height=9cm]{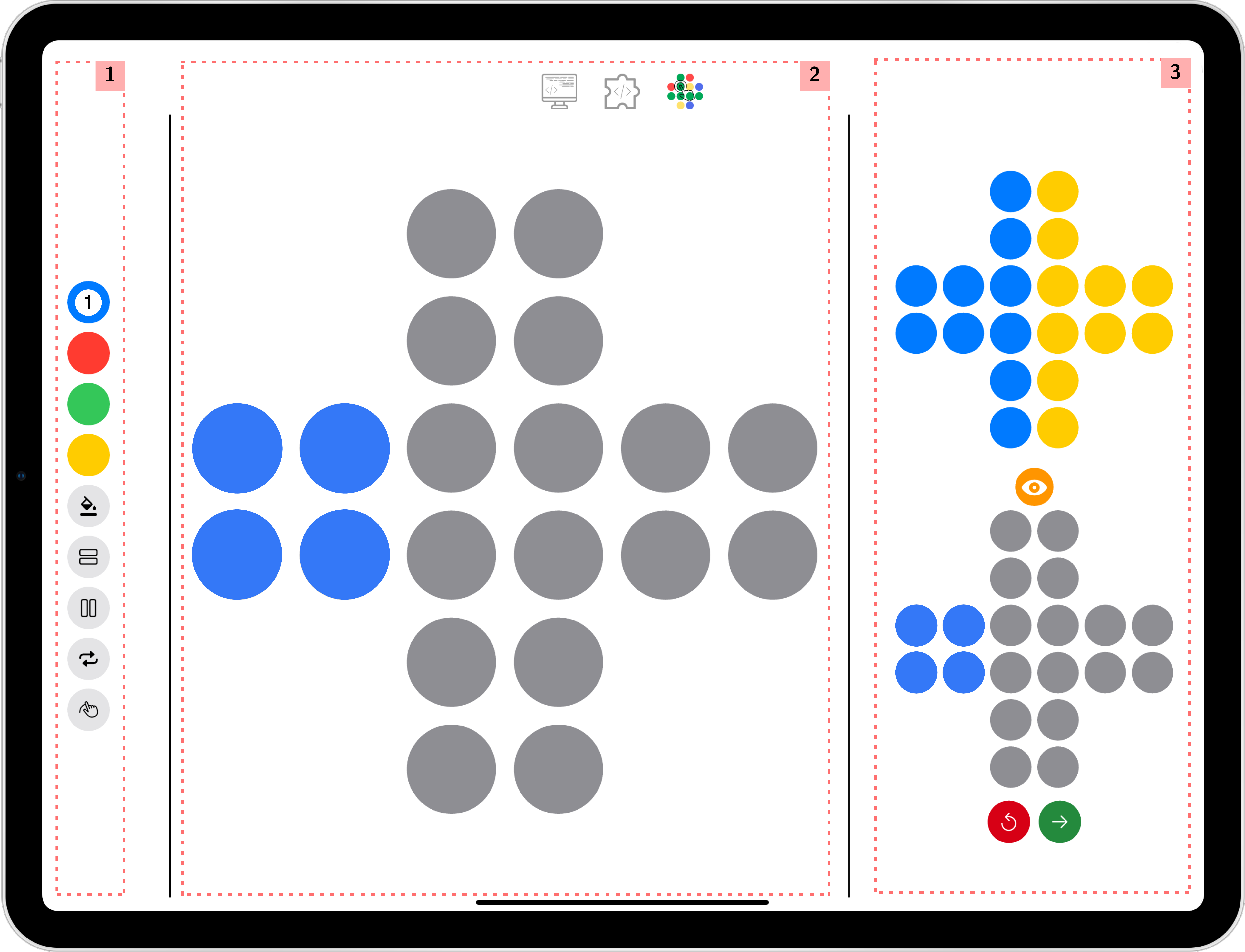}
    \caption{\textbf{Second prototype of the CAT-GI.}}
    \label{fig:2nd_prototype_CAT-GI}
\end{figure*}

% \begin{figure*}[h]
%     \centering
%     \includegraphics[height=9cm]{images/visual_programming_interface.pdf}
%     \caption{ \textbf{Second prototype of the CAT-VPI.}}
%     \label{fig:2nd_prototype_CAT-VPI}
%     \hfill
%     \begin{subfigure}[t]{0.1\columnwidth}
%         \centering
%         \includegraphics[height=9cm]{images/text.pdf}
%         \caption{\textbf{Textual.}}
%         \label{fig:text}
%     \end{subfigure}
%     \hfill
%     \begin{subfigure}[t]{0.1\columnwidth}
%         \centering
%         \includegraphics[height=9cm]{images/symbols.pdf}
%         \caption{\textbf{Symbolic.}}
%         \label{fig:symbols}
%     \end{subfigure}
%     \caption{\textbf{CAT-VPI blocks.}}
%     \label{fig:blocks}    
%     \vspace{.5cm}
%     \centering
%     \includegraphics[height=9.5cm]{images/gesture_interface.pdf}
%     \caption{\textbf{Second prototype of the CAT-GI.}}
%     \label{fig:2nd_prototype_CAT-GI}
% \end{figure*}

To transition to a working prototype for classroom testing, all interfaces now include three buttons at the top centre of the workspace \squared{2} that allow users to switch between interaction modes.

In response to the feedback from the UX expert, we redesigned the user interfaces to ensure consistency between them by adopting the same three-column layout used in the CAT-VPI, illustrated in \cref{fig:2nd_prototype_CAT-VPI}. 
The changes were applied solely to the CAT-GI, as shown in \cref{fig:2nd_prototype_CAT-GI}.
The predefined buttons to select colours and actions are now grouped in the left column \squared{1}.
Previously, the large cross served as both the workspace and the colouring schema. Now, the central section \squared{2} functions as the main workspace, while the right column \squared{3} displays the reference schema at the top and the colouring schema at the bottom, where users can enable visual feedback.
Additionally, new action buttons were added to the CAT-GI, aligning it with the commands available in the CAT-VPI. For example, the ``copy/repeat'' command, absent in the first prototype, has been included, ensuring both interfaces now offer the same set of functionalities.

A major overhaul was conducted to replace the existing icons with more intuitive and universally recognisable symbols.
In the CAT-VPI, the command blocks were simplified for greater clarity. 
The predefined building blocks now use a colour-coding system that groups similar commands together (e.g., indigo for colouring action, orange for the mirror function, etc.).
Most container blocks now come pre-loaded with the necessary components inside, so students don't need to decide which components to include. 
For example, in the case of the paint block, students no longer need to figure out whether to insert the colour or another component, as these are already provided, and they only need to select the component's specific detail, such as the colour. Instructions are provided to guide the student when a component is not pre-loaded.
Other mechanisms were simplified to make the tool more intuitive, streamlining the approach by reducing the steps required for the task and improving the user experience. 
For example, in the previous prototype, for colouring patterns with alternating colours, users had to insert multiple colour components into the paint container and specify the number of repetitions, or cells, to colour.
Now, a dedicated block is available for this operation, where users can select the colours, specify the number of repetitions, and choose a pattern.

In response to the recommendation on accessibility for younger students, we introduced two types of blocks in the CAT-VPI: textual and symbolic (see \cref{fig:2nd_prototype_CAT-VPI}).
Since most kindergarten pupils cannot yet read, this dual approach accommodates a wider range of users. Textual blocks provide instructions for those who can read, while symbolic blocks use intuitive symbols, offering a language-independent way to interact with the system.
This ensures the platform remains accessible to younger, multilingual, or pre-literate students, making the interface engaging for all learners.

% \begin{figure}[h]
%     \centering
% % \hfill
%     \begin{subfigure}[t]{0.485\columnwidth}
%         \centering
%         \includegraphics[height=9.5cm]{images/text.pdf}
%         \caption{\textbf{Textual.}}
%         \label{fig:text}
%     \end{subfigure}
%     \hfill
%     \begin{subfigure}[t]{0.485\columnwidth}
%         \centering
%         \includegraphics[height=9.5cm]{images/symbols.pdf}
%         \caption{\textbf{Symbolic.}}
%         \label{fig:symbols}
%     \end{subfigure}
%     \caption{\textbf{CAT-VPI blocks.}}
%         \label{fig:blocks}
% \end{figure}

Finally, based on feedback from pedagogical experts regarding the ``trial and error'' process, we implemented a ``retry'' button, represented by a red circular arrow at the bottom of the colouring schema in the right section of the interfaces \squared{3}.
This feature allows students to restart exercises anytime, encouraging them to revisit their mistakes, refine their solutions, and engage in iterative learning.

% These modifications (a seguito dell UX expert feedback) aimed to improve the overall user experience, making the platform more intuitive, consistent, and accessible, particularly for novice users.

% These revisions (a seguito dei pedagogical experts feedback)  aim to create a more flexible, supportive, and personalised learning environment that caters to diverse learning styles. 
% By giving students more control over their learning path, the platform encourages self-directed exploration, reduces frustration, and enhances overall engagement. 
% The modifications address the experts' feedback while striving to maintain a balance between guidance and independence, helping students learn more effectively and enjoyably.

% These modifications, informed by expert feedback, contributed to refining the platform and aligning it more closely with its pedagogical objectives.

% The final user interface of the CAT platform, employed in the participatory study detailed in Section~\ref{sec:pilot}, integrates both gesture and visual programming interfaces.
% \textcolor{red}{REVISE}

\subsection{Final application}\label{sec:third_prototype}

\begin{figure*}[!ht]
    \centering
    \includegraphics[height=9.5cm]{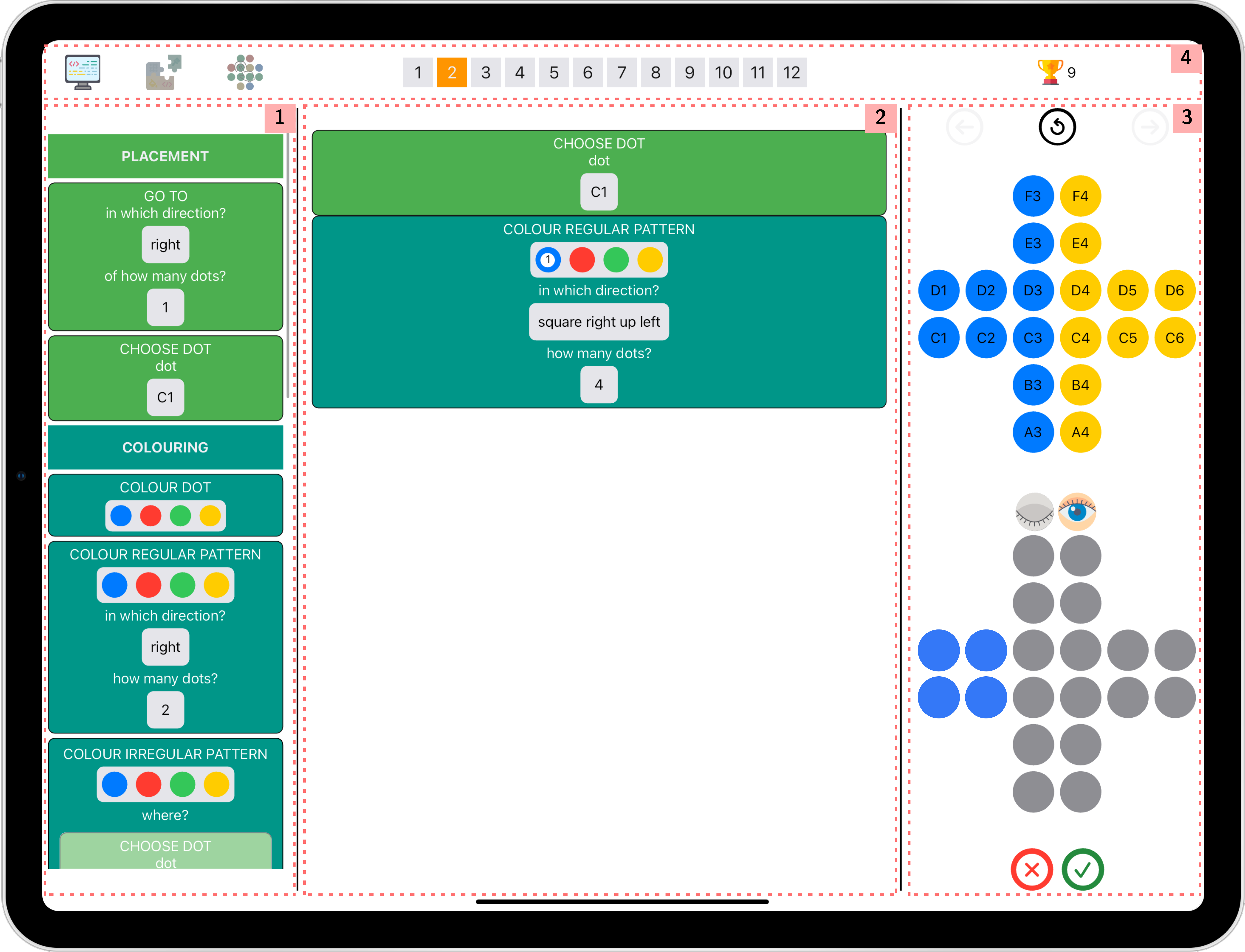}
    \caption{\textbf{Final CAT-VPI.}}
    \label{fig:3rd_prototype_CAT-VPI}
% \end{figure*}
% \begin{figure*}[b]
\vspace{10pt}
    \centering
    \includegraphics[height=9.5cm]{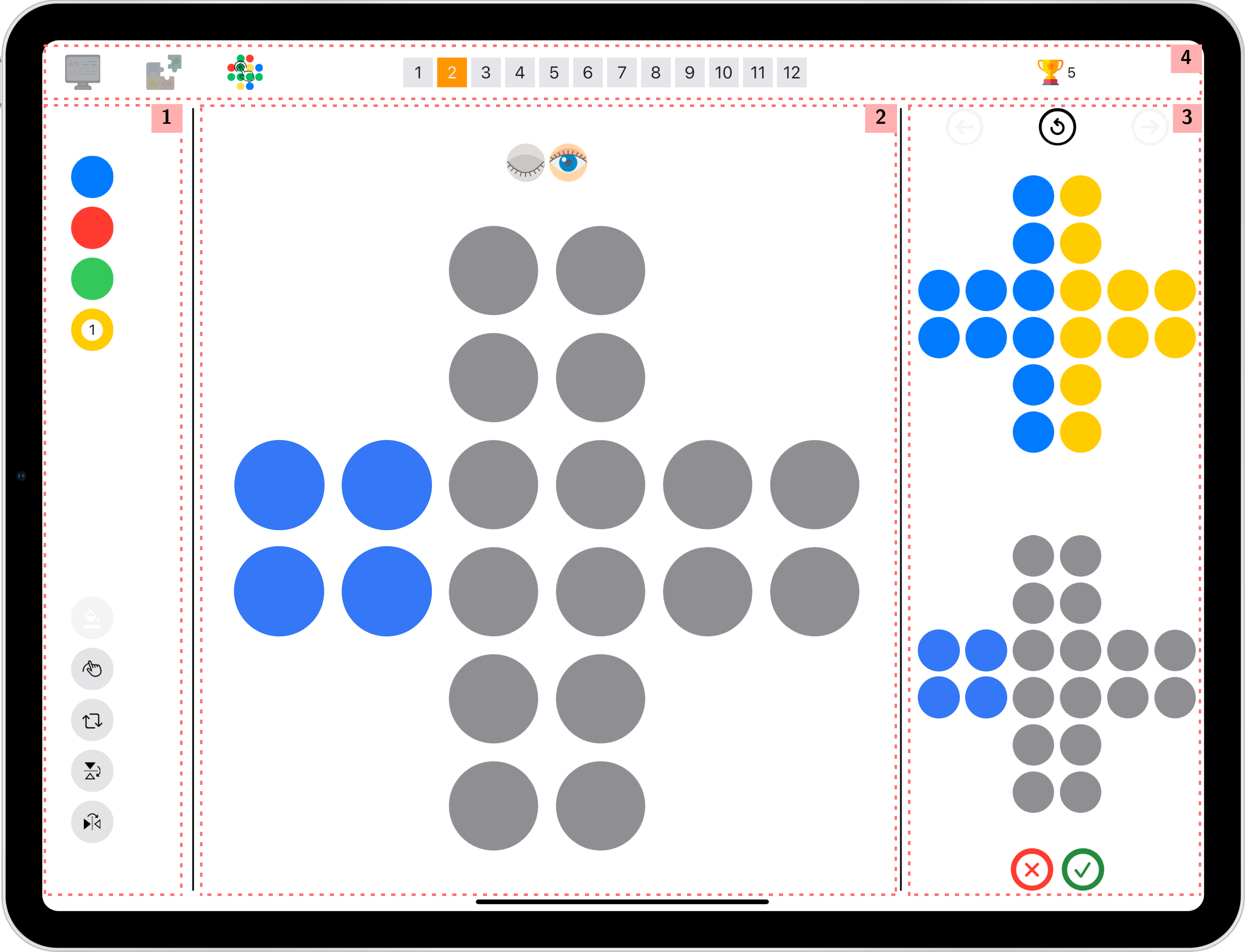}
    \caption{\textbf{Final CAT-GI.}}
    \label{fig:3rd_prototype_CAT-GI}
\end{figure*}

The third and final version of the application was developed through active collaboration with teachers and pupils during the pilot session of the participatory study.
After active collaboration with teachers and pupils from different age groups and schools during the pilot session of the participatory study. 
Feedback and observations of user interactions guided targeted refinements to the interface and functionality, addressing usability issues, enhancing accessibility, and better aligning the platform with the needs of students and educators.

Collaborating with students yielded invaluable insights that guided several critical changes to the platform. 
Initially, we observed that as time was running out, the need to confirm each schema at the end of the activity individually became cumbersome and unnecessary. 
Moreover, this process led to schemas being incorrectly marked as ``failed'' instead of ``not attempted''. To address this, we introduced a ``surrender'' button at the bottom of the right section of the interfaces \squared{3}, allowing users to skip specific schemas.
This feature also proved useful for students who felt stuck and wanted to move on. 

% visual feedback
Another modification stemmed from feedback about the visual feedback button, which some students found unclear. 
In response, we replaced the original button with two new icons: an open eye indicating active visual feedback and a closed eye symbolising that feedback was turned off. % (see Figure~\ref{fig:skipfeedback}).

% progress bar
As students progressed through the activity, some expressed interest in knowing how many schemas remained to be completed. To address this, we added a progress bar at the top of the central column in the interface \squared{4}.
This addition serves a dual purpose: it satisfies students' curiosity about their progress and resolves a limitation observed in the unplugged CAT activity, where the rigid sequencing of tasks restricted students' ability to navigate exercises flexibly. 
% The progress bar helps users stay organised, plan task sequencing, and maintain motivation throughout the activity cycle \citep{hartson2018ux}. 
Additionally, we included navigation arrows at the top of the right section of the interface \squared{3} to allow students to explore upcoming schemas, providing further flexibility and accommodating those who wish to skip ahead.
Further adjustments were made based on the researcher's observations during the study, which highlighted areas for improvement that were not always evident through feedback alone. 
% menus
While observing pupils interacting with the CAT-VPI, it became evident that they were not using all the available commands but only readily visible ones. This was because some commands were not immediately accessible and required scrolling down the column to see them. 
To address this, we grouped related commands into menus in the left column of the interface \squared{1} and revised their colours to improve visibility and accessibility. % (see Figure~\ref{fig:menu}).
% These changes align with usability principles, emphasising the organisation and grouping of design elements to simplify content complexity, improve visibility, accessibility, intuitiveness, and streamline user experience \citep{hartson2018ux}.

% paint pattern
% To provide users with more versatile options for customising basic paint colours, we also introduced a new block command that allows them to create and apply colour patterns to the schemas (see Figure~\ref{fig:custom_pattern}).

% colour parameter
Additionally, we observed that some pupils occasionally forgot to select the colour parameter within the paint blocks.
Thus, we enclosed all customisable parameters within shaded boxes to make it easier for users to identify and adjust them. % (see Figure~\ref{fig:custom_pattern}). 
% This design enhancement aligns with usability principles that recommend indicating active defaults to suggest choices and values and guiding data entry in formatted fields \citep{hartson2018ux}. 

% Nested 
Another observation concerned the use of nested blocks. Despite written instructions, some users struggled to fill these blocks correctly. To improve clarity, we added a transparent representation of the block types that could be inserted within nested blocks and provided more detailed instructions for each label.
% Finally, despite having written instructions, some users found it difficult to understand how to fill nested blocks. To improve clarity and enhance user understanding, we added a transparent representation of the type of blocks that could be inserted within nested blocks, accompanied by more detailed and clearer instructions for each label (see Figure~\ref{fig:nested}).
% This design improvement aligns with usability principles that emphasise assisting users in getting started with a task and providing them with comprehensive instructions to understand how to use different interface elements effectively \citep{hartson2018ux}. 

% impossible buttons
Finally, while observing users interact with the CAT-GI, we noticed issues with certain commands, such as the \texttt{fillEmpty} button, which was often used without selecting a colour first. To address this, we implemented conditional activation of buttons, enabling them only when appropriate for the given context.
% , aligning with usability principles that recommend disabling buttons or menu choices to prevent inappropriate choices and greying out unavailable options \citep{hartson2018ux}.
Additionally, we introduced a visual feedback mechanism, including a shaking effect on incorrect actions and flashing available commands when users deviate from the intended workflow. This feature aims to guide users towards the correct actions, improving the overall user experience.

Following teacher feedback, additional improvements were made to enhance the platform further, focusing on refining the user experience and ensuring the tool met pedagogical and functional needs.
One key suggestion from the teachers was to provide real-time feedback, allowing students to monitor their progress and performance during the activity. 
% Finally, teachers also offered valuable insights. One noteworthy suggestion was to provide real-time feedback to help students monitor their progress and performance during the assessment. 
In response, in the right part of the top bar \squared{4}, we included a display box showing the current score for the ongoing schema. % (see Figure~\ref{fig:skipfeedback}).

% dashboard
Additionally, we introduced a final dashboard that provides a comprehensive summary of student performance across all completed schemas (see Figure~\ref{fig:dashboard}). 
This feature not only aids teachers in quickly assessing student progress but also provides students with a clear overview of their performance, helping them identify areas where they may need to focus more effort.

% video tutorial
Finally, we redesigned the training module to enhance further the platform's feasibility for large-scale data collection and assessment. Instead of requiring an administrator to guide users through the app, we integrated in-app video tutorials, enabling users to navigate the platform independently (see Figure~\ref{fig:tutorial}). 
This change eliminates potential biases that could arise from researcher-led explanations.

% All these refinements aim to enhance the platform's usability, proficiency and suitability.
% We redesigned the training module to make the platform entirely feasible for large-scale data collection and assessment. 
% Rather than relying on a researcher to introduce the app, we've integrated in-app video tutorials to let users navigate the platform independently, thereby eliminating potential biases stemming from researcher explanations (see Figure~\ref{fig:tutorial}).

Our vision for future developments involves continuous refinement and expansion of the platform.
To assess user experience, we decided to incorporate a brief survey at the end of the validation module to gather pupils' subjective impressions and insights into their perceptions of the tool (see Figure~\ref{fig:survey}).
This survey aligns with established UX design techniques for data elicitation \citep{hanington2019universal,hartson2018ux} and the Technology Acceptance Model \citep{surendran2012,hartson2018ux}, assessing factors such as ease of use, perceived usefulness, attitude towards use, and behavioural intention to assess users' acceptance of a system. 
It explores various facets of user interaction, from the clarity of app rules and preferred interaction modes to the perceived difficulty of exercises and overall enjoyment. 
Additionally, it prompts participants to reflect on whether they would use the app again in the future.
% 
% The survey covers various aspects, including perceived usability, satisfaction, emotional impact, and overall acceptance of the platform. These factors are essential for understanding pupils' attitudes and experiences with technology in an educational context.
% It delves into various facets of the user experience, from the initial enjoyment and familiarity with such apps to the clarity of the app's rules and the user's preferred mode of interaction. It also gauges the difficulty level of the exercises and the time taken to complete them. Towards the end, participants are prompted to reflect on whether they'd revisit the app. 
To accommodate the diverse literacy levels and age groups of our users, the survey features an audio playback option for reading questions aloud, ensuring accessibility even for younger students. Responses are collected using a smileyometer scale (happy, neutral, sad), a child-friendly format shown to be effective in assessing children's attitudes toward interactive technologies \citep{Giannakos2022state,read2006,Guran2020}.
% We also considered that our users span different age groups and literacy levels. To accommodate this diversity, we have incorporated a feature that reads the survey questions aloud, ensuring that even younger students who may not be proficient readers can effectively participate.
% The survey format is designed to be engaging and accessible to students. They can respond using emoticons (happy, neutral, sad), which is consistent with research indicating that child-specific data collection methods, like smileyometers, are valuable for assessing children's subjective attitudes toward interactive products \citep{Giannakos2022state,read2006,Guran2020}. 
% By collecting feedback through this well-structured survey, we aim to gain a deeper understanding of pupils' perspectives and further enhance the platform to cater to their specific needs and preferences.

\section{Data analysis}\label{sec:data_analysis}

% \textcolor{red}{Detailed tables and figures illustrating the comprehensive results of the pilot study would greatly enhance its appropriateness and usefulness to readers. ADD STATISTICAL DATA?}

In this section, we present a preliminary analysis of the data collected during the pilot session of the participatory study with the virtual CAT application. 
The data were automatically logged as students interacted with the platform, completing tasks designed to assess their AT skills. 
These logs captured timestamped actions, such as adding, confirming, removing, or reordering commands, modifying command properties (e.g., adjusting colours or directions), and marking tasks as completed or abandoned.
The anonymised data were compiled into a publicly available dataset through Zenodo \citep{adorni_virtualCATdatasetpilot_2023}.

% The aim of this section is to analyse the data gathered during the pilot session to assess whether the application functions as intended and to identify any potential areas for improvement. Specifically, we will examine task completion rates, interaction patterns, and user engagement to evaluate the overall effectiveness of the application.

% \subsubsection{Preliminary data analysis and evaluation criteria}\label{sec:data_analysis}

In this analysis, we are particularly interested in how students engage with the platform. Specifically, we examine: (1) the tool usability across age groups -- whether students of different ages can use all available interfaces effectively; (2) proficiency and success rates -- whether students can work through tasks of varying complexity, demonstrating algorithmic skills from basic to advanced levels; (3) suitability for large-scale assessment -- whether the tool is suitable for large-scale automated assessment of AT skills in K-12 students. 
These factors will help determine if the application can be deployed effectively in educational settings on a larger scale. 
\begin{table}[h]
\footnotesize\centering
\setlength{\tabcolsep}{4.5mm}
\caption{\textbf{Analysis of activity completion time across interaction dimensions.} The table presents a comprehensive overview of the time taken by students to complete all schemas using different interfaces, gestures (G) or visual programming (P), with and without visual feedback (F). The average, minimum, and maximum completion times, in minutes, are reported for each interface.}\label{tab:time_by_artefact}
\begin{tabular}{lccc}
\toprule
 \textbf{Interface}  & \textbf{Avg time}   & \textbf{Min time}   & \textbf{Max time}   \\
\midrule
 GF & 16 min       & 4 min       & 29 min       \\
 G   & 13 min       & 4 min        & 29 min       \\
 PF   & 18 min       & 8 min       & 28 min       \\
 P  & 17 min       & 7 min       & 28 min       \\
 \cmidrule(lr){1-1}\cmidrule(lr){2-2}\cmidrule(lr){3-3}\cmidrule(lr){4-4}
 \textbf{Total}   & 16 min       & 4 min       & 29 min       \\
\bottomrule
\end{tabular}
\end{table}

% We evaluate usability by analysing the completion time across different interaction dimensions. 
Table~\ref{tab:time_by_artefact} reveals that the gesture interface, both with and without feedback, leads to quicker task completion times than the visual programming interface counterparts. 
This observation aligns with expectations, as the gesture interface represents a less complex dimension of the artefactual environment, making it more intuitive and efficient for students.
%MAXIMUM TIME GESTURE 
However, it's interesting to note that when considering the maximum completion times, the gesture interface, particularly with feedback, recorded the longest time. One possible explanation is that less proficient students may gravitate towards the gesture interface, which could lead to longer completion times. 
% This suggests that interface choice may not solely reflect usability but also user proficiency.
%FEEDBAK
It is important to note that students who rely on visual feedback take more time to complete their tasks on average. This could indicate that while feedback aids pupils in task comprehension, it might extend the overall interaction duration as they process and respond to the feedback.
% However, it's important to note that the visual programming interface with feedback exhibits the longest average completion time, possibly indicating that while feedback aids in task comprehension, it might extend the overall interaction duration. 
% This data demonstrates that all pupils, irrespective of age, could use all interfaces provided by the platform.

\begin{figure}[h]
    \centering
    \includegraphics[width=\linewidth]{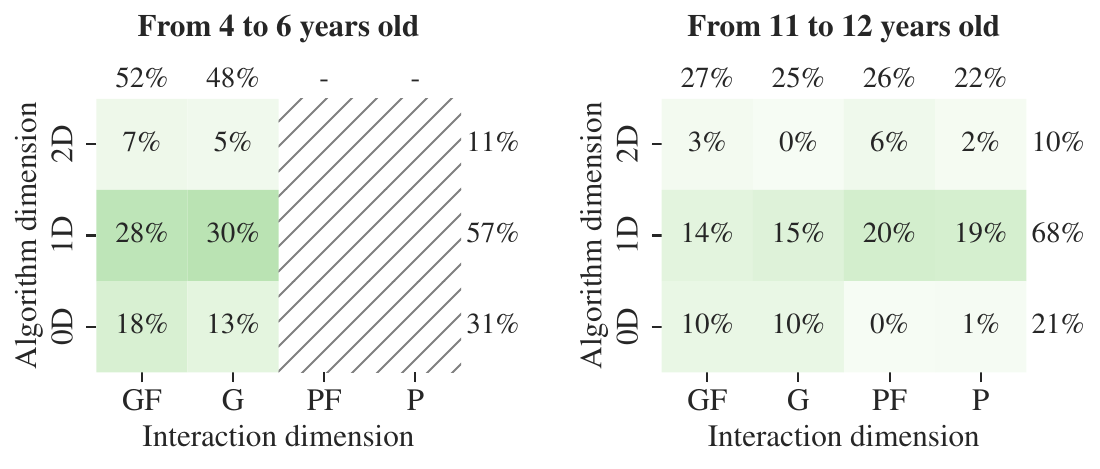}
    \caption{\textbf{Algorithmic and interaction strategies.} 
    The table illustrates the distribution of algorithmic dimensions -- 0D, 1D, 2D -- across interaction dimensions -- gesture interface and visual feedback (GF), gesture interface (G), visual programming interface and visual feedback (PF), and visual programming interface (P) -- for younger and older pupils. 
    Percentages represent the proportion of each combination within their respective age groups.
    It's worth noticing that the younger age category was not allowed to use the visual programming interfaces (PF and P).}
    \label{fig:overall_performance}
\end{figure}

Figure~\ref{fig:overall_performance} provides insights into younger and older pupils' strategies to solve the tasks based on the algorithmic and interaction dimensions.
Pupils across different age groups display a balanced usage of multiple interfaces and autonomy levels on the platform, underscoring the application's ability to accommodate diverse interaction preferences.
Moreover, both groups demonstrate proficiency in generating algorithms across all three algorithmic dimensions, with 1D algorithms being the most commonly used.
% Among younger pupils, about 31\% employed straightforward 0D algorithms, while only a tiny portion used 2D algorithms, indicating a preference for simplicity. 
% For older pupils, 1D algorithms were even more prevalent, with a higher percentage of pupils adopting this approach. Additionally, there was a reliance on 2D algorithms, demonstrating a combination of sequential and branching strategies in their problem-solving.
% non si capisce bene la transizione...
Interestingly, while younger pupils consistently use all available interfaces without significant variation across algorithmic complexity levels, older pupils exhibit distinct patterns in their choice of interaction modes based on the complexity of the task.
Specifically, simpler algorithms are predominantly constructed using the CAT-GI, which offers an intuitive and cognitively light interaction style. In contrast, the CAT-VPI is favoured for more advanced tasks, as it provides greater flexibility and supports the creation of complex algorithms. This adaptability among older pupils highlights their strategic use of the platform's features to address tasks of varying difficulty, emphasising the tool’s capability to support diverse skill levels and interaction strategies.

\Cref{tab:performance_by_schema_and_age} provides an overview of task success rates for the 12 schemas across two age groups.
Not all pupils complete every schema due to time constraints or classroom-related interruptions, while some younger pupils discontinued participation, likely due to attention span limitations typical for their age group \citep{Bennett2006,Mahone2012}.
The overall success rate of approximately 75\% suggests that pupils across age groups could engage effectively with the tasks.
% Schemas 9 and 10, in particular, recorded impressively high percentages of 92\% and 88\%, respectively. 
% These high success rates underscore the platform's effectiveness in supporting students in addressing a wide range of algorithmic challenges. It indicates that the tool is not only accessible but also proficient in guiding students through tasks of varying complexity.
% From the table, it is apparent that 
Older students generally performed better, while younger pupils showed more variation in success. [GIVE A REASON]
Even the most challenging tasks, designed to be intentionally difficult, were successfully completed by many pupils, demonstrating the platform's potential to support learners across varying skill levels through its flexibility and diverse interaction modes.

% These outcomes reflect the intentional design strategy to increase task difficulty as schemas progress. For example, Schema 7 posed a significant challenge across all age groups, registering the lower success rate.
% Interestingly, some students successfully tackled even more complex schemas. This demonstrates the platform's flexibility and adaptability, demonstrating its capacity to accommodate a range of difficulty levels.
% Nonetheless, the lower success rates on more challenging schemas highlight areas where students might benefit from additional support or guidance.
\begin{table}[ht]
\footnotesize
\centering\caption{\textbf{Analysis of student performance across age categories and schemas.} The table presents the number and percentage of students who attempted and solved each schema for each age category. 
The percentage of ``solved'' schemas is calculated only among pupils who attempted it.}\label{tab:performance_by_schema_and_age}
\setlength{\tabcolsep}{2.8mm}
    \begin{tabular}{crrr}
        \toprule
        \multirow{2}*{\textbf{Schema}}
         &   \multicolumn{3}{c}{\textbf{Num. pupils who solved the schema}} \\
        &  \multicolumn{1}{c}{3-6 years old} &  \multicolumn{1}{c}{10-13 years old} & \multicolumn{1}{c}{Total} \\
        \midrule
         1 &  3/6 \, (50\%) &  22/24 (92\%)  & 25/30 (83\%)  \\
         2 &  3/5 \, (60\%) &  21/24 (88\%)  & 24/29 (83\%)  \\
         3 &  4/6 \, (67\%) &  19/23 (83\%)  & 23/29 (79\%)  \\
         4 &  4/6 \, (67\%) &  17/20 (85\%)  & 21/26 (81\%)  \\
         5 &  6/6   (100\%) &  16/20 (80\%)  & 22/26 (85\%)  \\
         6 &  2/6 \, (33\%) &  21/22 (95\%)  & 23/28 (82\%)  \\
         7 &  1/5 \, (20\%) &  12/20 (60\%)  & 13/25 (52\%)  \\
         8 &  2/5 \, (40\%) &  18/21 (86\%)  & 20/26 (77\%)  \\
         9 &  3/4 \, (75\%) &  20/21 (95\%)  & 23/25 (92\%)  \\
         10 & 3/5 \, (60\%) &  18/19 (95\%)  & 21/24 (88\%)  \\
         11 & 2/4 \, (50\%) &  14/18 (78\%)  & 16/22 (73\%)  \\
         12 & 1/3 \, (33\%) &  14/17 (82\%)  & 15/20 (75\%)  \\
        \bottomrule
    \end{tabular}
\end{table}

Finally, to assess the suitability of the virtual CAT for large-scale assessment, as discussed earlier in \cref{sec:cat} comparing the virtual and unplugged CAT versions, the virtual CAT significantly improves efficiency and scalability for large-scale assessments.
Unlike the unplugged version, which required individual, time-consuming administration (36 hours of data collection for all 109 participants), the virtual CAT allows for simultaneous assessment across an entire class, provided each student has access to a device, opens up the possibility of conducting assessments across multiple class groups...
% a comparison with the unplugged CAT was necessary. 
% The original CAT, designed for one-on-one interactions between a student and a specialist, proved impractical for administering to an entire class with a single expert, resulting in a time-consuming and unsuitable choice for large-scale assessments.
% The virtual CAT introduced a pivotal shift in the administration process. Unlike the unplugged CAT, which necessitated a time-intensive individual administration process, totalling approximately 36 hours of data collection for all 109 participants, the virtual CAT's administration is contingent on device availability. 
% Providing each student with individual devices allows the activity to be orchestrated for the entire class simultaneously and seamlessly. This transition drastically reduces the overall time investment required. Moreover, it opens up the possibility of conducting assessments across multiple class groups.
Additionally, the virtual CAT offer a major advantage by automating data collection, eliminating the need for manual entry and further streamlining the overall process. 
However, the structure of the training module used in this pilot study, which required a human administrator to guide students through the platform, posed a challenge for large-scale implementation, as this reliance could introduce inconsistencies in the explanations given to different student groups, potentially affecting performance.
To address this limitation and enhance the platform's scalability for broader use, as discussed in \cref{sec:third_prototype}, we redesigned the training module by integrating standardised in-app video tutorials. This change ensures consistent instructions for all users, minimising potential biases introduced by varying researcher-led explanations, and supports more efficient large-scale data collection and assessment.

\section{Discussion and Conclusion}\label{sec:discussion}
%Start with a brief summary of the key results of your pilot study. This gives readers an overview before going into detail.

% \hl{RIVEDERE QUESTO ELENCO, FORSE RIPETE GLI EXPERIMENTAL.}
% % 1
% Feedback from students and teachers led to interface improvements, such as better command visibility and intuitive button placement.
% % 2
% The addition of a progress bar increased students' awareness of their task completion status.
% % 3 
% Clearer visual cues, like open and closed eye icons for visual feedback, were introduced to improve user understanding.
% % 4 
% The inclusion of a ``skip'' button provided students with additional flexibility during tasks.
% Noteworthy improvements included better command visibility, intuitive button placement, the addition of a progress bar for task completion status, and clearer visual cues like open and closed eye icons for visual feedback. Moreover, a "skip" button was introduced to provide students with additional flexibility during tasks.
In this paper, we presented the iterative design and evaluation of the virtual CAT platform, with a focus on assessing students' proficiency in algorithmic skills, as well as evaluating the tool's usability, accessibility, and suitability for large-scale assessment among K-12 students.

Each version of the platform was evaluated from different perspectives to ensure it met the needs of various stakeholders. The first prototype was assessed by UX design experts and pedagogists, who provided feedback on the platform’s user experience and educational relevance. The second prototype was tested by students from different age groups and teachers, ensuring that the platform was intuitive, engaging, and easy to integrate into educational contexts.

In terms of \textit{usability} and \textit{accessibility}, the platform was designed to be intuitive and accessible, allowing students from varying developmental stages to interact with it effectively.
By analysing the frequency of usage and task success rates across different age groups, we observed balanced interface usage, demonstrating the platform's versatility and ability to cater to learners from diverse backgrounds. This ensures that students can engage effectively with the assessment tasks, regardless of their age or prior experience \citep{Ghai2024,Ismail2024}.

Regarding \textit{proficiency} and \textit{success rates}, our analysis showed that students from different age groups approached algorithmic tasks with varying levels of complexity. The platform supports both basic and more advanced AT, facilitating a wide range of student abilities.
Additionally, the high engagement and success rates across age groups demonstrated that students of all ages were not only engaged but also able to complete the tasks successfully. This reflects the platform's effectiveness in motivating students and supporting their learning \citep{zhan2019,Singh2002,Caruth2018,Collie2019}.

Finally, we assessed the \textit{suitability} of the platform for large-scale assessments by evaluating the feasibility for widespread use.
We analysed time requirements, resource demands, and the potential for automation, ensuring the platform could scale to handle widespread use without compromising its effectiveness. 
Our findings show that the virtual CAT is well-equipped to handle extensive assessments efficiently. This aligns with prior research that emphasises the potential of technology-enhanced assessments to provide rich data and support formative assessment practices in educational settings \citep{Sweeney2017,Chiu2020}.

\subsection{Limitations and future works}
Several limitations should be considered in this study. 
First, the evaluation was conducted within a specific context, focusing on educational settings in Switzerland, and specifically in one canton. This limits the direct applicability of the findings to the broader Swiss context, as well as to other countries with different curricula and teaching approaches. 
Further research would benefit from a more diverse set of educational environments to assess the instrument's effectiveness across various contexts.

Additionally, the limited small sample size of 31 students, while appropriate for a pilot study, restricts the generalisability of the findings. 
Larger-scale studies with more participants have since been conducted, and the results are presented in a separate paper \citep{adorni_chbr}. 
This subsequent study includes a broader range of students from various regions, educational backgrounds, and age groups, offering more robust evidence of the platform's effectiveness. Furthermore, advanced statistical analyses were applied to validate the framework, moving beyond the pilot phase and providing a deeper assessment of the platform's impact and effectiveness.

Technical issues, such as server disconnections, data loss, and interruptions due to time constraints, class schedules, or student attention spans, might have impacted task attempts and success rates, particularly among younger pupils. 
For this reason, in the final application we included offline functionality and automatic saving of progress, and more flexible time management options to mitigate these interruptions.

Another limitation of this study is the lack of consideration for individual differences in learning styles and user satisfaction, which could significantly influence the results. To address this, we integrated a survey in the final version of the platform, as discussed in \cref{sec:third_prototype}, to capture user feedback on their learning experiences and satisfaction levels. 
Future studies could build on this by further examining how these individual differences affect engagement and performance outcomes. 
By correlating survey responses with task performance, researchers could gain a deeper understanding of the factors that influence user interaction with the platform, offering insights for further refining the tool to better meet the diverse needs of students across various age groups and backgrounds.

Limitations related to access to technology still pose a challenge, especially for students without regular access to electronic devices or stable internet connectivity, as well as those with limited technological skills. In the studies conducted so far, we have already provided the necessary devices and infrastructure to ensure all participants can engage with the platform. However, integrating the platform into regular classroom settings could present difficulties, particularly in schools where access to technology is limited or inconsistent. To address this, future studies could explore strategies for ensuring equitable access, such as collaborating with schools to provide devices or designing the platform to be more compatible with a variety of devices and internet conditions. Additionally, offering training to students with limited technological skills could help reduce disparities and facilitate more equitable participation in digital assessments.

Finally, a limitation of this study is the absence of an adaptive feedback or tutoring mechanism, which was suggested during the expert evaluation by pedagogists but has not yet been implemented.
This feature is crucial for personalising the learning experience and offering timely support to students. In future versions, we plan to integrate an adaptive feedback system that provides guidance after repeated failures or periods of inactivity, encouraging reflective problem-solving and offering constructive suggestions when necessary. Additionally, we will explore mechanisms to detect when students are not making progress and offer guidance to help them move forward. 
We have already attempted to build an Intelligent Tutoring and Assessment System (ITAS) based on Bayesian networks for more nuanced assessment, specifically focusing on the unplugged CAT \citep{adorni2023rubric, Antonucci2022Flairs, softcom}. 
This model can be easily integrated into the virtual CAT application and combined with tutoring features in future research for personalised feedback.

\section*{Ethical approval}
This study adhered to EPFL Human Research Ethics Committee's (HREC) ethical standards and received approval (HREC No: 048-2023).
Participants and, for those under 12, their parents or legal guardians provided informed consent; those over 12 also gave assent. Following Swiss and international guidelines, data was handled confidentially, with pseudonymisation for participant protection.

\section*{Data availability}
The data supporting this research is available in a Zenodo repository: \url{https://doi.org/10.5281/zenodo.10018292} \citep{adorni_virtualCATdatasetpilot_2023}.

\section*{Software availability}
The software components used in this study are open-source: virtual CAT platform (\url{https://doi.org/10.5281/zenodo.10027851}) \citep{adorni_virtualCATapp_2023}, 
virtual CAT programming language interpreter (\url{https://doi.org/10.5281/zenodo.10016535}) \citep{adorni_virtualCATinterpreter_2023}, 
virtual CAT data infrastructure (\url{https://doi.org/10.5281/zenodo.10015011}) \citep{adorni_virtualCATdatainfrastructure_2023}.

\section*{Funding}

This research was funded by the Swiss National Science Foundation (SNSF) under the National Research Program 77 (NRP-77) Digital Transformation (No: 407740\_187246).

\section*{Declaration of competing interest}
%%Non anonymised
The authors declare that they have no known competing financial interests or personal relationships that could have appeared to influence the work reported in this paper.

\section*{Acknowledgements}
We express our gratitude to Simone Piatti and Volodymyr Karpenko for contributing to the implementation of the CAT language interpreter and the virtual CAT application.

We extend our appreciation to the contributions of Lucio Negrini, Francesco Mondada, Francesca Mangili, Dorit Assaf, and Luca Maria Gambardella, members of this research project, and Jérôme Guillaume Brender and Giovanni Profeta.
They generously devoted their time and expertise to testing the application and providing valuable feedback. Their insights and support have been precious throughout the refinement and usability of the application.

% \section*{Declaration of generative AI and AI-assisted technologies in the writing process}
% During the preparation of this work, the authors used ChatGPT and Grammarly to enhance language and readability. 
% After using this tool/service, the author(s) reviewed and edited the content as needed and take(s) full responsibility for the content of the publication.

\printcredits

% \newpage
% \appendix
\newpage
% \mbox{~}
% \newpage

%\acresetall
% \clearpage
% \appendixpageoff
% \appendixtitleoff

% \renewcommand{\appendixname}{Supplementary material}
\appendix
% \renewcommand{\thesection}{\arabic{section}}    %%%% but here

%\section*{Supplementary Information}
\renewcommand{\thefigure}{\thesection.\arabic{figure}}
\setcounter{figure}{0}
\renewcommand{\thetable}{\thesection.\arabic{table}}
\setcounter{table}{0}
    
    % \renewcommand{\theHtable}{Appendix.\thetable}
    % \renewcommand{\theHfigure}{Appendix.\thefigure}
% \onecolumn
\section{\textbf{Screens of the final application}}\label{appendix:appA}
\begin{appendixfig}
    \centering
    \includegraphics[width=.75\textwidth]{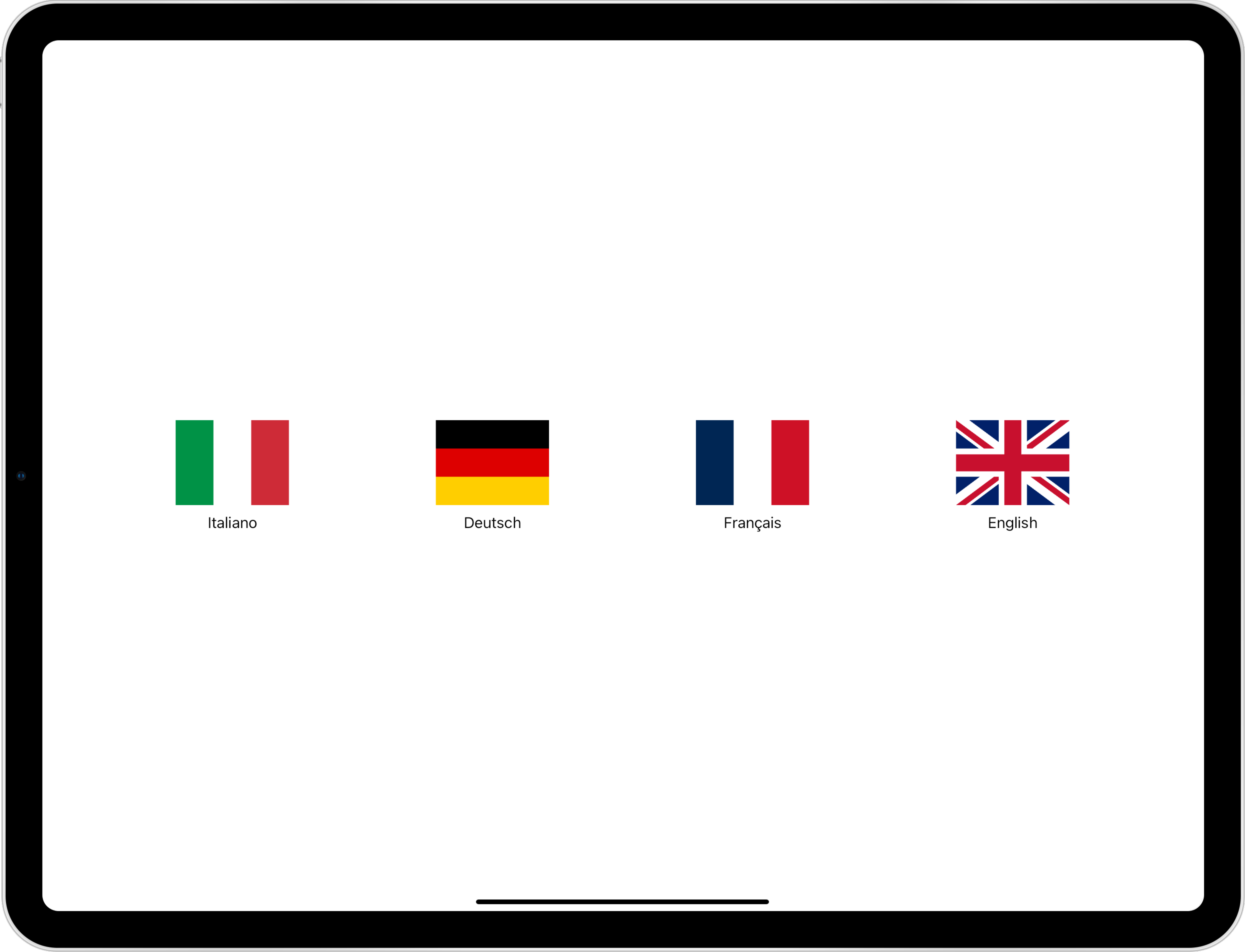}
        \captionof{figure}{\textbf{Language selection.}}
        \label{fig:language}
\end{appendixfig}

\begin{appendixfig}
    \centering
    \includegraphics[width=.75\textwidth]{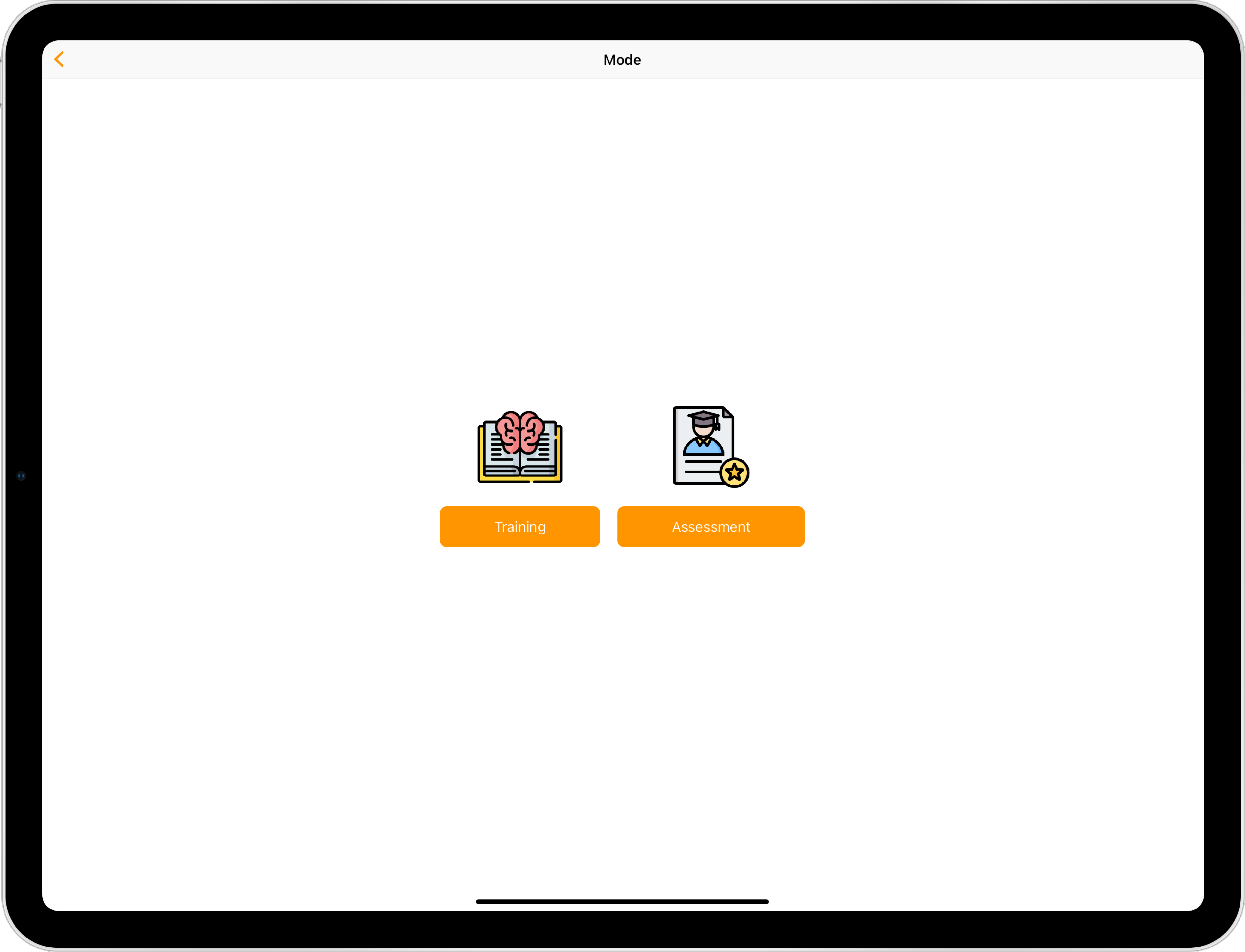}
        \captionof{figure}{\textbf{Module selection.}}
        \label{fig:mode}
\end{appendixfig}

\newpage
\mbox{~}
\newpage

\begin{appendixfig}
    \centering
    \includegraphics[width=.75\textwidth]{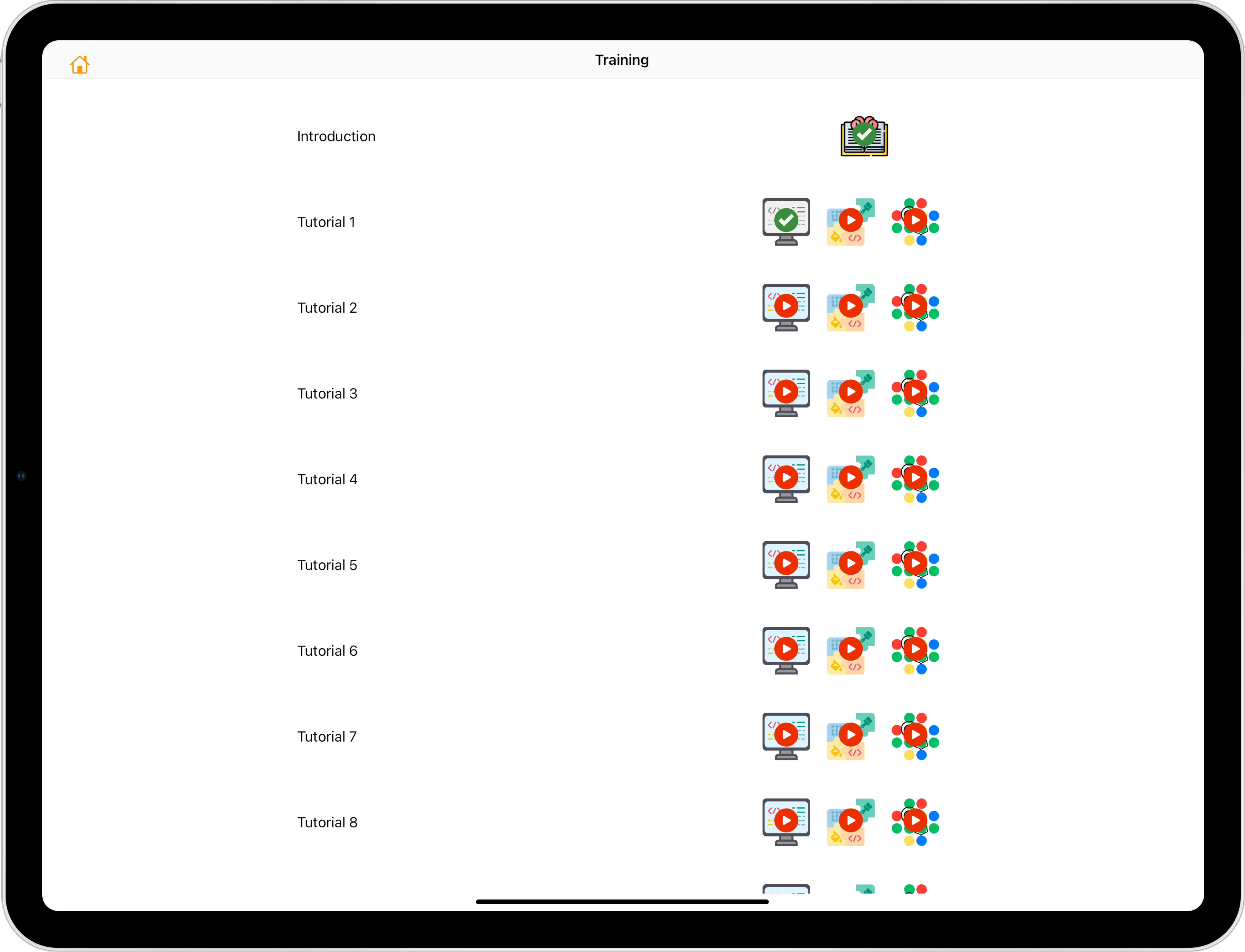}
        \captionof{figure}{\textbf{Training module.} An introductory video about the application is provided on the training screen, followed by a series of explanatory videos for all practice tasks in each interface. 
        After watching the video, users can attempt to solve the schema using the provided instructions. 
        When a schema is successfully solved, the video icon is marked with a green checkmark.}
        \label{fig:tutorial}
\end{appendixfig}

\begin{appendixfig}
    \centering
    \includegraphics[width=.75\textwidth]{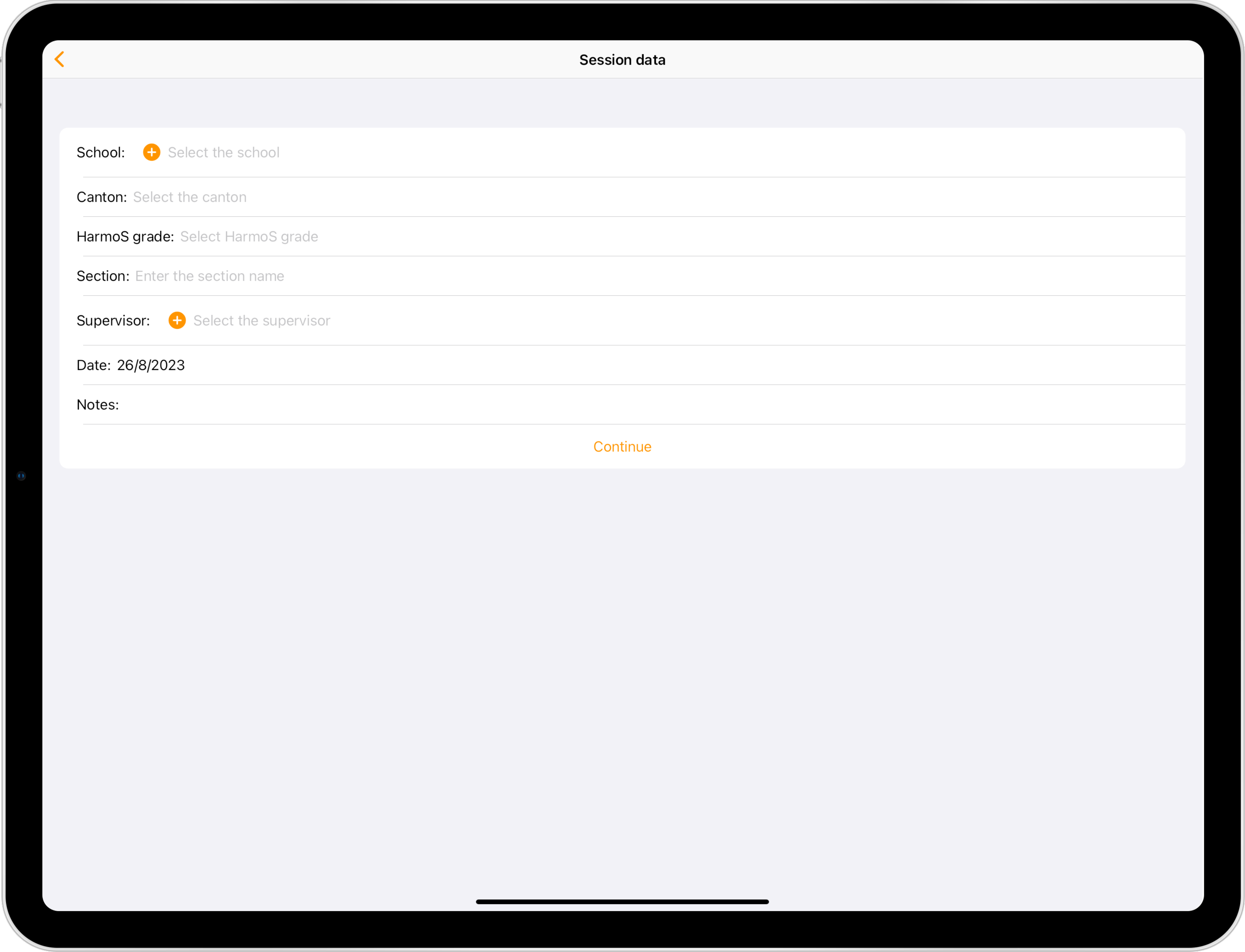}
        \captionof{figure}{\textbf{Session form in the validation module.}}
        \label{fig:session}
\end{appendixfig}

\newpage
\mbox{~}
\newpage

\begin{appendixfig}
    \centering
    \includegraphics[width=.75\textwidth]{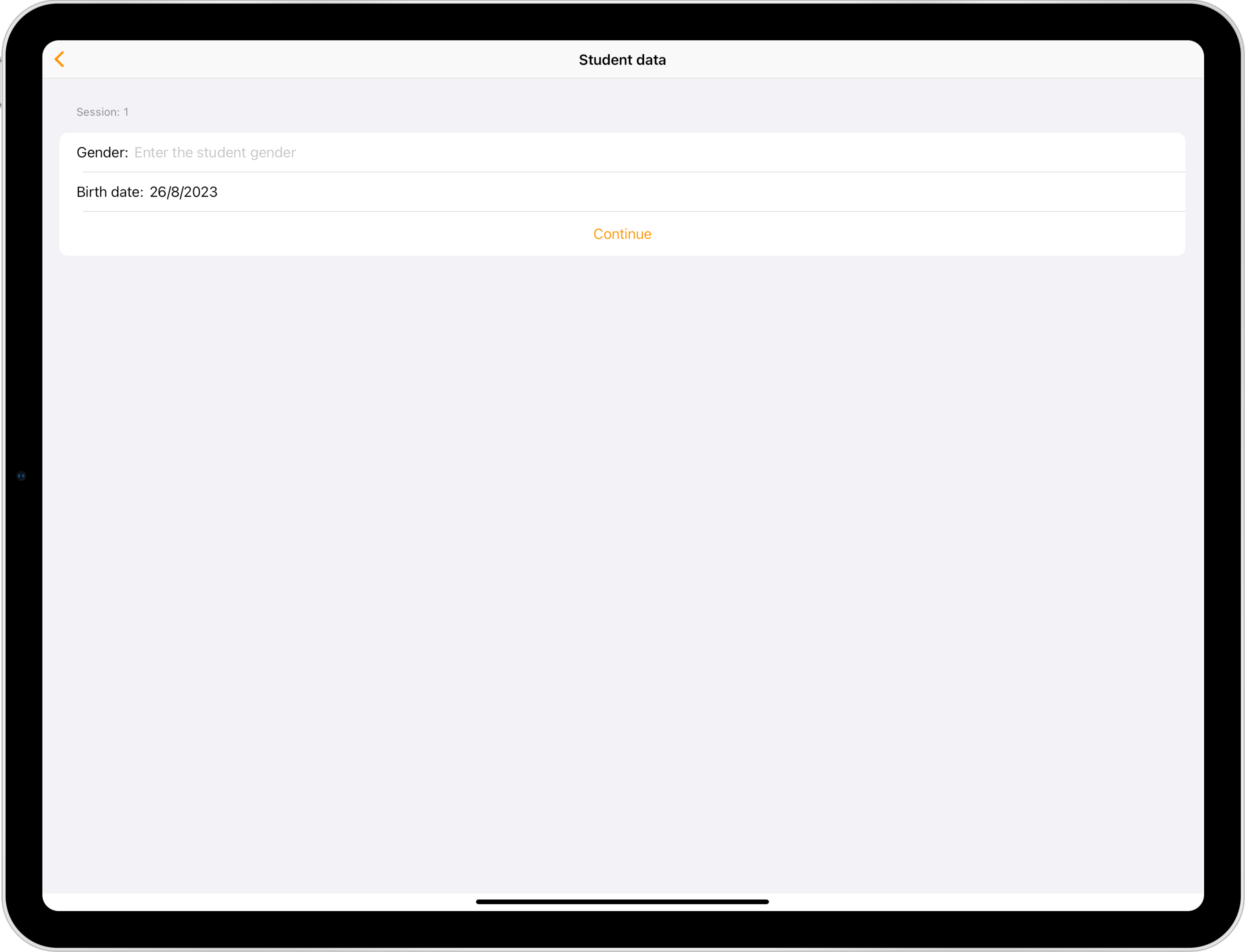}
        \captionof{figure}{\textbf{Student form in the validation module.}}
        \label{fig:student}
\end{appendixfig}

\begin{appendixfig}
    \centering
    \includegraphics[width=.75\textwidth]{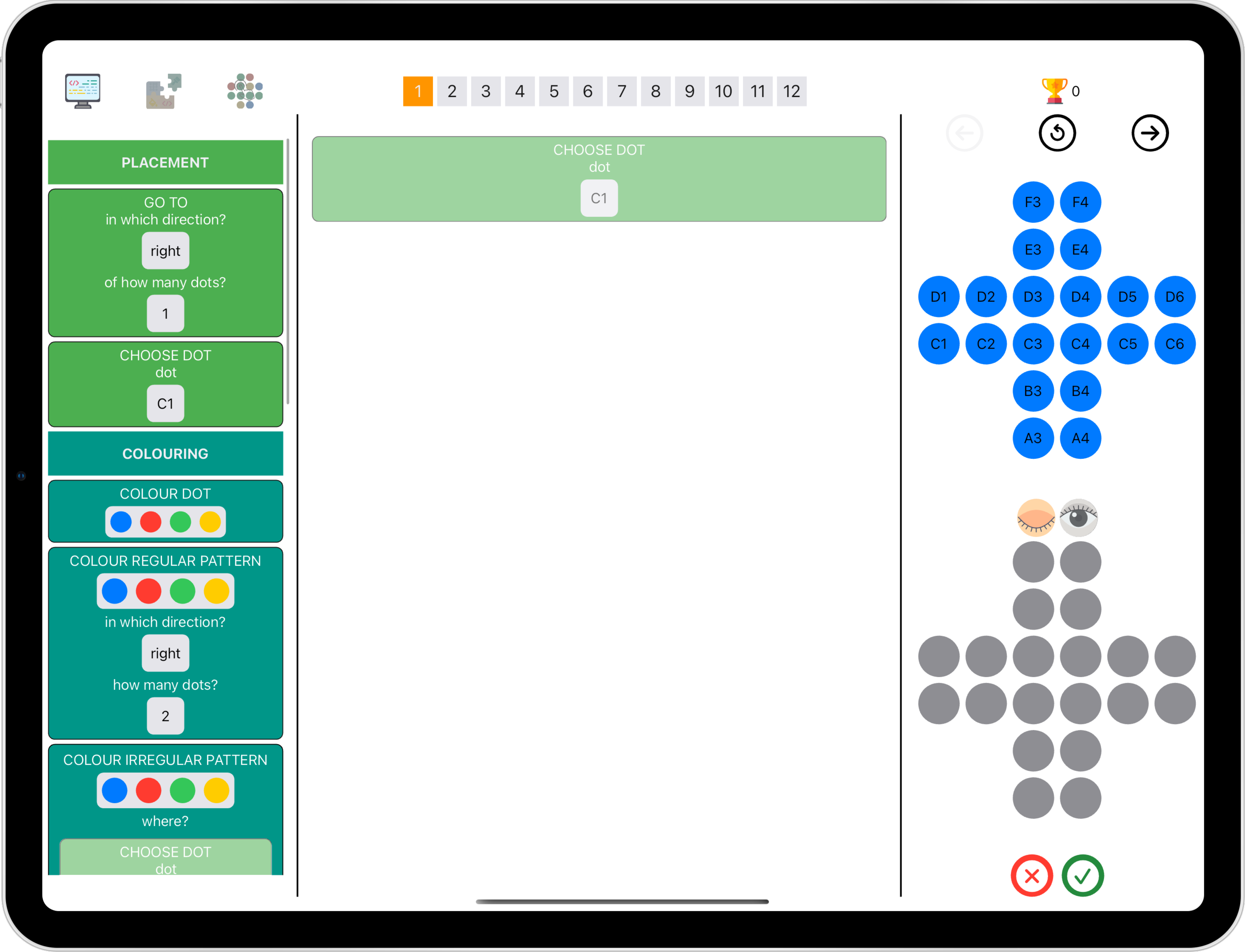}
    \captionof{figure}{\textbf{CAT visual programming interface (CAT-VPI) with textual commands.} }
    \label{fig:programming_final}
\end{appendixfig}
\newpage
\mbox{~}
\newpage

\begin{appendixfig}
    \centering
    \includegraphics[width=.75\textwidth]{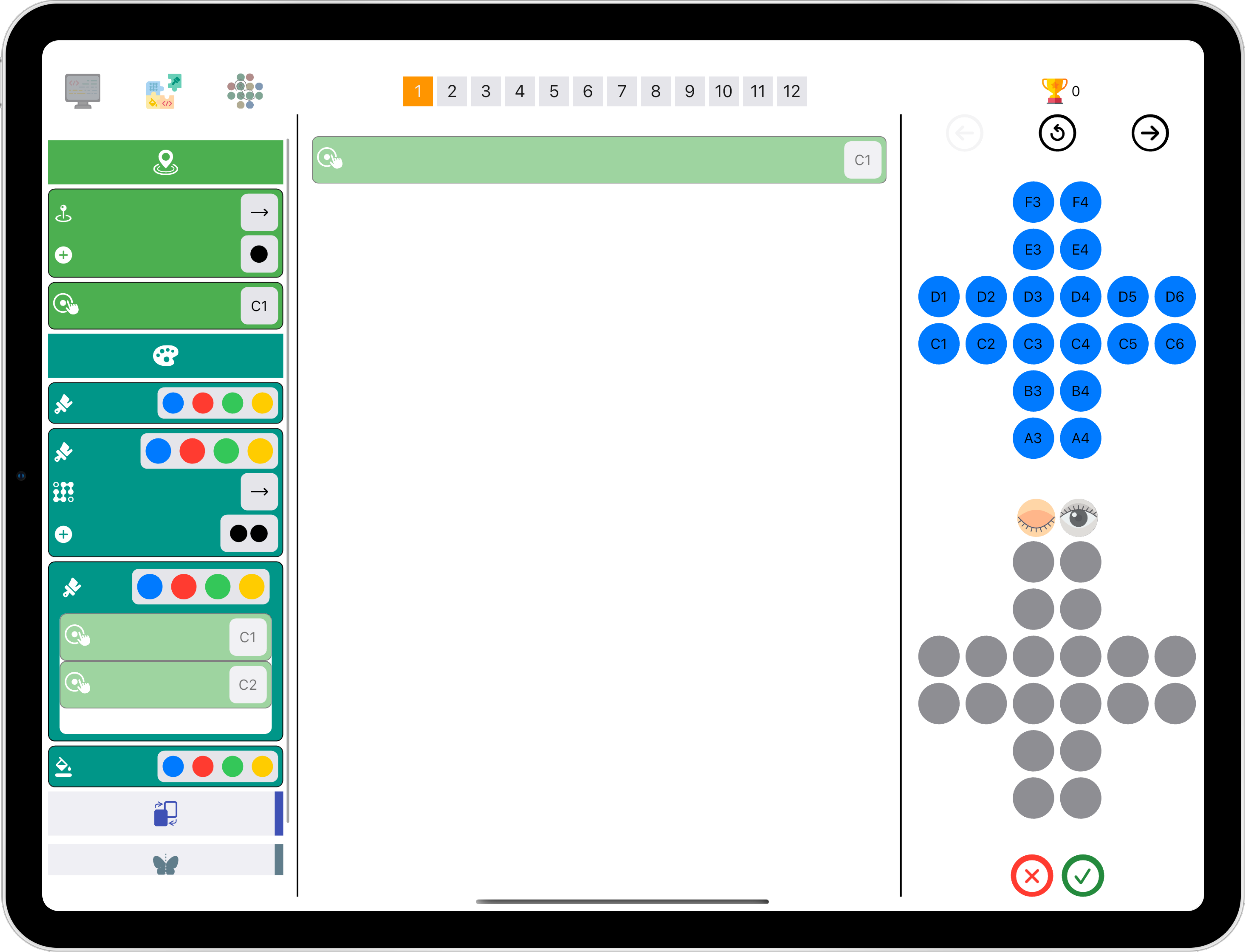}
    \captionof{figure}{\textbf{CAT visual programming interface (CAT-VPI) with symbolic commands.} }
    \label{fig:programming_symbols_final}
\end{appendixfig}

\begin{appendixfig}
    \centering
    \includegraphics[width=.75\textwidth]{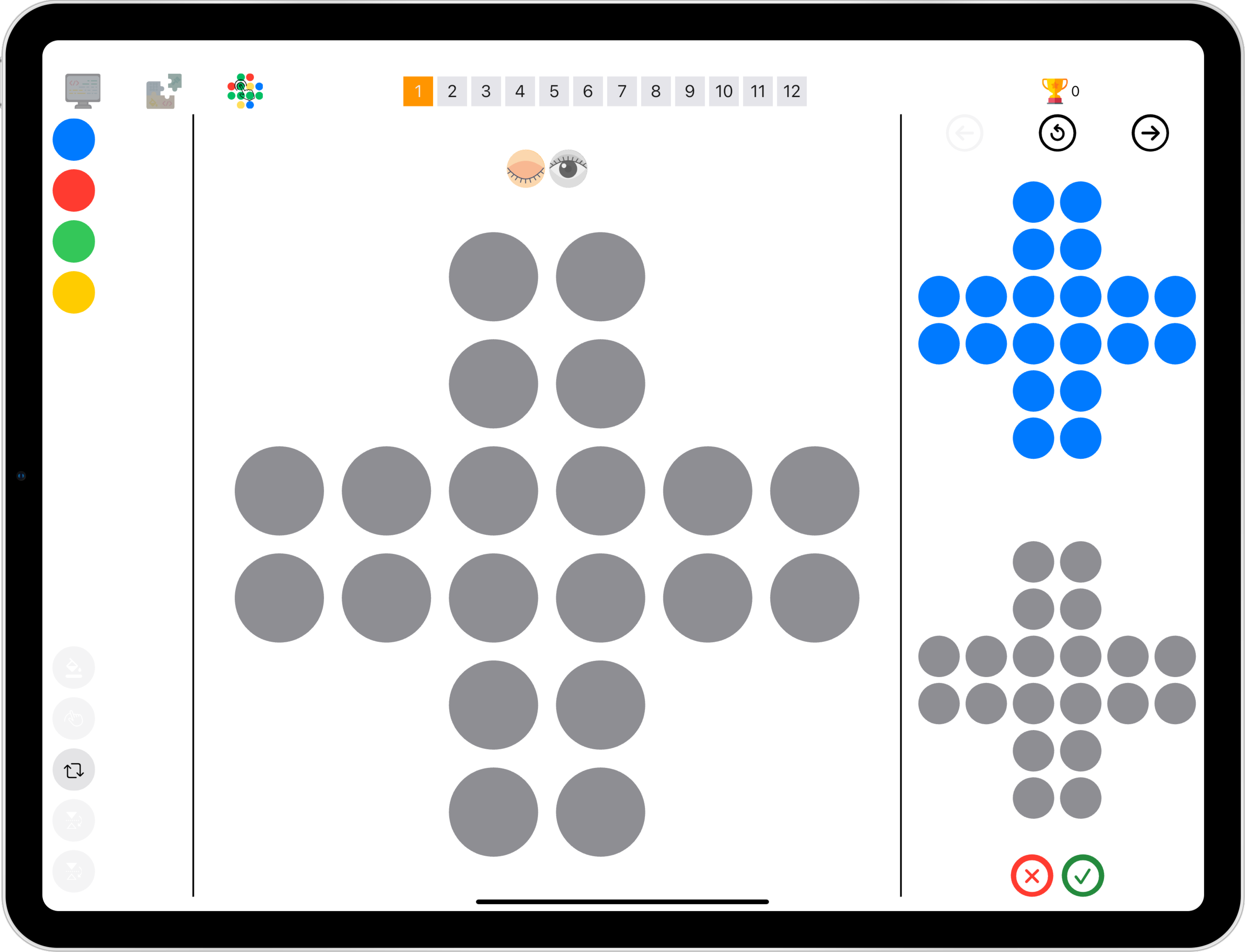}
        \captionof{figure}{\textbf{CAT gesture interface (CAT-GI).}}
        \label{fig:gestures_final}
\end{appendixfig}

\newpage
\mbox{~}
\newpage

\begin{appendixfig}
    \centering
    \includegraphics[width=.75\textwidth]{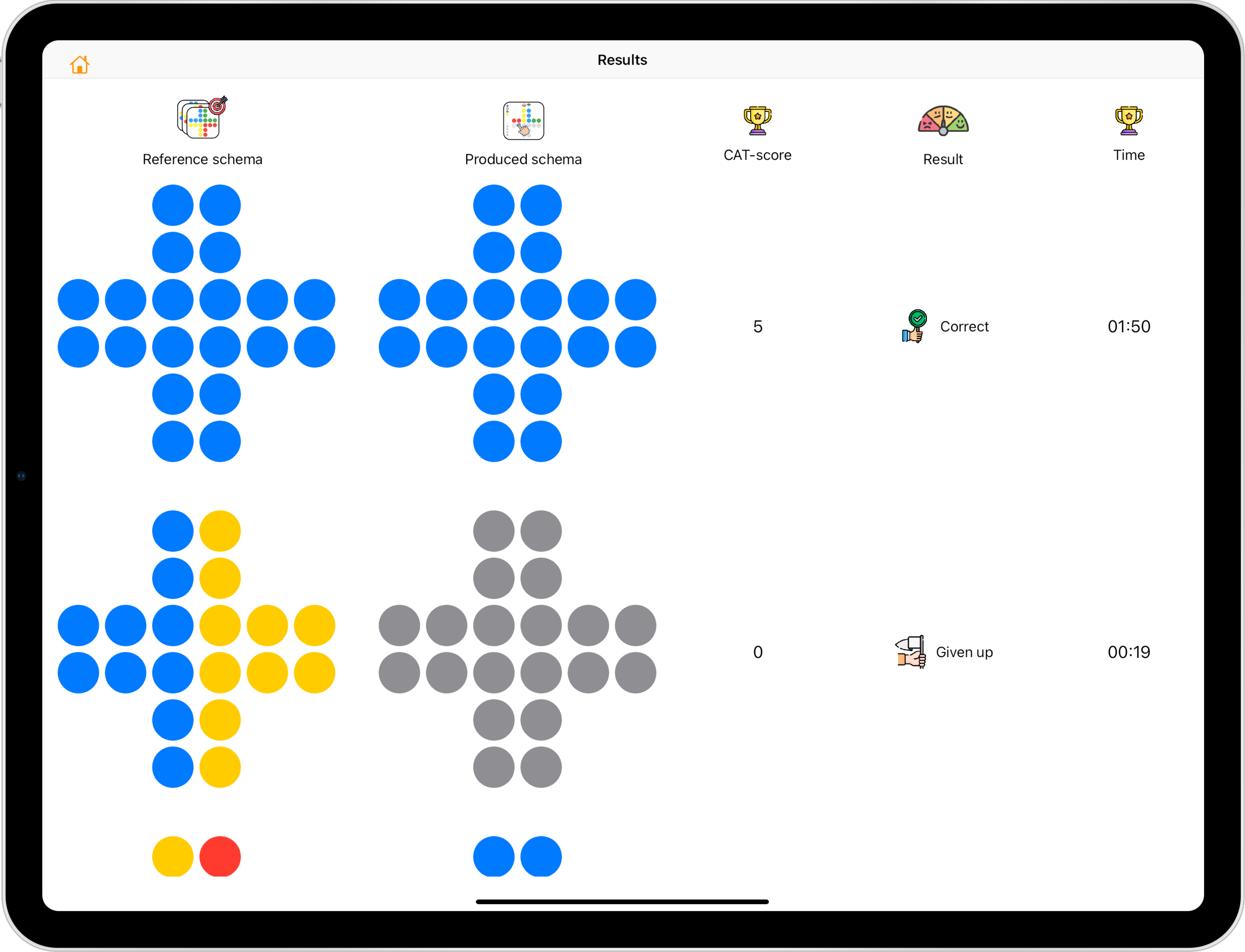}
        \captionof{figure}{\textbf{Results dashboard.} It comprehensively summarises pupils' performance across all schemas. 
        This dashboard includes a visual representation of reference schemas alongside those resulting from student instructions, the pupil's score, an indication of whether each schema was completed correctly, incorrectly, or skipped, and the time taken to complete the schema.}
        \label{fig:dashboard}
\end{appendixfig}

\begin{appendixfig}
    \centering
    \includegraphics[width=.75\textwidth]{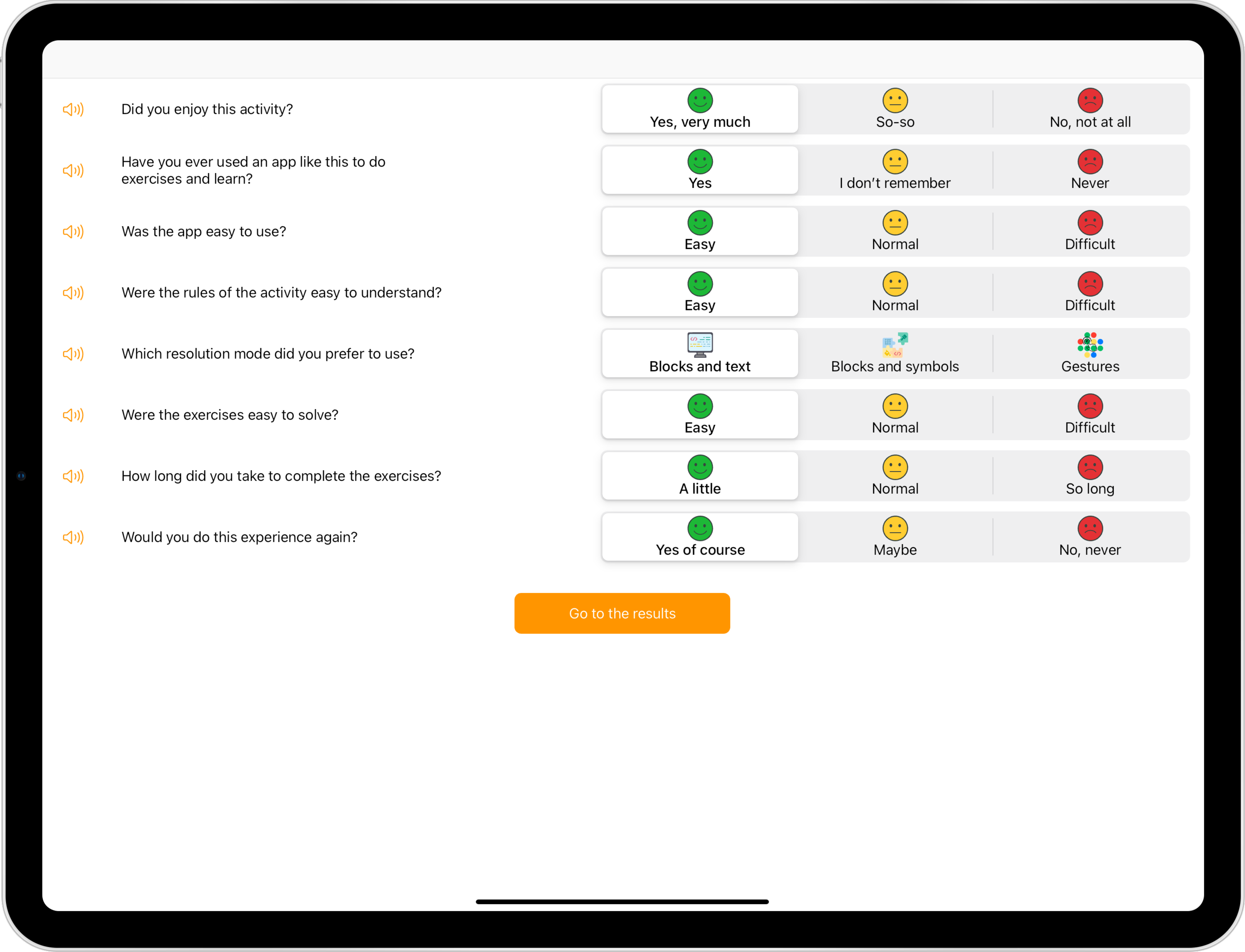}
        \captionof{figure}{\textbf{Pupil feedback survey.} The voice-assisted questions 
        evaluate user interactions with the app. Each question is accompanied by three distinct emoticon-style response options: a contented smiling face, a neutral face, and a discontented frowning face. 
        A concluding button invites users to view aggregated results.}
        \label{fig:survey}
\end{appendixfig}

\newpage
\mbox{~}
\newpage

\newpage

% \clearpage
% Add individuals who provided help during the research

\newpage
% References
% \bibliographystyle{model5-names}
% \bibliographystyle{cas-model2-names} % Loading bibliography style file
% \bibliographystyle{apacite}
% \bibliographystyle{acm}
\bibliographystyle{abbrvnat} 
\bibliography{bibliography} % Loading bibliography database

% \bibliographystyleWeb{model5-names}
% \bibliographyWeb{web}

% \newpage
% \dotfill End of the article \dotfill 

% \subsubsection*{Counts of words} 
% \wordcount
%TC:endignore

\end{document}